\documentclass{article}

\usepackage[numbers]{natbib}

\usepackage[preprint]{neurips_2025}

\usepackage[utf8]{inputenc} %
\usepackage[T1]{fontenc}    %
\usepackage{hyperref}       %
\usepackage{url}            %
\usepackage{booktabs}       %
\usepackage{amsfonts}       %
\usepackage{nicefrac}       %
\usepackage{microtype}      %
\usepackage{xcolor}         %

\usepackage{soul}
\usepackage{url}
\usepackage{caption}
\usepackage{graphicx}
\usepackage{amsthm,amsmath,amssymb,mathtools}
\usepackage{diagbox}

\usepackage{newfloat}
\usepackage{listings}
\usepackage{subfigure}
\usepackage{hhline,booktabs,colortbl,multirow,tabularx,diagbox,tablefootnote,threeparttable} %
\usepackage{multirow}
\usepackage{algorithm2e}
\usepackage{prettyref}
\usepackage{xcolor}
\usepackage[shortlabels]{enumitem}
\usepackage{float}
\usepackage{wrapfig}
\usepackage{makecell}
\usepackage{pifont}        %
\usepackage{bbding}        %
\usepackage[most]{tcolorbox}
\usepackage{fontawesome5}        %
\usepackage{minitoc}

\DeclareTCBListing{mylisting}{}{}

\definecolor{mybackground}{RGB}{241, 238, 252}
\definecolor{mybackground2}{RGB}{239, 246, 246}
\definecolor{mybackground3}{RGB}{236, 242, 254}

\definecolor{mycyan}{gray}{.7}

\newtcolorbox{quotebox}{
    colback=lightpurple,
    colframe=black!75,
    boxrule=0pt,
    top=5pt,
    bottom=5pt,
    left=8pt,
    right=8pt,
    arc=8pt,
    boxsep=0pt,
    toptitle=2pt,
    bottomtitle=2pt,
    fonttitle=\bfseries,
}
\definecolor{lightpurple}{rgb}{0.69, 0.61, 0.85} %

\newrefformat{fig}{Figure~\ref{#1}}
\newrefformat{tab}{Table~\ref{#1}}
\newrefformat{sec}{Section~\ref{#1}}
\newrefformat{app}{Appendix~\ref{#1}}
\newrefformat{alg}{Algorithm~\ref{#1}}
\newrefformat{property}{Property~\ref{#1}}
\newrefformat{theorem}{Theorem~\ref{#1}}
\newrefformat{corollary}{Corollary~\ref{#1}}
\newrefformat{proposition}{Proposition~\ref{#1}}
\newrefformat{def}{Definition~\ref{#1}}
\newrefformat{eq}{equation~(\ref{#1})}
\newrefformat{app}{Appendix~\ref{#1}}
\newrefformat{code}{Code Snippet~\ref{#1}}
\newcommand{\pref}{\prettyref}

\def\our{\texttt{OmniGenome}}

\def\our{\texttt{OmniGenBench}}

\def\og{OmniGenome}

\def\insilico{\textit{in-silico}}
\def\invivo{\textit{in-vivo}}

\title{\our: A Modular Platform for Reproducible Genomic Foundation Models Benchmarking}

\author{Heng Yang$^1$, Jack Cole$^1$, Yuan Li$^2$, Renzhi Chen$^3$, Geyong Min$^1$, Ke Li$^1$ \\
$^1$Department of Computer Science, University of Exeter, Exeter, UK\\
$^2$National University of Defense Technology, Changsha, China \\
$^3$Qiyuan Lab, Beijing, China \\
\texttt{\{hy345,j.cole1,g.min,k.li\}@exeter.ac.uk} \\
\texttt{\{liyuan22\}@nudt.edu.cn}
\texttt{\{chengrenzhi1989\}@gmail.com}
}

\begin{document}

\maketitle

\begin{abstract}
The code of nature, hidden in DNA and RNA genomes since the evolution of living systems, holds immense potential for impacting humans and ecosystems through genome modeling. Genomic Foundation Models (GFMs) have been proposed for genomic modeling as they hold transformative promise in genome deciphering. 
As GFMs scales up and reshape the landscape of AI-driven genomics, the field faces a growing need for rigorous, reproducible evaluation. We introduce \our{}, a modular benchmarking platform designed to unify data, model, benchmark, and interpretability layers across GFMs. \our{} enables standardized, one-command evaluation of any GFM on five benchmark suites, with seamless integration of $31+$ open-source models. Through automated pipelines and community-extensible features, the platform addresses reproducibility gaps in data transparency, model interoperability, benchmark fragmentation, and black-box interpretability. \our{} aspires to be a foundational infrastructure for reproducible genomic AI research, accelerating trustworthy discovery and collaborative innovation in the era of genomic-scale modeling.
\end{abstract}

\section{Introduction}
\label{sec:intro}

All living systems encode information in their genomic sequences. Just as Watson and Crick’s helical structure cracked the DNA code~\citep{WatsonC53}, AI is now learning the chemical language of life to decode the regulatory grammar hidden within  genomes~\cite{Tang23}. Foundation models (FMs), also known as large language models (LLMs), are generative AI systems that understand and generate human language. Prominent examples like \href{https://chatgpt.com/}{OpenAI's ChatGPT} and \href{https://gemini.google.com/}{Google's Gemini} have already transformed sectors like education~\citep{Nature23education,Adeshola24opportunities}, entertainment~\citep{Lv23generative,Chamola24beyond} and business~\citep{Edelman23Business,George23review}. Analogous to LLMs learning text, genomic FMs (GFMs) learn biological insights directly from extensive genomic data~\citep{Consens25transformers}. The potential impact of GFMs in life sciences is substantial. McKinsey estimates that generative AI could unlock $\$60$–$\$110$bn in annual economic value within life science industries\footnote{\url{https://tinyurl.com/4yhwhzfd}}. Echoing the latest PCAST report in the USA\footnote{\url{https://tinyurl.com/2dfj68ne}}, FMs offer unprecedented potential to drive a new era of digital life sciences, supercharging scientific progress, from drug discovery to super-personalized medicine.

\vspace{-.5em}
\paragraph{Reproducibility Crisis}
FMs have sparked immense enthusiasm for their potential in genomics and the broader life sciences. However, their real-world uptake remains surprisingly slow compared to their rapid adoption in fields like natural language processing and computer vision. This slower uptake primarily arises from four critical challenges, each directly tied to \textit{reproducibility}.
\begin{enumerate}[leftmargin=.7cm]
	\item\textbf{Data availability}
A major barrier to reproducibility in GFM research is limited access to datasets. \ding{238} \underline{\textit{Lack of shared training data:}} Many open-sourced GFMs do not release the exact datasets used for pre-training or fine-tuning. In our analysis of $31$ GFMs (as shown in \pref{app:gfm_details}) integrated into \our, only $16$ of them are provided with the curated datasets. This lack of transparency directly restricts reproducibility and meaningful model comparisons. \ding{238} \underline{\textit{Absence of standardized task data:}} Genomics modeling lacks standardized formats and central repositories for task-specific data. Consequently, researchers often rely on ad-hoc preprocessing. This results in opaque benchmarks that others cannot reliably replicate. \ding{238} \underline{\textit{Limited scope of existing benchmark datasets:}} Existing genomic benchmarks typically focus on narrow tasks (e.g., single data modality). Researchers thus repeatedly assemble custom datasets, causing fragmented and non-standardized evaluations.

	\item\textbf{Model accessibility} Inconsistencies in model implementations hinder widespread adoption and benchmarking. \ding{238} \underline{\textit{Inconsistent model implementations:}} GFMs are often released with highly customized codebases, architectures, and tokenization schemes. Such variability creates interoperability barriers. For example, some GFMs use 6-mer tokenization~\citep{Ji21dnabert}, while others use byte-pair encoding~\citep{Zhou23DNABERT2}, resulting in incompatible formats. \ding{238} \underline{\textit{Lack of standardized interfaces:}} Many GFMs are unavailable through common model repositories or standardized programming interfaces. Thus, integrating new models requires substantial model-specific technical efforts. For instance, the recent Plant Genome Benchmark (PGB)~\citep{Mendoza23}, despite its valuable datasets, lacked generic interfaces for incorporating new GFMs, limiting its utility for comparative evaluation.

	\item\textbf{Benchmarking standardization} Disjointed evaluation practices prevent meaningful comparisons across models. \ding{238} \underline{\textit{Inconsistent evaluation practices:}} Different groups evaluate GFMs using distinct tasks and metrics. Such variation prevents fair model comparisons and complicates verification of reported results. \ding{238} \underline{\textit{Barriers to universal benchmarking:}} Evaluating diverse GFMs across multiple genomic tasks remains challenging due to varying data modalities (e.g., DNA vs. RNA) and incompatible model implementations. As a result, universally benchmarking \lq\textit{any model on any genomic task}\rq\ remains impractical under current fragmented approaches. \ding{238} \underline{\textit{Limitations of existing benchmarks:}} Current benchmark suites like Genomic Benchmarks (GB)~\citep{Grevsova23} and GUE~\citep{Zhou23DNABERT2} narrowly focus on specific DNA classification tasks. Others benchmarks focus exclusively on niche domains (e.g., plant-specific genomics~\citep{Mendoza23}). These benchmarks often operate independently, complicating evaluations across different suites. Even comprehensive benchmarks like BEACON~\citep{Ren24BEACON} for RNA has limited portability due to specialized environments. Without standardization, researchers must repeatedly reconstruct evaluation pipelines. %

	\item\textbf{Interpretability} The opaque nature of GFMs limits their practical usability in biology and medicine. \ding{238} \underline{\textit{Lack of mechanistic insight:}} GFMs typically generate predictions without clear biological explanations. This opacity prevents reliable biological validation, causing skepticism among clinicians and biologists. \ding{238} \underline{\textit{Ad-hoc interpretability practices:}} Interpretability analyses, such as motif discovery or feature attribution, usually rely on manually designed, post-hoc methods. These approaches vary widely across studies, limiting consistency and reproducibility. \ding{238} \underline{\textit{Inconsistent scientific conclusions:}} Such inconsistent practices often yield divergent interpretations of model capabilities, even for identical tasks. 
  The absence of routine, reproducible interpretability methods not only reduces user trust but complicates validation of biologically meaningful insights (e.g., detecting known regulatory motifs).
\end{enumerate}

\vspace{-.5em}
\paragraph{Our Solution \our}
To explicitly address this reproducibility crisis, we introduce \our, a unified and modular benchmarking platform. \our\ consists of four core modules, each specifically designed to address one of the critical barriers identified above.
\begin{itemize}[leftmargin=.7cm]
    \item \textbf{Data module:} \our\ provides centralized access to $123$ carefully curated genomic datasets, accompanied by clear documentation and standardized formats. It simplifies data reuse across pre-training and fine-tuning tasks and ensures consistency across different studies. This module also facilitates easy community sharing of genomic datasets.    

    \item \textbf{Model module:} The platform hosts a standardized model hub featuring unified wrappers and application programming interfaces (APIs). This hub integrates $31$ GFMs to date and simplifies the integration of diverse GFMs by eliminating interoperability barriers from customized codebases, varied architectures, or incompatible tokenization methods. As a result, researchers and practitioners can significantly reduce model integration efforts.
    
    \item \textbf{Benchmark module:} \our\ offers a comprehensive, automated benchmarking suite that covers diverse genomic tasks across multiple modalities (DNA and RNA). It includes $123+$ datasets with $58+$ metrics for integrated GFMs. Its extensible design and consistent evaluation protocols explicitly enable researchers to benchmark \lq\textit{any model on any genomic task}\rq\ within a unified framework. This approach ensures fair, transparent, and verifiable performance comparisons.
    
    \item \textbf{Interpretability module:} Recognizing the need to \lq\textit{open the black box}\rq, \our\ integrates three standardized interpretability tools into its evaluation pipelines. Researchers can routinely conduct reproducible analyses such as motif discovery and feature attribution mapping. This integration provides clear, mechanistic insights into GFM predictions, significantly enhancing model transparency and trustworthiness.
\end{itemize}
Through these interconnected components, \our\ aims to serve as a foundational platform that fosters a reproducible, transparent, and collaborative ecosystem, which in term accelerating meaningful progress in GFM research, scientific discovery, and beyond.

\section{System Design of \our}
\label{sec:framework}

\begin{figure*}[t!]
    \centering
    \includegraphics[width=\linewidth]{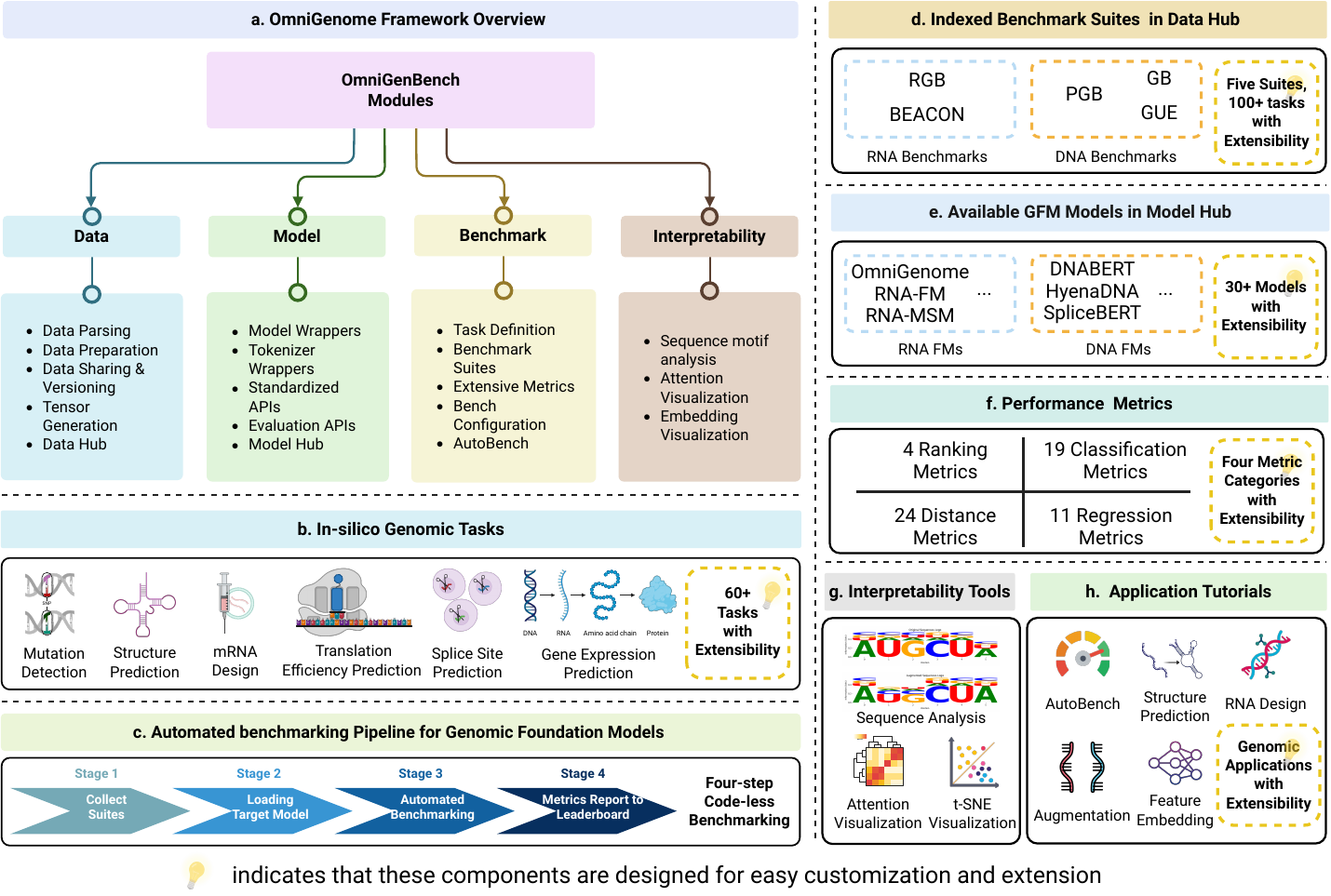}
    \caption{Overview of the \our\ framework.
    $\textbf{a)}$~\our\ consists of four core modules covering data, model, benchmark, and interpretability aspects.
    $\textbf{b)}$~The current release includes $60+$ \insilico{} genomic tasks covering diverse biological processes.
    $\textbf{c)}$~A four-stage code-less benchmarking pipeline that can automate the end-to-end evaluation.
    $\textbf{d)}$~Five benchmark suites (containing $123+$ datasets) are indexed in the \texttt{Data Hub}.
    $\textbf{e)}$~\texttt{Model Hub} hosts $31+$ GFMs, supporting simple deployment, to streamline applications for hosted GFMs from usually several weeks to one day.
    $\textbf{f)}$~A library of four different types of $58+$ evaluation metrics covering ranking, classification, regression, and distance.
    $\textbf{g)}$~Interpretability tools such as sequence-level motif analysis and embedding space analysis.
    $\textbf{h)}$~Eight pre-compiled pedagogical tutorials with specific applications (\pref{app:tutorials}). }
    \label{fig:framework}
    \vspace{-15pt}
\end{figure*}

\pref{fig:framework}.\textbf{a} gives the hierarchical structure of \our\ system. Our framework comprises several clearly defined modules, each explicitly linked to one of the reproducibility challenges identified in~\pref{sec:intro}. \our\ is designed as both a \textbf{ready-to-use} and \textbf{ready-to-expand} platform. As a \textit{ready-to-use} system, \our\ enables users to apply existing GFMs directly to specific biological tasks, such as predicting translation efficiency or performing sequence generation. As a \textit{ready-to-expand} system, \our\ provides clear APIs and comprehensive tutorials. These resources allow AI researchers, domain scientists, and stakeholders of varying expertise levels to: $i)$ upload their own datasets, which are automatically standardized into formats compatible with FM training, and $ii)$ easily customize or build innovative FMs guided by robust documentation and standardized protocols. To illustrate specific implementation details, we provide an auxiliary website showcasing relevant code snippets. In addition, we offer pedagogical tutorials in~\pref{app:tutorials}, helping users practically understand and effectively interact with the platform's core functionalities. The following paragraphs introduce the functional design of each module step by step. To elaborate on the module implementation details, we introduce the inputs, outputs and description for each module with the source code examples at this \href{https://github.com/COLA-Laboratory/OmniGenBench/blob/master/ModuleAppendix}{online page}\footnote{\url{https://github.com/COLA-Laboratory/OmniGenBench/blob/master/ModuleAppendix}}. 

\subsection{Data Module}
This module is engineered with four key functionalities to address the data availability challenge. \ding{46} \underline{\textit{Standardized data parsing:}} \our\ provides \texttt{data parsing} tools to convert genomic data from diverse formats (e.g., FASTA, JSON) into clearly documented and standardized datasets, each accompanied by detailed metadata. \ding{46} \underline{\textit{Flexible data preparation:}} Researchers can easily utilize built-in capabilities for dataset preparation. These include \texttt{adaptive sequence manipulation} (truncation and padding), \texttt{instance filtering} via established tools like CD-HIT-EST\footnote{\url{https://bioinformatics.org/cd-hit/}} to prevent label leakage in RNA-structure tasks, and \texttt{sequence augmentation} techniques~\citep{DevlinCLT19}. These tools simplify the generation of researcher-curated datasets for subsequent modeling. \ding{46} \underline{\textit{Community data sharing \& versioning:}} Once curated, datasets can be versioned and shared within the centralized \our\ \texttt{Data Hub} (see~\pref{fig:framework}.\textbf{d}). This capability directly addresses the current lack of accessible genomic training datasets and promotes broader coverage beyond existing specialized benchmarks. \ding{46} \underline{\textit{Standardized tensor generation:}} Finally, we transform diverse genomic inputs, from existing benchmarks or community contributions, into standardized \texttt{tensor data}. These tensor datasets are directly compatible with downstream GFM modeling, significantly simplifying experimentation workflows.
\begin{quotebox}
    \noindent
    \faQuoteRight\, \textit{{By standardizing data processing, simplifying dataset creation, and encouraging community sharing, this module resolves the field's reliance on fragmented, ad-hoc preprocessing methods.}}
\end{quotebox}
\vspace{-.7em}

\subsection{Model Module}
It addresses the challenges of model accessibility and usability by standardization and simplified integration. \ding{46} \underline{\textit{Unified model wrappers \& wrappers:}} \our\ offers a \texttt{base model template} featuring unified APIs to address interoperability barriers from diverse GFM architectures (e.g., transformer~\citep{VaswaniSPUJGKP17,Lin23esm2}, Hyena~\citep{PoliMNFDBBER23,Nguyen23,Nguyen24}, Mamba~\citep{GuD23mamba,SchiffKGDGK24}) and customized tokenization methods. The platform employs standardized \texttt{model wrappers} and \texttt{tokenizer wrappers}. These wrappers abstract away specific architectural details, providing consistent input-output handling whether a GFM uses $k$-mer or byte-pair encoding. \ding{46} \underline{\textit{Standardized APIs for core operations:}} The \texttt{model template} provides standardized, universal APIs covering critical operations. It includes three clearly defined training options: $i)$ a \texttt{basic native trainer}, $ii)$ a \texttt{scalable Hugging Face trainer}, and $iii)$ a \texttt{distributed accelerate trainer}. We also standardize \texttt{evaluation} and \texttt{inference} APIs, which significantly reduce the technical effort typically needed to integrate and use diverse GFMs. \ding{46} \underline{\textit{Centralized \& extensible model Hub:}} Built on Hugging Face infrastructure, the \our\ \texttt{Model Hub} (Figure~\ref{fig:framework}.\textbf{e}) hosts an extensive collection of $31+$ pre-trained GFMs. This includes specialized DNA models (e.g., DNABERT~\cite{Ji21dnabert,Zhou23DNABERT2}, HyenaDNA~\cite{Nguyen23}) and RNA models (e.g., \og{}~\cite{YangL24omnigenome}, RNA-FM~\cite{Chen22}). This centralized hub ensures models are readily accessible, with clearly documented interfaces. Additionally, the platform actively supports community contributions. Researchers can easily upload and share new GFMs and datasets (see Appendices~\ref{app:benchmark_details} and~\ref{app:gfm_details}), further expanding the ecosystem and promoting collaborative advancement.
\begin{quotebox}
    \noindent
    \faQuoteRight\, \textit{{By providing standardized interfaces, unified wrappers, and a centralized hub, this module directly resolves key accessibility barriers, greatly simplifying the integration, comparison, and use of diverse genomic foundation models.}}
\end{quotebox}
\vspace{-.7em}

\subsection{Benchmark Module}
This module addresses inconsistent evaluation practices and the limitations of existing disjointed benchmarks by four clearly defined functionalities. \ding{46} \underline{\textit{Curated benchmark suites \& Data Hub integration:}} The \our\ \texttt{Data Hub} (\pref{fig:framework}.\textbf{d}) provides unified access to five systematically curated benchmark suites, covering $123+$ diverse genomic dataset. These include RNA-focused benchmarks (RGB~\cite{YangL24omnigenome}, BEACON~\cite{Ren24BEACON}) and DNA-focused benchmarks (PGB~\cite{Mendoza23}, GB~\cite{Grevsova23}, GUE~\cite{Zhou23DNABERT2}). This comprehensive, centralized collection, detailed further in Appendix~\ref{app:benchmark_details}, facilitates broad assessments of GFM generalization beyond narrowly defined tasks. The \our\ \texttt{Data Hub} is designed explicitly for ongoing community-driven expansion. \ding{46} \underline{\textit{Flexible task definition \& configuration:}} At the core of \our\ is a robust \texttt{task module}. It enables clear, structured definition and configuration of genomic tasks. Researchers can explicitly specify data preprocessing steps, model architectures, loss functions, and evaluation metrics. Currently, the module includes $60+$ pre-defined genomic tasks (\pref{fig:framework}.\textbf{b}), such as RNA secondary structure prediction~\citep{TanFSM17,DanaeeRWDHH18,Mathews19,Kalvari21} and mRNA design~\cite{Corbett20}. The module also supports the straightforward addition of new community-defined tasks, ensuring consistent evaluation setups across studies. \ding{46} \underline{\textit{Centralized \& extensible metric registry:}} To guarantee fair and consistent model assessments, \our\ includes a dedicated \texttt{metric module}. This module provides a comprehensive, task-agnostic registry of over $58+$ standardized metrics. Metrics span various categories, including classification and ranking (e.g., $F_1$, ROC-AUC), regression (e.g., RMSE, $R^2$), and distance or similarity measures (e.g., cosine similarity). These metrics are largely integrated from robust libraries such as \textsc{scikit-learn}\footnote{\url{https://scikit-learn.org/}}. Importantly, new or custom genomic metrics (e.g., structural alignment scores) can be easily integrated using task-level JSON configuration files, balancing standardization with flexibility. \ding{46} \underline{\textit{Automated \& reproducible evaluation workflow:}} Finally, the framework supports automated evaluation through streamlined task compilation, integrating data, model configurations, and evaluation metrics. The integrated \texttt{AutoBench} engine significantly simplifies benchmarking workflows. For example, evaluating GFMs like OmniGenome on benchmarks (e.g., RGB) can be executed in a single command, e.g., \texttt{\textcolor{teal}{[autobench --model OmniGenome-52M --benchmark RGB]}}. 
\begin{quotebox}
    \noindent
    \faQuoteRight\, \textit{{By providing integrated benchmarks, standardized task definitions, flexible metrics, and automated benchmark workflows, \our\ promotes fair comparisons, reliable result verification, and rigorous scientific inquiry within the GFM research community.}}
\end{quotebox}
\vspace{-.7em}

\subsection{Interpretability Module}

To address the transparency challenges associated with the \lq black-box\rq\ nature of GFMs, this module provides standardized workflows and integrated tools to systematically examine and validate GFM predictions through three main functionalities (\pref{fig:framework}.{g}). \ding{46} \underline{\textit{Built-in tools for mechanistic insight:}} \our\ includes built-in \texttt{interpretability tools}, eliminating the reliance on inconsistent and manually designed post-hoc analyses. Current integrated tools are broadly applicable across various GFMs and genomic tasks, including: $i)$ \texttt{sequence-level motif analysis} for identifying learned sequence patterns (demonstrated in~\pref{sec:seq_motif_preservation}); $ii)$ \texttt{embedding space analysis} for enabling visual exploration of how models represent genomic features (see Appendix~\ref{app:feature_embedding_analysis}); and $iii)$ \texttt{Attention map visualization} that reveals model focus and decision-making processes during sequence analysis (see \pref{app:attention_inspection}). Integrating these tools directly within the evaluation pipeline ensures their routine and consistent use. \ding{46} \underline{\textit{Promoting reproducible scientific conclusions:}} By standardizing interpretability analyses, this module ensures the reproducibility and comparability of the biological insights derived from GFMs. Utilizing clearly documented, common methodologies (see tutorials in~\pref{app:interpretability}), researchers can more reliably interpret and validate biological significance (e.g., detecting known regulatory motifs). This systematic approach reduces the risk of inconsistent conclusions associated with varied interpretability techniques. \ding{46} \underline{\textit{Extensibility for future methods:}} This module is designed for extensibility, allowing for the future integration of new and emerging interpretability techniques.
\begin{quotebox}
    \noindent
    \faQuoteRight\, \textit{{By embedding systematic and standardized interpretability analyses directly within the GFM evaluation process, \our\ enhances transparency, improves user trust, and facilitates robust, biologically meaningful insights from GFMs.}}
\end{quotebox}

\section{Overall Benchmark Results}
\label{sec:experiments}

This section presents pivotal performance of integrated GFMs benchmarked via \our{}. The evaluations span four major benchmark suites: the RNA Genomic Benchmark (RGB; see \pref{app:rgb}), the Plant Genomic Benchmark (PGB; \pref{app:pgb}), the Genomic Understanding Evaluation (GUE; \pref{app:gue}), and the Genomics Benchmark (GB; \pref{app:gb}). For fair comparisons, data splits and primary metrics follow their original publication settings or established best practices. \textbf{Detailed numerical results and in-depth task-specific discussions are provided in \pref{app:benchmark_details}.}

\subsection{State-of-the-Art (SoTA) Performance}
\label{sec:sota_performance_analysis}

\begin{wrapfigure}{r}{0.5\textwidth}
\centering
\includegraphics[width=\linewidth]{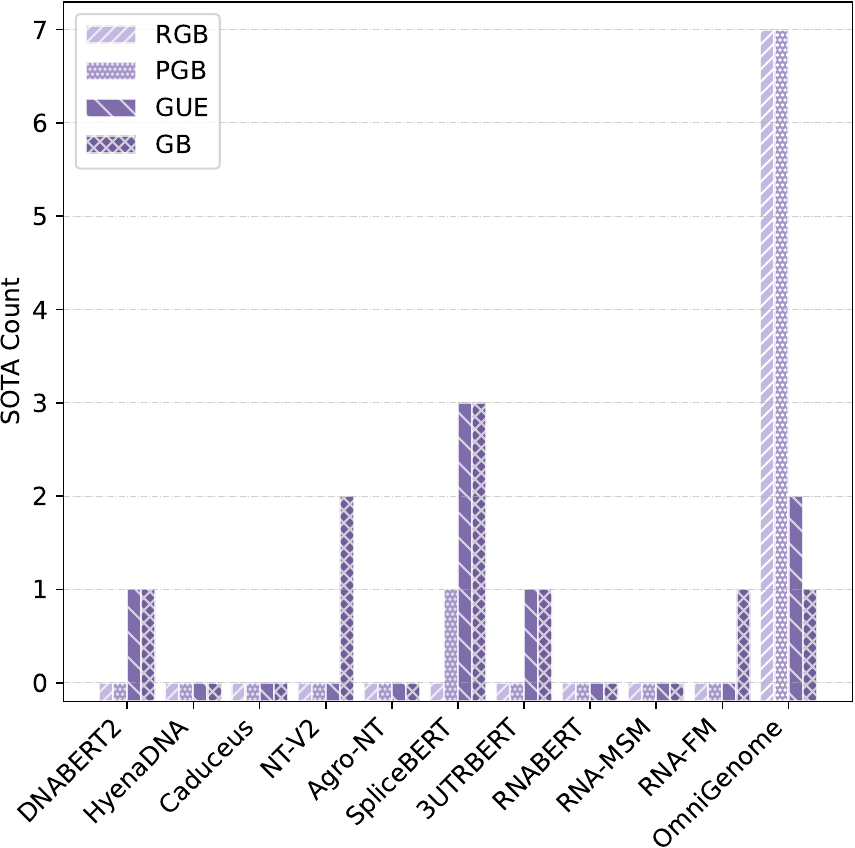}
\caption{State-of-the-Art (SoTA) achievements of public GFMs across tasks within the four primary benchmark suites.}
\label{fig:overall_performance}
\vspace{-10pt}
\end{wrapfigure}

We first quantify the SoTA achievements for 11 public GFMs across all evaluated tasks within the aforementioned benchmark suites. A model achieves SoTA by attaining the top performance on a task's primary metric in our evaluation. This SoTA count intuitively measures cross-scenario generalization and excellence.
\pref{fig:overall_performance} summarizes these SoTA counts. \og{} notably leads in SoTA achievements, particularly within the RGB and PGB suites. SpliceBERT~\cite{ChenZD23} and 3UTRBERT also secure several SoTAs, primarily in PGB, GUE, and GB. Conversely, models such as HyenaDNA, Caduceus~\cite{Schiff24caduceus}, Agro-NT~\cite{Mendoza23}, RNABERT~\cite{Sato2021rna}, RNA-MSM~\cite{ZhangLJGXLCSHXS24}, and RNA-FM~\cite{ChenZD23} did not achieve any SoTAs in our current evaluation setup.

\paragraph{Discussion.}
The distribution of SoTA achievements in \pref{fig:overall_performance} offers initial insights into GFM capabilities. \og{}'s superiority, especially in RGB, suggests its pre-training techniques are well-suited for RNA tasks and generalize effectively. The absence of SoTAs for models like HyenaDNA, Caduceus, Agro-NT, RNABERT, RNA-MSM, and RNA-FM in this evaluation (while they may be strong in their original contexts) could indicate that their pre-training objectives (e.g., standard masked language modeling without genomics-specific knowledge) are less effective for the complex tasks in these suites (e.g., RNA secondary structure prediction). This observation may underscore the importance of GFMs incorporating more structure-aware~\citep{YangL24omnigenome}, multi-species pre-training strategies for specialized genomic tasks. However, SoTA counts are aggregate measures and do not capture task-level nuances or the specific strengths of individual GFMs, which are further explored in \pref{app:benchmark_details}.

\subsection{Rank-Based Performance}
\label{sec:rank_based_performance}

\begin{figure}[t!]
\centering
\includegraphics[width=0.999\linewidth]{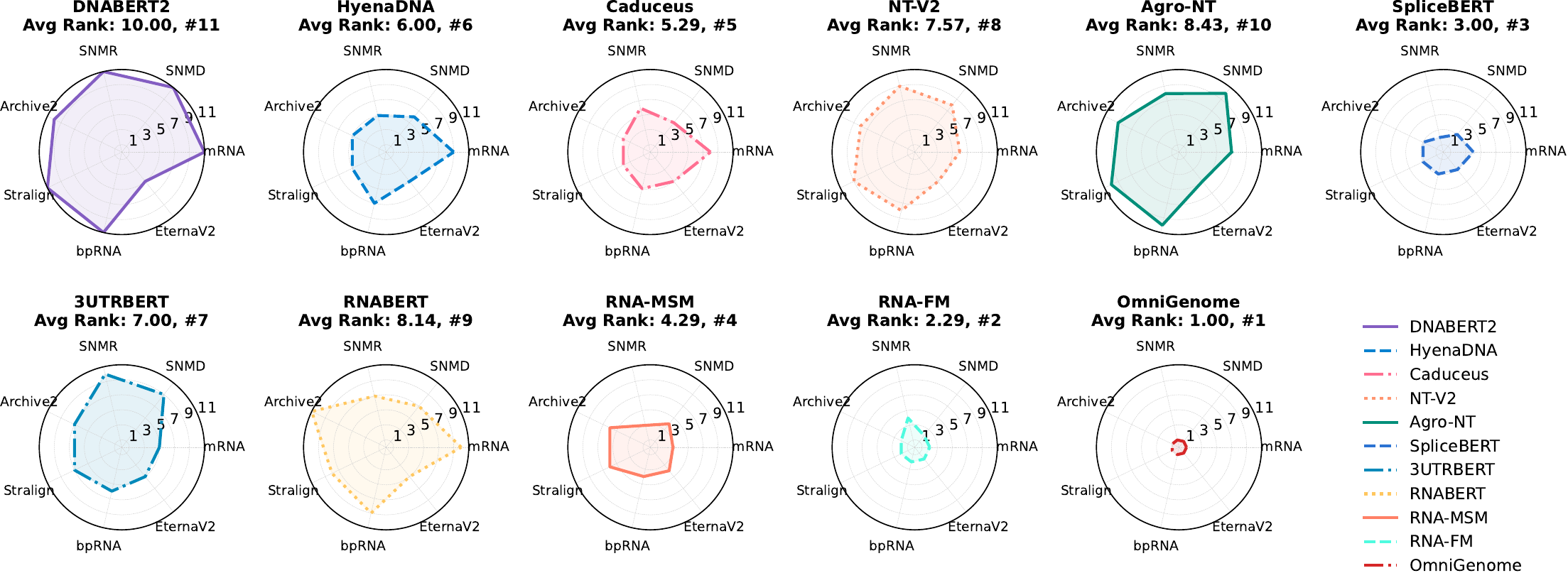}
\caption{Rank-based radar charts comparing eleven GFMs on the RGB suite. Each small plot represents a model, with axes corresponding to different RNA task categories. Lower ranks (closer to the center) indicate better performance. The average rank for each model on RGB is displayed above its plot.}
\label{fig:radar_rgb}
\vspace{-10pt}
\end{figure}

To provide a more nuanced view of where each GFM excels or struggles, we present rank-normalized radar charts. For each benchmark suite, an individual radar chart is generated for each GFM, where axes represent distinct task categories. A model's performance on each axis is its average rank (rank 1 = best). Smaller, more centrally-focused polygons indicate stronger, more balanced performance. The average rank of each model within a suite is also noted. Due to space constraints, we illustrate the RGB radar charts (\pref{fig:radar_rgb}) in this section; comprehensive radar charts for all suites are in \pref{app:rank_based_performance}.

\paragraph{Discussion.}
Focusing on the RGB suite (\pref{fig:radar_rgb}), \og{} (avg. rank 1.00) demonstrates clear dominance, its radar polygon nearly collapsing to the center. RNA-FM (avg. rank 2.29) follows as a strong runner-up, with SpliceBERT~\cite{ChenZD23} (avg. rank 3.00) also showing competitive performance. DNA-centric models like DNABERT-2 (avg. rank 10.00) exhibit significantly larger polygons, underscoring their weaker RNA task transferability. Structure-aware models, particularly \og{} and RNA-FM, excel in structure-related task categories such as secondary structure prediction (e.g., the SSP task).

Broader observations from the comprehensive radar charts across all four benchmark suites (see \pref{fig:radar_combined}) reveal several key trends regarding GFM behaviours:
\begin{itemize}
    \item \og{} leverages RNA structure-aware pre-training, significantly boosting its performance on structure modeling tasks like secondary structure prediction and RNA design. Its pre-training on multi-species plant genomes also translates to strong performance in PGB on tasks such as PolyA site prediction. Notably, \og{} demonstrates robust generalization capabilities even on out-of-domain benchmarks (GUE and GB), where pre-training data has less direct overlap with downstream tasks. This suggests that incorporating structural information into pre-training is a promising direction for future GFM research.
    \item RNA-FM, another RNA-focused model, also achieves commendable results on the RGB suite, particularly in structure prediction tasks. However, its generalization ability appears less pronounced, with suboptimal performance on out-of-domain benchmarks like PGB. Its reliance on structural information for RNA also means it is not directly applicable to DNA sequence modeling using the same framework.
    \item SpliceBERT, a DNA-centric GFM, excels in DNA-related tasks such as DNA sequence modeling within GUE and GB. However, its performance can be inconsistent across different tasks, possibly due to the specificity of its training data, indicating a need for broader pre-training data for more robust, generalizable DNA models.
    \item Models employing single nucleotide tokenization (SNT) generally exhibit strong performance across both fine-grained RNA modeling and broader DNA sequence tasks, showcasing a degree of flexibility. However, SNT can lead to longer effective sequence lengths, potentially increasing computational demands and impacting modeling efficiency for very long sequences.
    \item Generative models like HyenaDNA and Caduceus generally show modest performance across most benchmarks in this evaluation. This may reflect the current developmental stage of generative GFMs, whose sequence understanding capabilities may not yet match those of discriminative models optimized for specific tasks. We anticipate that continued advancements in generative GFM architectures will lead to improved performance in precise sequence understanding and generation.
\end{itemize}

In summary, these rank-based visualizations effectively highlight model-specific strengths and weaknesses across diverse genomic task categories beyond what aggregate SoTA scores reveal. This detailed insight is crucial for informed GFM selection for specific downstream applications and for guiding future GFM development strategies, including choices regarding pre-training data, model architecture, and tokenization schemes.

\section{Interpretability Analysis of Genomic Foundation Models}
\label{sec:interpretability_main} %

While quantitative performance metrics are essential for evaluating GFMs, they often fall short of addressing critical concerns regarding the trustworthiness and biological relevance of the model predictions. Therefore, this section presents an interpretability case study focused on sequence motif preservation to assess the fidelity of GFMs when used for sequence augmentation. Further interpretability case studies, including feature embedding analysis and attention mechanism inspection, are detailed in \pref{app:interpretability}.

\subsection{Sequence Motif Preservation in GFM-based Augmentation}
\label{sec:seq_motif_preservation}

\paragraph{Motivation.}
A critical question for the practical application of GFMs in sequence analysis is their ability to preserve evolutionarily conserved contexts encoded within biological sequences. Wet-lab scientists often express skepticism regarding whether sequences generated or augmented by these models authentically reflect native sequence characteristics, such as those captured in a Multiple Sequence Alignment (MSA). This study investigates whether GFMs, when used for sequence augmentation, can faithfully reproduce the position-specific motifs characteristic of conserved RNA families.

\begin{figure}[t!]
    \centering
    \begin{minipage}[t]{0.48\textwidth}
        \centering
        \includegraphics[width=\linewidth]{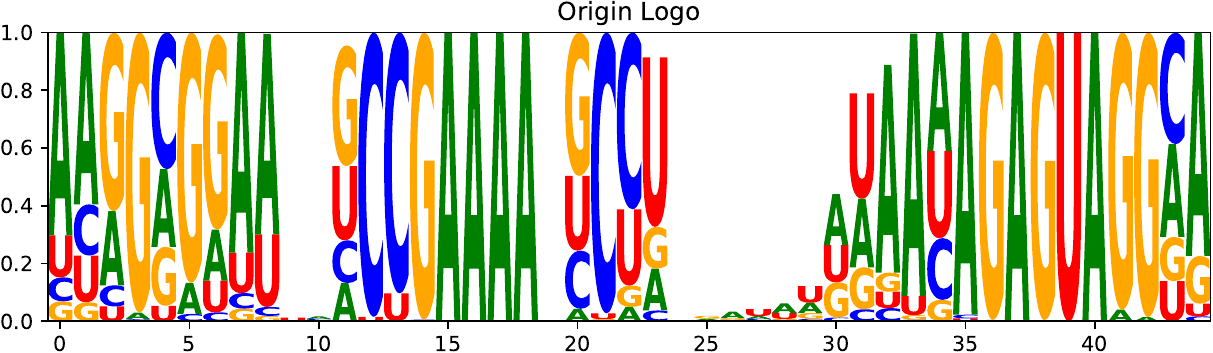}
        \includegraphics[width=\linewidth]{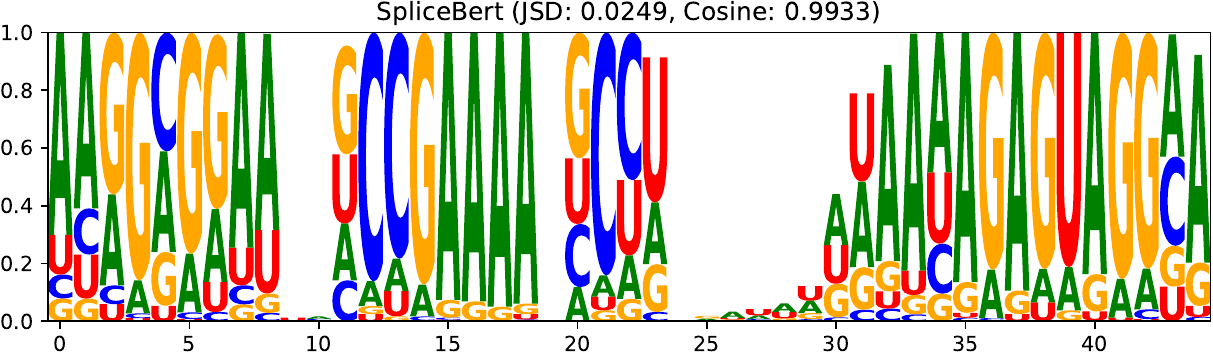}
        \includegraphics[width=\linewidth]{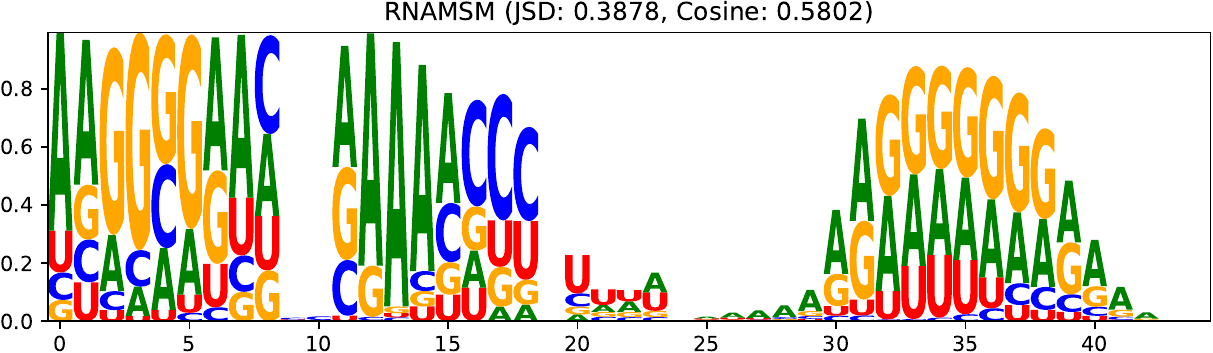}
        \includegraphics[width=\linewidth]{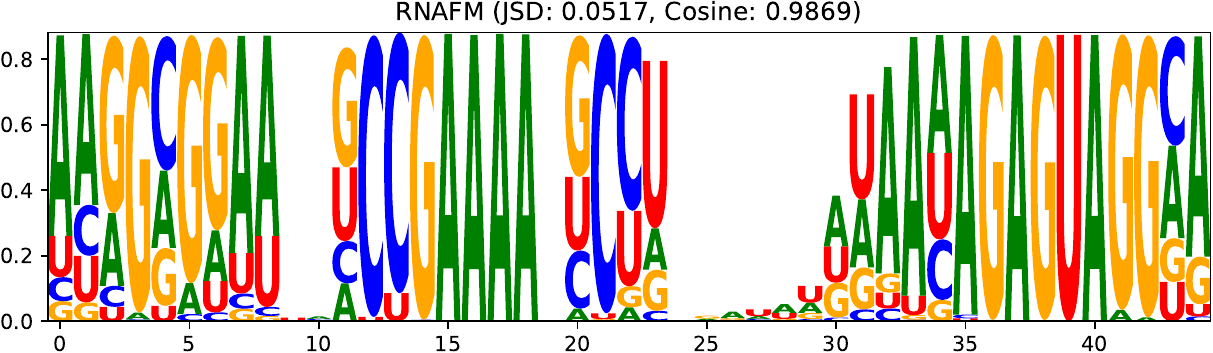}
        \includegraphics[width=\linewidth]{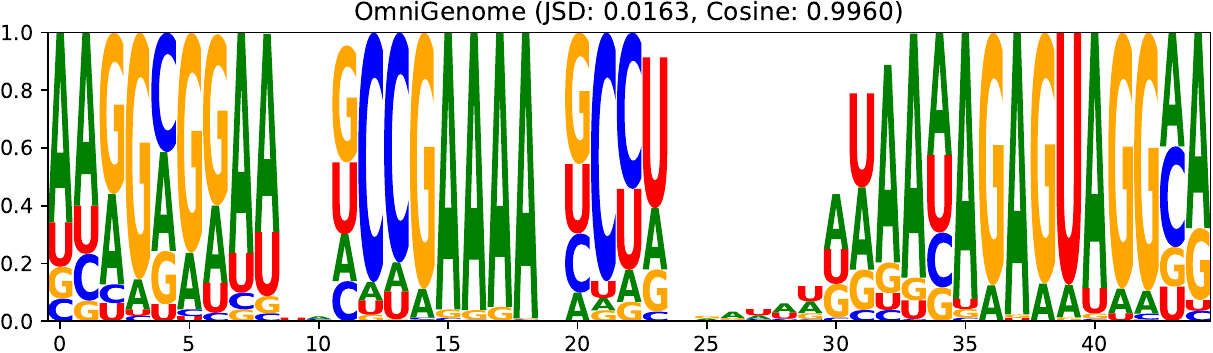}
        \caption*{(a) Sequence logos for RF02914}
    \end{minipage}
    \hfill
    \begin{minipage}[t]{0.48\textwidth}
        \centering
        \includegraphics[width=\linewidth]{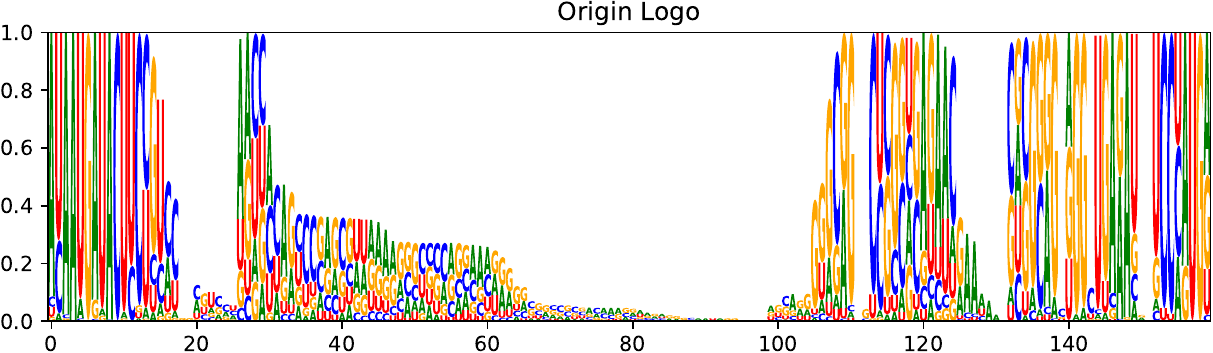}
        \includegraphics[width=\linewidth]{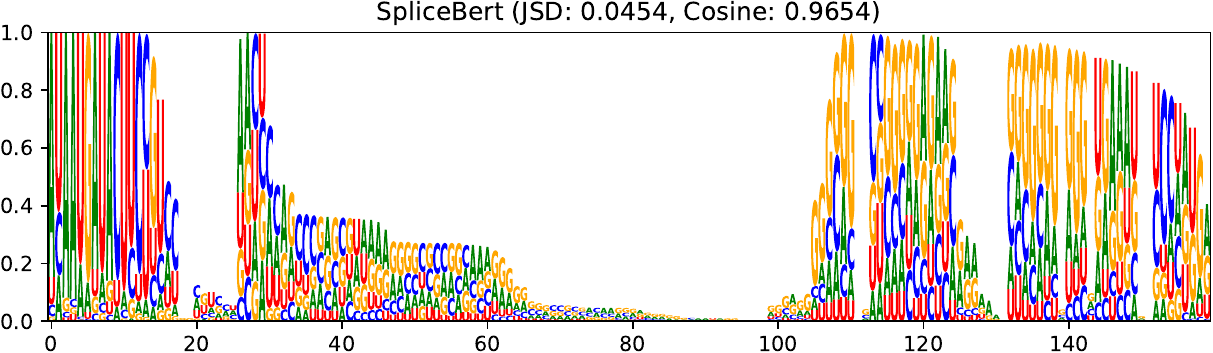}
        \includegraphics[width=\linewidth]{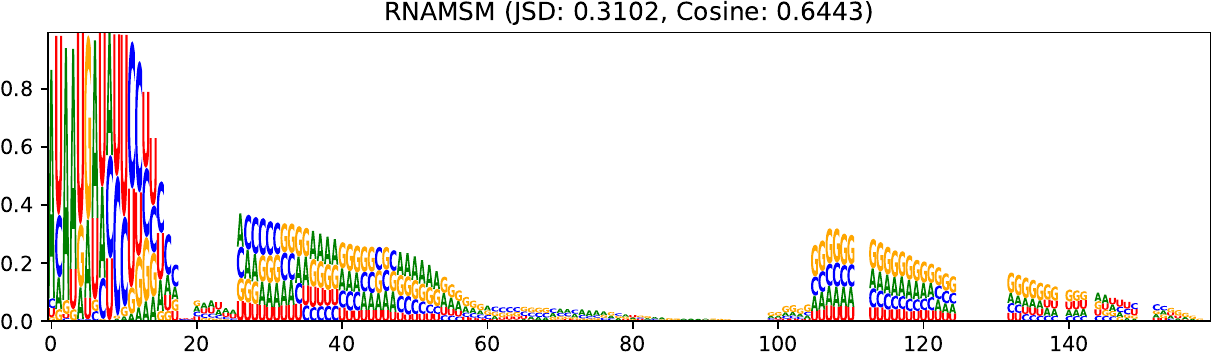}
        \includegraphics[width=\linewidth]{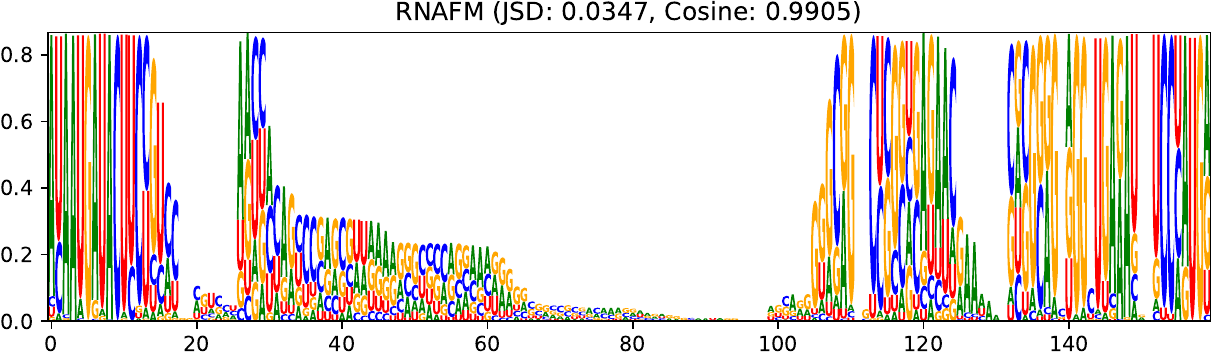}
        \includegraphics[width=\linewidth]{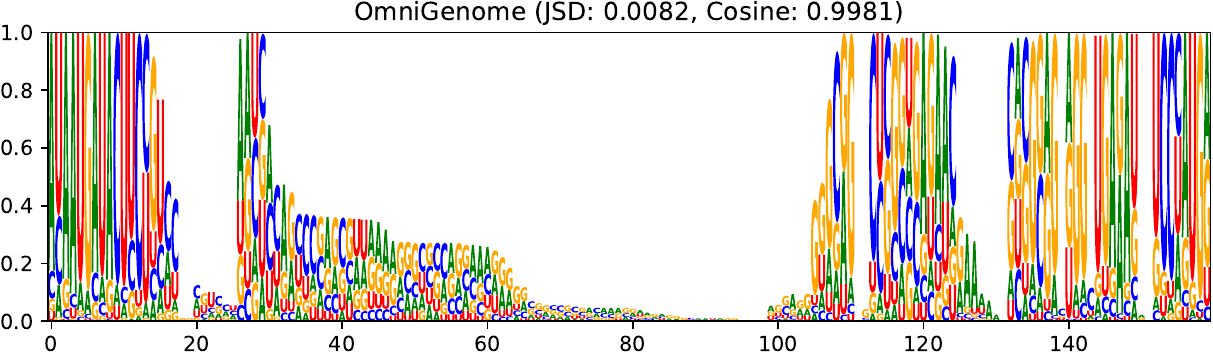}
        \caption*{(b) Sequence logos for RF02913}
    \end{minipage}
    \caption{Sequence logo comparison for RNA families after MLM-based augmentation with different GFMs. }
    \label{fig:rf02914_rf02913_column}
    \vspace{-10pt}
\end{figure}

\vspace{-10pt}
\paragraph{Experimental Design.}
To evaluate this fidelity, we employ visual inspection of sequence logos complemented by quantitative, information-theoretic scores.
We selected two distinct, well-characterized RNA families from the Rfam database~\citep{Kalvari21} for this study: RF02914 (DUF805 motif)\footnote{\url{https://rfam.org/family/RF02914}} and RF02913 (pemK motif)\footnote{\url{https://rfam.org/family/RF02913}}. These families represent conserved RNA structures initially discovered through bioinformatics methods and possess curated seed alignments representing conserved primary sequences and, implicitly, structural information.
The core methodology employs GFMs in a Masked Language Modeling (MLM) task to generate augmented sequences. The experimental steps were as follows:
\begin{enumerate}
\item \textbf{Sequence Selection}: Sequences were drawn from the seed MSA of each RNA family.
\item \textbf{Masking}: In each selected sequence, 15\% of nucleotides were randomly masked.
\item \textbf{Prediction and Augmentation}: Four GFMs capable of single nucleotide prediction, \og{}, SpliceBERT~\citep{ChenZD23}, RNAFM~\citep{Chen22}, and RNA-MSM~\citep{ZhangLJGXLCSHXS24}, were used to predict the masked nucleotides, generating ten augmented variants per original sequence.
\item \textbf{Aggregation}: The collection of augmented sequences for each GFM was treated as a new MSA group for comparison against the original seed MSA.
\item \textbf{Visualization}: Sequence logos were generated from the augmented MSAs using Logomaker\footnote{\url{https://logomaker.readthedocs.io/en/latest/}} to visualize position-specific nucleotide frequencies.
\item \textbf{Quantification}: The distributional similarity between the nucleotide frequencies of the GFM-augmented MSA subset and the original seed MSA subset (evaluated at unmasked positions for direct comparison, and across the entire sequence for overall fidelity) was quantified using \textbf{Jensen-Shannon Divergence ($D_{JS}$)} (lower values indicate higher similarity) and \textbf{Cosine Similarity} (higher values indicate greater similarity).
\end{enumerate}
The visual and quantitative results for the RF02914 and RF02913 families are presented in \pref{fig:rf02914_rf02913_column}.

\vspace{-10pt}
\paragraph{Results.}
As shown in \pref{fig:rf02914_rf02913_column}, the fidelity of motif preservation varies notably across the evaluated GFMs. \og{} achieves the best performance, with minimal divergence from native distributions ($D_{JS}$ of 0.0163 and 0.0082; Cosine Similarity of 0.9960 and 0.9981, for RF02914 and RF02913 respectively), producing sequence logos nearly indistinguishable from the original seeds. SpliceBERT shows moderate fidelity ($D_{JS}$ up to 0.0454 for RF02913, Cosine Similarity down to 0.9654 for RF02913), retaining core features but with some over-smoothing evident in less conserved regions. RNAFM performs similarly or slightly worse than SpliceBERT in this context, with $D_{JS}$ values between 0.0347–0.0517 and Cosine Similarities around 0.9869–0.9905 across the two families, reflecting reasonable but imperfect reconstruction. In stark contrast, RNA-MSM diverges significantly from the native patterns ($D_{JS}$ of 0.3102–0.3878; Cosine Similarity of 0.5802–0.6443), substantially distorting conserved motifs and introducing artifacts, likely due to its architectural design or specific pre-training objectives not being optimized for this type of fine-grained motif preservation.

\vspace{-10pt}
\paragraph{Conclusion.}
\og{} demonstrates a strong ability to preserve the position-specific motifs of conserved RNA families, indicating that its pre-training objective (potentially incorporating sequence-structure alignment information) effectively captures and respects the complex dependencies inherent in RNA sequences. This comparative analysis underscores the critical role of model architecture and pre-training strategies in a GFM's capacity to generate biologically plausible sequence variants. Such sequence-centric interpretability analyses, focused on motif fidelity, provide a crucial framework for validating the reliability of GFMs as sequence augmentation engines, identifying their potential biases, and guiding the selection of appropriate models for downstream tasks that depend on high sequence fidelity, ultimately informing the development of future GFMs that better capture the nuanced features of biological sequences.

\section{Conclusion}

We present \our{}, a modular, extensible platform that consolidates data hubs, model repositories, automated benchmarking, and interpretability tools into a cohesive ecosystem. By enabling reproducible evaluation and streamlined application across diverse genomic tasks, \our{} provides an infrastructure backbone for accelerating research, standardization, and real-world deployment of GFMs. This initiative marks a step forward in maturing genome-scale AI from isolated prototypes to robust, interpretable, and scalable systems for computational biology.

\bibliographystyle{unsrtnat}

\bibliography{ref}

\appendix

\section*{Table of Contents in Appendices}
\label{sec:table_of_content}

\begin{itemize}[label={}, leftmargin=0pt, itemsep=0.5ex]
    \item \textbf{A \quad Related Works \dotfill 16}
    \begin{itemize}[label={}, leftmargin=2em, itemsep=0.3ex]
        \item A.1 \quad Benchmarking Platforms and Tools for Genomic Models \dotfill 16
        \item A.2 \quad Evolution of Genomic Foundation Models \dotfill 16
    \end{itemize}

    \item \textbf{B \quad Extended Overall Benchmark Results \dotfill 17}
    \begin{itemize}[label={}, leftmargin=2em, itemsep=0.3ex]
        \item B.1 \quad Rank-based Performance \dotfill 17
    \end{itemize}

    \item \textbf{C \quad Integrated Benchmark Suites in \our{} \dotfill 19}
    \begin{itemize}[label={}, leftmargin=2em, itemsep=0.3ex]
        \item C.1 \quad RNA Genomic Benchmark (RGB) \dotfill 19
        \item C.2 \quad Plant Genomic Benchmark \dotfill 19
        \item C.3 \quad Genomic Understanding Evaluation \dotfill 20
        \item C.4 \quad Genomic Benchmarks \dotfill 21
        \item C.5 \quad BEACON Benchmark \dotfill 22
        \item C.6 \quad Data Filtering in Benchmarking \dotfill 23
    \end{itemize}

    \item \textbf{D \quad Detailed Benchmark Performance Report \dotfill 23}
    \begin{itemize}[label={}, leftmargin=2em, itemsep=0.3ex]
        \item D.1 \quad Evaluation Settings in Benchmarking \dotfill 23
        \item D.2 \quad Evaluation GFMs in Benchmarks \dotfill 24
        \item D.3 \quad RNA Genomic Benchmark (RGB) \dotfill 25
        \item D.4 \quad Plant Genomic Benchmark (PGB) \dotfill 26
        \item D.5 \quad Genomic Understanding Evaluation (GUE) \dotfill 27
        \item D.6 \quad Genomic Benchmarks (GB) \dotfill 27
        \item D.7 \quad BEACON Results \dotfill 28
        \item D.8 \quad Overall Discussion \dotfill 28
    \end{itemize}

    \item \textbf{E \quad Integrated Genomic Foundation Models in OmniGenBench \dotfill 29}
    \item \textbf{F \quad Public Leaderboard \dotfill 31}

    \item \textbf{G \quad Tutorials \dotfill 31}
    \begin{itemize}[label={}, leftmargin=2em, itemsep=0.3ex]
        \item G.1 \quad Automated Benchmarking with AutoBench \dotfill 31
        \item G.2 \quad Fine-tuning for RNA Secondary Structure Prediction \dotfill 33
        \item G.3 \quad Zero-Shot RNA Secondary Structure Prediction \dotfill 35
        \item G.4 \quad Generating RNA Embeddings \dotfill 35
        \item G.5 \quad Computational RNA Sequence Design \dotfill 36
        \item G.6 \quad Sequence Augmentation via Masked Language Modeling \dotfill 37
    \end{itemize}

    \item \textbf{H \quad Interpretability Cases for Genomic Foundation Models \dotfill 38}
    \begin{itemize}[label={}, leftmargin=2em, itemsep=0.3ex]
        \item H.1 \quad Feature-Embedding Analysis \dotfill 38
        \item H.2 \quad Attention Representation Inspection \dotfill 40
        \item H.3 \quad Development Environment \dotfill 43
    \end{itemize}

    \item \textbf{I \quad Ethical Considerations \dotfill 43}
    \item \textbf{J \quad Societal Impact \dotfill 43}
    \item \textbf{K \quad Limitations \dotfill 44}
\end{itemize}

\section{Related Works}
\label{app:related_works}

The development of robust platforms for GFMs intersects with advancements in genomic benchmarking tools and the evolution of GFMs themselves.

\subsection{Benchmarking Platforms and Tools for Genomic Models}
\label{sec:related_benchmarking}

Effective evaluation is crucial for advancing genomic models. Several benchmarking tools and platforms have emerged, yet most exhibit limitations in scope, extensibility, or GFM-specific support.

\paragraph{Genomic Benchmarking Suites.}
Early efforts like Kipoi~\citep{Avsec19kipoi} focused on standardizing access to trained models for genomic sequence analysis, primarily classic predictive models rather than modern GFMs, and offered limited benchmarking capabilities. More recent suites have targeted specific modalities or tasks. For instance, RNABench~\citep{Runge24rnabench} provides benchmarks for RNA-centric tasks like secondary structure prediction but lacks comprehensive support for evaluating diverse pre-trained GFMs. GenBench~\citep{Liu24Genbench} offers a modular framework for DNA sequence evaluation but does not extend to RNA and can be challenging for users not deeply familiar with its architecture. BEACON~\citep{Ren24BEACON} is a notable recent effort for RNA foundation models, introducing several RNA evaluation datasets. However, our experience indicates that its complex environment setup can hinder model portability and benchmarking scalability. DEGB~\citep{West24DGEB} evaluates genomic embeddings for both nucleic acids and amino acids but is restricted by the small scale of its benchmarks and does not support the evaluation of GFMs in downstream application contexts. Broader platforms like Galaxy~\citep{Galaxy2022galaxy} democratize pipeline execution but lack the specific abstractions and integrated GFM support necessary for streamlined GFM research. Commercial cloud platforms (e.g., DNAnexus, Seven Bridges) provide computational infrastructure but typically treat GFMs as external entities rather than integrated components.

\paragraph{Protein Language Model Benchmarks.}
The protein language modeling domain has also seen the development of specialized benchmarks, such as ProteinGym~\citep{Notin24proteingym} for fitness prediction, FLIP~\citep{Dallago21Flip} for sequence-function relationships, and PEER~\citep{Xu22Peer} for diverse protein-related tasks. While these tools are valuable in their specific area, their focus is on protein sequences and associated tasks, and they are not directly applicable to the unique challenges of DNA and RNA GFM benchmarking.

\paragraph{Limitations of Existing Genomic Benchmarks.}
A significant gap remains: existing tools often provide narrow task coverage (e.g., DNA-only or RNA-only), lack extensibility for new datasets and models, or do not support the comprehensive end-to-end evaluation of GFMs, from pre-training to downstream application and interpretability. The diverse architectures and tokenization strategies of GFMs further complicate direct comparisons using these fragmented tools. This landscape underscores the need for a unified and extensible platform like \our{}, which is designed to address these limitations by offering broad support for both DNA and RNA GFMs, flexible benchmark integration, and ease of use for the genomics community.

\subsection{Evolution of Genomic Foundation Models}
\label{sec:related_gfms}

The application of foundation models to biological sequences, inspired by successes in natural language processing (NLP), has rapidly advanced, though progress in DNA and RNA modeling has historically lagged behind protein modeling (e.g., AlphaFold~\citep{Jumper21,Evans21,Abramson24}, ESM~\citep{Lin22}).

\paragraph{DNA Foundation Models.}
Early DNA GFMs adapted NLP architectures. DNABERT~\citep{JiZLD21} applied the BERT~\citep{DevlinCLT19} architecture to DNA, with DNABERT2~\citep{Zhou23DNABERT2} later enhancing performance by adopting byte-pair encoding (BPE) over k-mer tokenization. Subsequently, models like Nucleotide Transformers V2 (NT-V2)~\citep{Dalla23NT}, AgroNT~\citep{Mendoza23} (focused on plant DNA), and SegmentNT~\citep{De24} explored scaling to billions of parameters, achieving strong results on various DNA understanding tasks. However, specialized models like AgroNT have shown limited transferability to other modalities like RNA. To handle the long-sequence nature of genomes, auto-regressive models such as HyenaDNA~\citep{Nguyen23} and Evo~\citep{Nguyen24} have also been introduced, emphasizing long-range dependency modeling.

\paragraph{RNA Foundation Models.}
RNA GFM development has faced challenges due to the relative scarcity of large-scale, annotated RNA datasets. Initial models like scBERT~\citep{YangWWFTHLY22} (for single-cell RNA), RNABERT~\citep{Akiyama22informative}, RNA-FM~\citep{Chen22}, RNA-MSM~\citep{ZhangLJGXLCSHXS24}, %
and RNAErnie~\citep{WangBLLMKX24} were often trained on smaller databases. Some GFMs target specific RNA types, such as coding sequences (CDSBERT~\citep{HalleeRG23}), 5'UTRs (5UTR-LM~\citep{Chu24}), 3'UTRs (3UTRBERT~\citep{YangLP23}), or precursor mRNAs (SpliceBERT~\citep{ChenZD23}), which can limit their generalizability across the diverse RNA landscape. While models like Uni-RNA~\citep{WangGCLJKW23} have reported strong performance due to large-scale pre-training, their closed-source nature restricts comparative analysis and community adoption.

\paragraph{Multimodal and Conversational Genomic Agents.}
More recently, models like ChatNT~\citep{Richard24chatnt} have emerged as multimodal conversational agents capable of assisting with tasks across DNA, RNA, and protein sequences. These tools aim to integrate various AI capabilities to facilitate broader research in genomics and proteomics.

\paragraph{Positioning of \our{}.}
Despite the proliferation of GFMs, their practical application and comparative evaluation are hindered by the lack of a unified platform that can accommodate diverse model architectures, data types, and downstream tasks. \our{} aims to fill this void by providing the necessary infrastructure to integrate, benchmark, and apply these varied GFMs effectively, thereby fostering a more cohesive and rapidly advancing GFM ecosystem.

\section{Extended Overall Benchmark Results}

\subsection{Rank-based Performance}
\label{app:rank_based_performance}

\begin{figure}[htbp]
  \centering
  \begin{minipage}[t]{\textwidth}
    \centering
    \includegraphics[width=0.999\linewidth]{figs/radar/RGB-test.pdf}
    \caption*{(a) RGB}
  \end{minipage}
  \begin{minipage}[t]{\textwidth}
    \centering
    \includegraphics[width=0.999\linewidth]{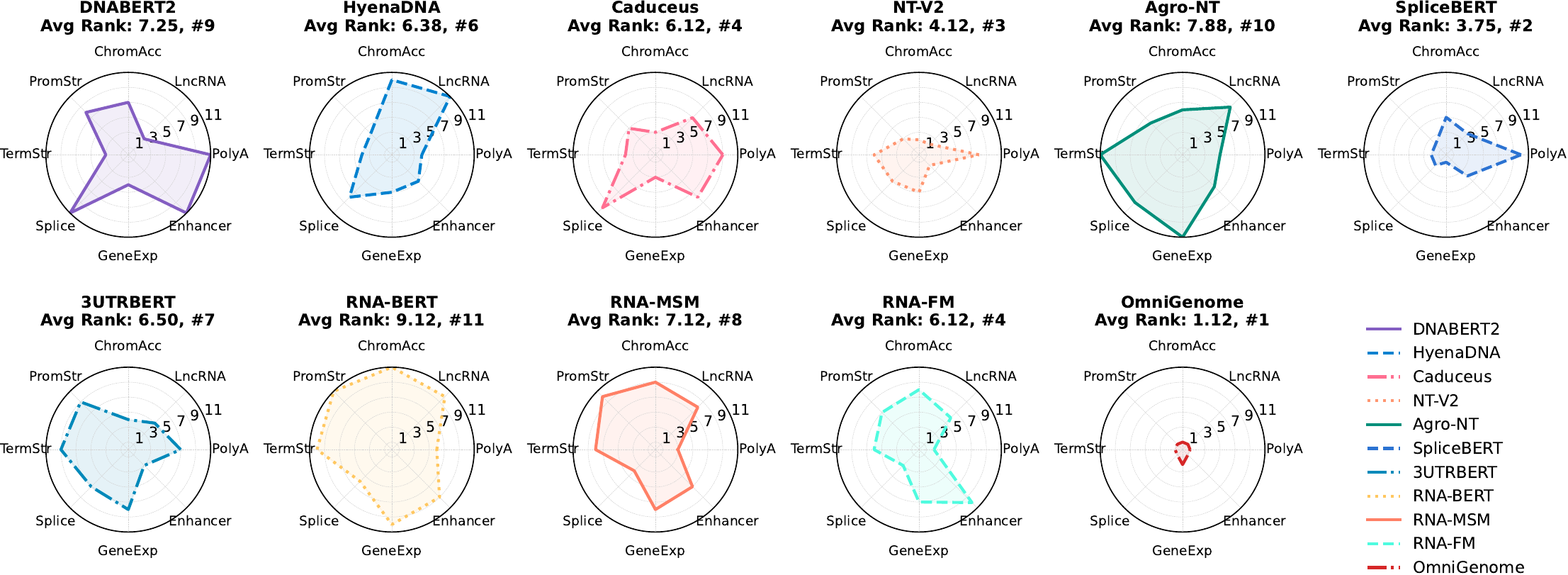}
    \caption*{(b) PGB}
  \end{minipage}

  \begin{minipage}[t]{\textwidth}
    \centering
    \includegraphics[width=0.999\linewidth]{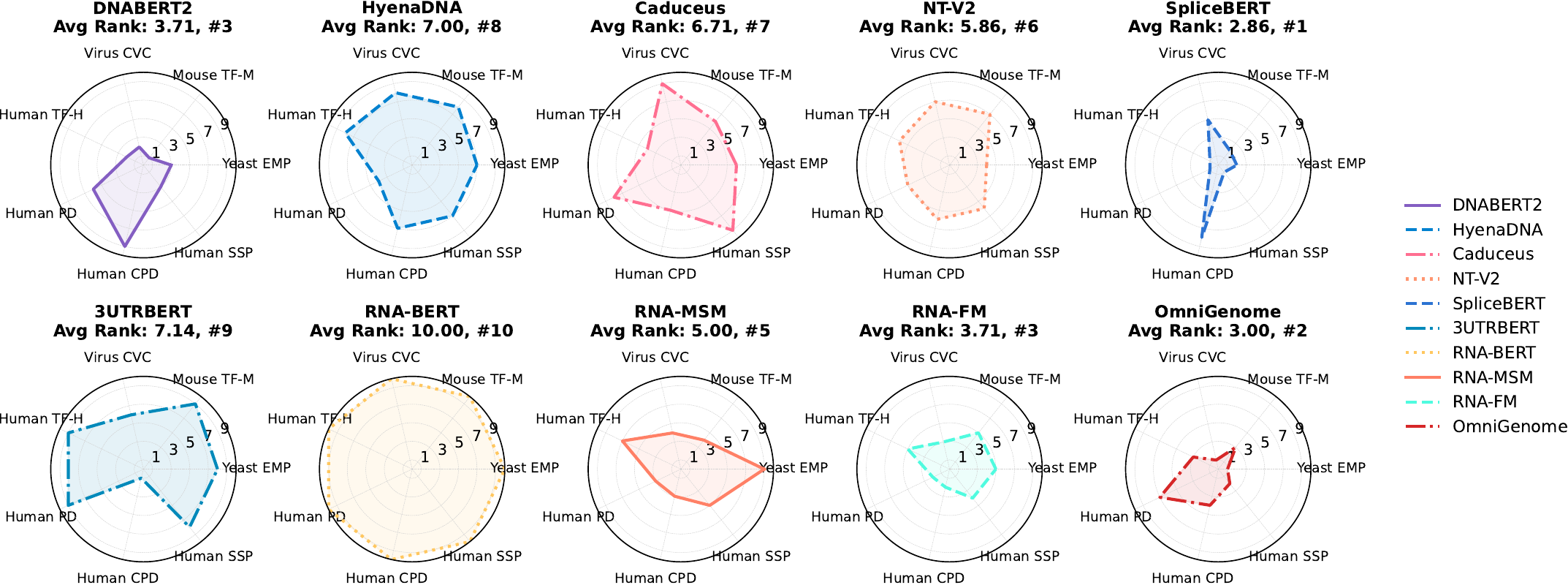}
    \caption*{(c) GUE}
  \end{minipage}
  \begin{minipage}[t]{\textwidth}
    \centering
    \includegraphics[width=0.999\linewidth]{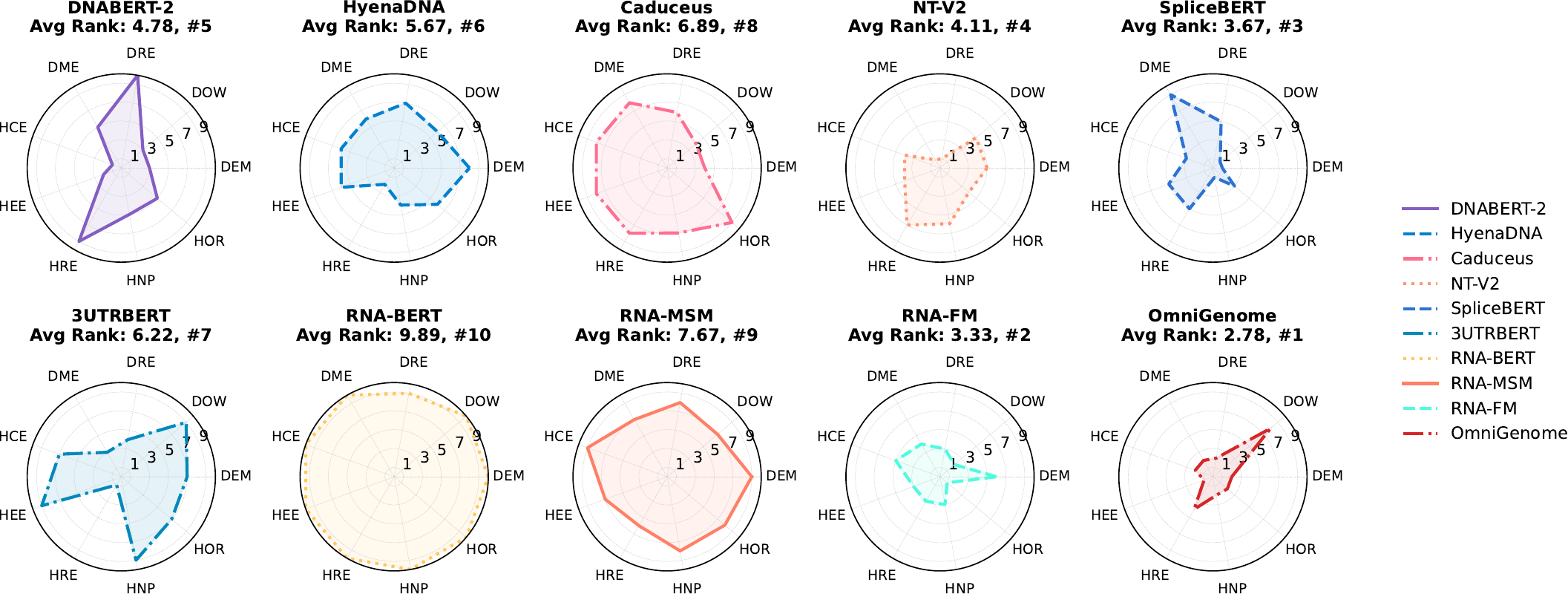}
    \caption*{(d) GB}
  \end{minipage}

  \caption{Rank-based radar charts comparisons between GFMs across four genomic benchmark suites.}
  \label{fig:radar_combined}
  \vspace{-10pt}
\end{figure}

To provide a more nuanced view of where each GFM excels or struggles, we present rank-normalized radar charts in \pref{fig:radar_combined}. For each benchmark suite, an individual radar chart is generated for each of the baseline GFMs. The axes of these charts represent distinct task categories within the respective suite. A model's performance on each axis is its average rank for tasks in that category (rank 1 = best). Consequently, a smaller and more centrally-focused polygon for a model indicates stronger and more balanced performance across the task categories of that benchmark suite. The average rank of each model across all task categories within a suite is also noted on its respective plot.

\vspace{-10pt}
\paragraph{Results.}
The rank-normalized radar charts (\pref{fig:radar_combined}) reveal distinct GFM performance profiles across the four benchmark suites. Each GFM's individual radar plot, with its average rank displayed, visualizes strengths and weaknesses across task categories (axes). In these plots, smaller and more central polygons indicate superior, balanced performance across the task categories of that benchmark suite. The average rank of each model across all task categories within a suite is also noted on its respective plot.
For RGB in \pref{fig:radar_combined}a, \og{} (avg. rank 1.00) demonstrates clear dominance with a near-central polygon. RNA-FM (avg. rank 2.29) is a strong second, followed by SpliceBERT (avg. rank 3.00). DNA-centric models like DNABERT-2 (avg. rank 10.00) exhibit larger polygons, underscoring weaker RNA task transferability. Structure-aware models, particularly \og{} and RNA-FM, excel in structure-related task categories (e.g.,  SSP axis).
As for the PGB in \pref{fig:radar_combined}b, \og{} (avg. rank 1.12) again shows exceptional generalization with a compact plot, notably on PolyAP and ProStrP axes, despite no specific plant pre-training. SpliceBERT (avg. rank 3.75) and the specialized NT-v2 (avg. rank 4.12) are also top performers.  HyenaDNA (avg. rank 6.38) and Caduceus (avg. rank 6.12) have larger, skewed plots, suggesting variable performance across PGB task categories. 
For the GUE in \pref{fig:radar_combined}c, performance is more varied. \og{} (avg. rank 3.00) and SpliceBERT (avg. rank 2.86) show the best overall balance. DNABERT-2 (avg. rank 3.71) and NT-V2 (avg. rank 5.86) are also highly competitive. Specific strengths are visible, e.g., SpliceBERT excels on the Human SSP axis, DNABERT-2 on Mouse TF-M. \og{}, while strong overall, displays a comparatively larger rank on the Mouse TF-M axis, indicating a specific performance gap.
Within the GB in \pref{fig:radar_combined}d, \og{} (avg. rank 2.78) leads with a compact radar. NT-V2 (avg. rank 4.11), RNA-FM (avg. rank 3.33), and SpliceBERT (avg. rank 3.67) follow closely. DNABERT-2's plot highlights its strength on the HCE axis but relative weakness on DRE/HRE, potentially due to tokenizer differences. SpliceBERT and NT-V2 show strong performance on their respective favored task categories (e.g., DOW/HNP and DME/DRE).

These granular radar visualizations effectively highlight model-specific strengths and weaknesses across diverse genomic task categories beyond what aggregate scores reveal. This detailed insight is crucial for informed GFM selection for downstream tasks and for guiding future model development.

\section{Integrated Benchmark Suites in \our{}}
\label{app:benchmark_details}

There are five benchmark collections in this paper: RGB, PGB, GUE, GB, and BENCON, with 123 tasks\footnote{We define a task as a downstream task compiled from with a dataset. For example, the tasks in RGB are compiled from the datasets in RGB.} complied in total.

\subsection{RNA Genomic Benchmark (RGB)}
\label{app:rgb}

RGB now comprises \textbf{12} single-nucleotide (SN)-level tasks spanning mutation analysis, RNA degradation, secondary-structure prediction, \emph{de-novo} RNA design, plant-specific genic‐region annotation, and translation-efficiency (TE) modelling.  
Sequences longer than~512 nt (\,$<\!3\%$ of bpRNA/ArchiveII/Stralign) are discarded, leaving a length range of 107-512 nt, which is sufficient for most RNA-understanding workloads.  
The benchmark therefore probes both cross-species generalisation (human, plant, virus) and fine-grained SN-level reasoning.  Brief task synopses follow; full statistics appear in \pref{tab:rgb_detail}.  

\begin{itemize}[leftmargin=*]
\item \textbf{SN Mutation Detection (SNMD)}  
      Binary token classification that flags mutated positions in synthetic plant RNAs (up to ten random SNVs per sequence).  
      Loss: cross-entropy.  (eg, used to screen deleterious variants).

\item \textbf{SN Mutation Repair (SNMR)}  
      Four-way token classification (\texttt{A/U/C/G}) that proposes the corrected base at each mutated position; shares the SNMD splits.

\item \textbf{mRNA Degradation Rate Prediction (mRNA)}  
      Token-level regression (MSE) on the Stanford \emph{OpenVaccine} dataset that measures in-line hydrolysis of 102-130 nt constructs.

\item \textbf{RNA Secondary-Structure Prediction}  
      bpRNA-1m \citep{DanaeeRWDHH18}, ArchiveII \citep{Mathews19}, and Stralign \citep{TanFSM17}: three-label (\texttt{`(', `.', `)' }) token classification; metric = macro-F1.

\item \textbf{RNA Design (\textit{EternaV2})}  
      Sequence-level accuracy on \emph{EternaV2}, which contains\,\(\sim27\,\mathrm{k}\) player-designed RNAs with desired secondary structures.  
      Models must predict whether a candidate folds into the target structure, mirroring practical design pipelines.

\item \textbf{Genic-Region Classification}  
      Two plant species from PlantRNA-FM \citep{Yu2024interpretable}:  
      \begin{itemize}
        \item \textbf{Region-\textit{Ara}} (Arabidopsis)  
        \item \textbf{Region-\textit{Rice}} (Oryza sativa)  
      \end{itemize}
      Token-level 3-way classification of \texttt{5'UTR}, \texttt{CDS}, \texttt{3'UTR}; metric = macro-F1.  
      These tasks test whether GFMs capture plant-specific transcript organisation \citep{Yu2024interpretable}.  

\item \textbf{Translation-Efficiency Prediction}  
      Again using PlantRNA-FM data \citep{Yu2024interpretable}:  
      \begin{itemize}
        \item \textbf{TE-\textit{Ara}} (Arabidopsis)  
        \item \textbf{TE-\textit{Rice}} (Rice)  
      \end{itemize}
      Sequence-level binary classification (high vs.\ low TE) on \(5'\)-UTRs; metric = AUC.  
      Successful models must integrate subtle sequence cues governing ribosome loading \citep{Yu2024interpretable}.  
\end{itemize}

\begin{table*}[t]
\centering
\caption{Statistics of RGB subtasks.  ``Cls'' = classification, ``Reg'' = regression.  `—' denotes values not explicitly reported in source papers.}
\label{tab:rgb_detail}
\resizebox{\linewidth}{!}{%
\begin{tabular}{lcccccc}
\toprule
\textbf{Task} & \textbf{Type} & \textbf{\#Train/Val/Test} & \textbf{Classes} & \textbf{Metric} & \textbf{Len.\ (max/mean)} & \textbf{Source} \\
\midrule
SNMD                  & Token Cls & 8\,000/1\,000/1\,000 & 2  & AUC        & 200/200         & This work \\
SNMR                  & Token Cls & 8\,000/1\,000/1\,000 & 4  & macro-F1   & 200/200         & This work \\
mRNA                  & Token Reg & 1\,735/193/192       & —  & RMSE       & 107/107         & \href{https://www.kaggle.com/competitions/stanford-covid-vaccine}{Kaggle} \\
bpRNA-1m              & Token Cls & 10\,814/1\,300/1\,305& 3  & macro-F1   & 512/134         & \citep{DanaeeRWDHH18} \\
ArchiveII             & Token Cls & 2\,278/285/285       & 3  & macro-F1   & 500/151         & \citep{Mathews19} \\
RNAStralign           & Token Cls & 17\,483/2\,186/2\,185& 3  & macro-F1   & 500/142         & \citep{TanFSM17} \\
\midrule
EternaV2              & Seq Cls   & 20\,430/3\,607/3\,607$^{*}$ & 2 & Accuracy   & 180/101         & \citep{Lee14} \\
Region-\textit{Ara}   & Token Cls & 12\,838/1\,604/1\,604 & 3 & macro-F1   & 1024/1024        & \citep{Yu2024interpretable} \\
Region-\textit{Rice}  & Token Cls & 11\,412/1\,426/1\,426 & 3 & macro-F1   & 1024/1024        & \citep{Yu2024interpretable} \\
TE-\textit{Ara}       & Seq Cls   &  9\,644/1\,206/1\,206 & 2 & macro-F1   & 500/500        & \citep{Yu2024interpretable} \\
TE-\textit{Rice}      & Seq Cls   &  8\,102/1\,013/1\,013 & 2 & macro-F1   & 500/500        & \citep{Yu2024interpretable} \\
\bottomrule
\end{tabular}}
\end{table*}

\pref{tab:examples} shows the virtual examples of different datasets in RGB. Please refer to our supplementary materials to find the datasets for more details.

\begin{table}[htbp]
  \centering
  \caption{The virtual input and output examples in the four benchmarks. The ``$\dots$'' represents the sequences that are omitted for better presentation and the \textcolor{red}{red} color indicates the wrong prediction in classification tasks. In the mRNA dataset, all single nucleotides have three values to predict. Note that ``\texttt{T}'' and ``\texttt{U}'' can be regarded as the same symbol in RNA sequences and depend on different datasets. }
  \resizebox{.8\linewidth}{!}{
    \begin{tabular}{lccc}
    \toprule
    \textbf{Genome Type} & \textbf{Dataset} & \textbf{Column}  & \textbf{Examples} \\
    \midrule
    \multirow{12}[2]{*}{RNA}   
    & \multirow{3}[2]{*}{SNMD} & Input Sequence & G~A~G~T~A~$\dots$~T~T~G~A~G \\
    &     & True Label & 0~~0~~1~~0~~0~$\dots$~0~~0~~1~~0~~0 \\
    &     & Prediction & 0~~0~~\textcolor{red}{0}~~0~~0~$\dots$~0~~0~~1~~0~~0 \\
    \cmidrule{2-4}          
    & \multirow{3}[2]{*}{SNMR} & Input Sequence & T~A~C~G~A~~$\dots$~C~T~G~A~T \\
    &     & True Label & T~A~C~A~A~$\dots$~G~T~A~A~T \\
    &     & Prediction & T~A~C~A~A~$\dots$~\textcolor{red}{C}~T~G~A~T \\
    \cmidrule{2-4}          
    & \multirow{3}[2]{*}{mRNA} & Input Sequence & G~G~$\dots$~A~C \\
    &     & True Label & [0.1,0.3,0.2]~[0.8,0.4,0.1]$\dots$[0.9,0.4,0.3]~[0.5,0.2,0.6] \\
    &     & Prediction & [0.1,0.3,0.2]~[0.8,0.4,0.1]$\dots$[0.9,0.4,0.3]~[0.5,0.2,0.6] \\
    \cmidrule{2-4}          
    & \multirow{3}[2]{*}{bpRNA} & Input Sequence & G~G~C~G~A~$\dots$~C~U~U~U~U \\
    &     & True Label & (~~~(~~~(~~~$\cdot$~~~$\cdot$~$\dots$~$\cdot$~~~$\cdot$~~~)~~~)~~~) \\
    &     & Prediction & (~~~(~~~(~~~\textcolor{red}{(}~~~$\cdot$~$\dots$~$\cdot$~~~\textcolor{red}{)}~~~)~~~)~~~) \\
    \midrule
    \multirow{9}[2]{*}{DNA}   
    & \multirow{3}[2]{*}{Classification} & Input Sequence & A~T~C~G~A~$\dots$~T~A~G \\
    &     & True Label & 1 \\
    &     & Prediction & \textcolor{red}{0} \\
    \cmidrule{2-4}          
    & \multirow{3}[2]{*}{Regression} & Input Sequence & G~C~C~A~T~$\dots$~G~C~T \\
    &     & True Label & 2.56 \\
    &     & Prediction & 2.45 \\
    \bottomrule
    \end{tabular}%
    }
   \label{tab:examples}%
\end{table}%

\subsection{Plant Genomic Benchmark}
\label{app:pgb}
PGB~\citep{Mendoza23} provides a comprehensive suite of datasets designed to evaluate and improve the predictive capabilities of GFMs in plant biology. This benchmark, as shown in \pref{tab:pgb_detail}, encompasses a range of critical genomic tasks, including binary classification, single and multi-variable regression, and multi-label classification, addressing various aspects of plant genomics such as RNA processing, gene expression, and chromatin accessibility. By integrating diverse genomic tasks, the PGB aims to facilitate advanced research and development in plant genomics, offering a robust platform for the assessment and enhancement of model performance across different plant species. To obtain a detailed description of PGB, please refer to Agro-NT~\citep{Mendoza23}.

\begin{table*}[t!]
    \centering
    \caption{The genomic tasks in the Plant Genomic Benchmark. This table briefly enumerates each task by name, the number of datasets available, the type of classification or regression analysis required, the range of sequence lengths, and the total number of samples in each dataset. ``Cls.'' indicates classification. Please find the dataset details of PGB in Agro-NT.}
    \resizebox{\linewidth}{!}{
        \begin{tabular}{lcccccc}
            \toprule
             \textbf{Task}  & \multicolumn{1}{c}{\textbf{\# of datasets}} & \multicolumn{1}{c}{\textbf{Task Type}} & \multicolumn{1}{c}{\textbf{Total \# of examples}} & \multicolumn{1}{c}{\textbf{\# of classes}} & \multicolumn{1}{c}{\textbf{Metric}} & \multicolumn{1}{c}{\textbf{Sequence length}} \\
            \midrule
            Polyadenylation & $6$ & Classification & $738,918$  & $2$ & macro F1 & $400$  \\
            Splice site  & $2$  & Classification & $4,920,835$  & $2$ & macro F1 & $398$  \\
            LncRNA &  $2$   & Classification & $58,062$ & $6$ & macro F1 & $101-6000$  \\
            Promoter strength & $2$  & Regression & $147,966$   & --- & RMSE & $170$  \\
            Terminator strength & $2$  & Regression & $106,818$   & --- & RMSE & $170$  \\
            Chromatin accessibility & $7$   & Multi-label Cls. & $5,149,696$  & $9$-$19$ (multi-label) & macro F1 & $1,000$  \\
            Gene expression & $6$  & Multi-variable Reg. & $206,358$   & --- & RMSE & $6,000$  \\
            Enhancer region & $1$  & Classification  & $18,893$  &  $2$ & macro F1 & $1,000$  \\
            \bottomrule
        \end{tabular}
     }
    \label{tab:pgb_detail}
    
\end{table*}

\subsection{Genomic Understanding Evaluation}
\label{app:gue}

GUE~\citep{Zhou23DNABERT2} serves as a DNA genomic benchmark, encompassing 36 datasets across nine crucial genome analysis tasks applicable to a variety of species. Similar to PGB and GB, it is used for evaluating the generalizability of \our\ on DNA genome benchmarking. To thoroughly assess the capabilities of genome foundation models across sequences of varying lengths, tasks have been chosen with input lengths spanning from $70$ to $10,000$. The brief statistics for each dataset included in the GUE benchmark are displayed in \pref{tab:gue_detail}, and the task descriptions are available in \citet{Zhou23DNABERT2}. Due to resource limitations, we do not include large-scale FMs in this benchmark, e.g., Agro-NT. Besides, all reported scores are on a randomly stratified 10 k-sample subset per split; hence they are \emph{not} directly comparable to the original GUE leaderboard.

\begin{table*}[ht]
    \centering
    \caption{Statistics of tasks in the GUE, these details can be found in Section B.2. from \citet{Zhou23DNABERT2}.}
    \resizebox{\linewidth}{!}{
    \begin{tabular}{lccccc}
        \toprule
        \textbf{Task} & \textbf{Metric} & \textbf{Datasets} & \textbf{Training} & \textbf{Validation} & \textbf{Testing} \\ \midrule
        \multirow{3}[1]{*}{Core Promoter Detection} & \multirow{3}[1]{*}{macro F1} & tata & $4,904$ & $613$ & $613$ \\
         &  & notata & $42,452$ & $5,307$ & $5,307$ \\
         &  & all & $47,356$ & $5,920$ & $5,920$ \\ \midrule
        \multirow{3}[1]{*}{Promoter Detection} & \multirow{3}[1]{*}{macro F1} & tata & $4,904$ & $613$ & $613$ \\
         &  & notata & $42,452$ & $5,307$ & $5,307$ \\
         &  & all & $47,356$ & $5,920$ & $5,920$ \\ \midrule
        \multirow{5}[1]{*}{Transcription Factor Prediction (Human)} & \multirow{5}[1]{*}{macro F1} & wgEncodeEH000552 & $32,378$ & $1,000$ & $1,000$ \\
         &  & wgEncodeEH000606 & $30,672$ & $1,000$ & $1,000$ \\
         &  & wgEncodeEH001546 & $19,000$ & $1,000$ & $1,000$ \\
         &  & wgEncodeEH001776 & $27,497$ & $1,000$ & $1,000$ \\
         &  & wgEncodeEH002829 & $19,000$ & $1,000$ & $1,000$ \\ \midrule
        Splice Site Prediction & macro F1 & reconstructed & $36,496$ & $4,562$ & $4,562$ \\ \midrule
        \multirow{5}[1]{*}{Transcription Factor Prediction (Mouse)} & \multirow{5}[1]{*}{macro F1} & Ch12Nrf2\textbackslash iggrab & $6,478$ & $810$ & $810$ \\
         &  & Ch12Zrf384hpa004051\textbackslash iggrab & $5,395$ & $674$ & $674$ \\
         &  & MelJun\textbackslash iggrab & $2,620$ & $328$ & $328$ \\
         &  & MelMafkDm2p5dStd & $1,904$ & $239$ & $239$ \\
         &  & MelNelf\textbackslash iggrab & $15,064$ & $1,883$ & $1,883$ \\ \midrule
        \multirow{10}[1]{*}{Epigenetic Marks Prediction} & \multirow{10}[1]{*}{macro F1} & H3 & $11,971$ & $1,497$ & $1,497$ \\
         &  & H3K14ac & $26,438$ & $3,305$ & $3,305$ \\
         &  & H3K36me3 & $29,704$ & $3,488$ & $3,488$ \\
         &  & H3K4me1 & $25,341$ & $3,168$ & $3,168$ \\
         &  & H3K4me2 & $24,545$ & $3,069$ & $3,069$ \\
         &  & H3K4me3 & $29,439$ & $3,680$ & $3,680$ \\
         &  & H3K79me3 & $23,069$ & $2,884$ & $2,884$ \\
         &  & H3K9ac & $22,224$ & $2,779$ & $2,779$ \\
         &  & H4 & $11,679$ & $1,461$ & $1,461$ \\
         &  & H4ac & $27,275$ & $3,410$ & $3,410$ \\ \midrule
        Covid Variant Classification & macro F1 & Covid & $77,669$ & $7,000$ & $7,000$ \\ \midrule
        \multirow{6}[1]{*}{Enhancer Promoter Interaction} & \multirow{6}[1]{*}{macro F1} & GM12878 & $10,000$ & $2,000$ & $2,000$ \\
         &  & HeLa-S3 & $10,000$ & $2,000$ & $2,000$ \\
         &  & HUVEC & $10,000$ & $2,000$ & $2,000$ \\
         &  & IMR90 & $10,000$ & $2,000$ & $2,000$ \\
         &  & K562 & $10,000$ & $2,000$ & $2,000$ \\
         &  & NHEK & $10,000$ & $2,000$ & $2,000$ \\ \midrule
        \multirow{2}[1]{*}{Species Classification} & \multirow{2}[1]{*}{macro F1} & fungi & $8,000$ & $1,000$ & $1,000$ \\
         &  & virus & $4,000$ & $500$ & $500$ \\ \midrule
    \end{tabular}
    }
    \label{tab:gue_detail}
\end{table*}

\subsection{Genomic Benchmarks}
\label{app:gb}
GB is also a DNA-oriented FM benchmark suite, which can be used for generalizability evaluation of \og. It contains a well-curated collection of datasets designed for the classification of genomic sequences, focusing on regulatory elements across multiple model organisms. This collection facilitates robust comparative analysis and development of genomic FMs. The task names in the original repository are complex, we abbreviate the names as follows: 
\begin{itemize}[leftmargin=*]
    \item DEM corresponds to "Demo Coding vs Intergenomic Seqs"
    \item DOW is for "Demo Human or Worm"
    \item DRE represents "Drosophila Enhancers Stark"
    \item HCE is short for "Human Enhancers Cohn"
    \item HEE denotes "Human Enhancers Ensembl"
    \item HRE abbreviates "Human Ensembl Regulatory"
    \item HNP shortens "Human Nontata Promoters"
    \item HOR is an abbreviation for "Human Ocr Ensembl" 
    \item DME simplifies "Dummy Mouse Enhancers Ensembl"
\end{itemize}
The brief statistics for each dataset included in the GUE benchmark are displayed in \pref{tab:gue_detail}.
Similar to GUE, we run the evaluation on a subset of GB, where for each task we randomly select at most 10k samples from the original splits, e.g., training, testing and validation (if any) sets.

\begin{table*}[htbp]
    \centering
    \caption{The brief statistics of datasets reported in the genomic benchmark~\citep{Grevsova23}.}
    \resizebox{.7\linewidth}{!}{
    \begin{tabular}{lcccccc}
        \toprule
        \textbf{Task} & \textbf{\# of Sequences} & \textbf{\# of Classes} & \textbf{Class Ratio} & \textbf{Median Length} & \textbf{Standard Deviation} \\
        \midrule
        DME & $1,210$ & $2$ & $1.0$ & $2,381$ & $984.4$ \\
        DEM & $100,000$ & $2$ & $1.0$ & $200$ & $0.0$ \\
        DOW & $100,000$ & $2$ & $1.0$ & $200$ & $0.0$ \\
        DRE & $6,914$ & $2$ & $1.0$ & $2,142$ & $285.5$ \\
        HCE & $27,791$ & $2$ & $1:9$ & $500$ & $0.0$ \\
        HEE & $154,842$ & $2$ & $1:5$ & $269$ & $122.6$ \\
        HRE & $289,061$ & $3$ & $1.2$ & $401$ & $184.3$ \\
        HNP & $36,131$ & $2$ & $1.2$ & $251$ & $0.0$ \\
        HOR & $174,456$ & $2$ & $1.0$ & $315$ & $108.1$ \\
        \bottomrule
    \end{tabular}
    }
    \label{tab:gb_detail}
\end{table*}

\subsection{BEACON Benchmark}
\label{app:beacon}

To address the lack of standardized benchmarks for RNA foundation models, \textbf{BEACON} (BEnchmark for Comprehensive RNA Task and Language Models) was proposed as the first large-scale and multi-faceted benchmark tailored for RNA understanding. It includes 13 well-curated tasks spanning three major domains: \textit{Structural Analysis}, \textit{Functional Studies}, and \textit{Engineering Applications}. Comprising over 967,000 sequences ranging from 23 to 1,182 nucleotides in length, BEACON provides a diverse landscape to evaluate RNA-based Genomic Foundation Models (GFMs). Notably, pre-trained RNA language models have surpassed previous task-specific state-of-the-art (SOTA) performance on 8 of the 13 tasks.

According to~\citep{Ren24BEACON}, tasks are grouped as follows, detailed data statistics, metrics, and sources are listed in \pref{tab:beacon_detail}.

\begin{itemize}[leftmargin=*]

\item \textbf{Structural Analysis Tasks}: These tasks probe secondary and tertiary RNA structure, which is fundamental to molecular function and therapeutic applications.
\begin{itemize}
    \item \textbf{Secondary Structure Prediction (SSP)}: Predicts paired (stems) and unpaired (loops, bulges) regions using data from the bpRNA-1m dataset. Evaluated via F1 score.
    \item \textbf{Contact Map Prediction (CMP)}: Identifies nucleotide pairs in spatial proximity ($<$8Å). Evaluation metric: Top-$L$ precision.
    \item \textbf{Distance Map Prediction (DMP)}: Predicts pairwise nucleotide distances from 3D structure data. Metric: $R^2$.
    \item \textbf{Structural Score Imputation (SSI)}: Imputes missing experimental structural scores (e.g., icSHAPE signals). Evaluated using $R^2$.
\end{itemize}

\item \textbf{Functional Studies Tasks}: These evaluate the regulatory and functional roles of RNA in gene expression and biological processes.
\begin{itemize}
    \item \textbf{Splice Site Prediction (SPL)}: Classifies each nucleotide as donor, acceptor, or neither. Evaluated with Top-$k$ accuracy.
    \item \textbf{APA Isoform Prediction (APA)}: Predicts usage ratios of alternative polyadenylation sites in the 3' UTR. Metric: $R^2$.
    \item \textbf{Non-coding RNA Function Classification (ncRNA)}: Classifies ncRNAs into categories (e.g., miRNA, lncRNA). Metric: sequence-level accuracy.
    \item \textbf{Modification Prediction (Modif)}: Predicts presence of 12 RNA modification types. Metric: AUC.
    \item \textbf{Mean Ribosome Loading (MRL)}: Estimates translation efficiency of mRNA sequences. Metric: $R^2$.
\end{itemize}

\item \textbf{Engineering Applications Tasks}: These highlight RNA's role in synthetic biology and therapeutic engineering.
\begin{itemize}
    \item \textbf{Vaccine Degradation Prediction (VDP)}: Predicts nucleotide-level degradation rates of vaccine candidates. Evaluated by Mean Columnwise RMSE (MCRMSE).
    \item \textbf{Programmable RNA Switches (PRS)}: Predicts ON/OFF states of synthetic conformational switches. Metric: $R^2$.
    \item \textbf{CRISPR On-Target Prediction (CRI-On)}: Estimates gene-editing efficiency at intended target sites. Metric: weighted Spearman correlation.
    \item \textbf{CRISPR Off-Target Prediction (CRI-Off)}: Predicts editing efficiency at unintended loci. Uses same metric as CRI-On.
\end{itemize}

\end{itemize}

\begin{table*}[t!]
\centering
\caption{Summary of tasks in the BEACON benchmark. "Cls" = classification; "Reg" = regression. Sources as cited in \citet{Ren24BEACON}.}
\label{tab:beacon_detail}
\resizebox{\linewidth}{!}{
\begin{tabular}{lcccccc}
\toprule
\textbf{Task} & \textbf{Train/Val/Test} & \textbf{Metric} & \textbf{Task Type} & \textbf{Level} & \textbf{Max/Mean Length} & \textbf{Source} \\
\midrule
\multicolumn{7}{l}{\textit{Structural Analysis}} \\
SSP & 10,814 / 1,300 / 1,305 & F1 & Multi-label Cls & Nucleotide & 499 / 133.8 & bpRNA \\
CMP & 188 / 23 / 80 & Top-$L$ Precision & Multi-label Cls & Nucleotide & 960 / 110.3 & RNAcontact \\
DMP & 188 / 23 / 80 & $R^2$ & Regression & Nucleotide & 960 / 110.3 & RNAcontact \\
SSI & 14,049 / 1,756 / 3,095 & $R^2$ & Regression & Nucleotide & 100 / 100 & StructureImpute \\
\midrule
\multicolumn{7}{l}{\textit{Functional Studies}} \\
SPL & 144,628 / 18,078 / 16,505 & Top-$k$ ACC & Multi-class Cls & Nucleotide & 100 / 100 & SpliceAI \\
APA & 145,463 / 33,170 / 49,755 & $R^2$ & Regression & Sequence & 186 / 186 & APARENT \\
ncRNA & 5,679 / 650 / 2,400 & ACC & Multi-class Cls & Sequence & 1,182 / 158.4 & GENCODE+Rfam \\
Modif & 304,661 / 3,599 / 1,200 & AUC & Multi-label Cls & Sequence & 101 / 101 & MultiRM \\
MRL & 76,319 / 7,600 / 7,600 & $R^2$ & Regression & Sequence & 100 / 61.5 & Optimus \\
\midrule
\multicolumn{7}{l}{\textit{Engineering Applications}} \\
VDP & 2,155 / 245 / 629 & MCRMSE & Multi-label Reg & Nucleotide & 130 / 118.5 & OpenVaccine \\
PRS & 73,227 / 9,153 / 9,154 & $R^2$ & Multi-label Reg & Sequence & 148 / 148 & Angenent-Mari \\
CRI-On & 1,453 / 207 / 416 & Spearman Corr & Regression & Sequence & 23 / 23 & DeepCRISPR \\
CRI-Off & 14,223 / 2,032 / 4,064 & Spearman Corr & Regression & Sequence & 23 / 23 & DeepCRISPR \\
\bottomrule
\end{tabular}
}
\end{table*}

\subsection{Data Filtering in Benchmarking}
\label{app:filtering}
The pertaining involves RNA sequences and structures prediction, we take the data and annotation leakage problem seriously.
\begin{itemize}[leftmargin=*]
    \item To avoid structure annotation leakage of downstream benchmarks, the secondary structure predictors for all FMs were randomly initialized for fair comparisons, which means the pre-trained structure predictor of \our\ was not used in benchmarks, except for zero-shot SSP experiments. Please find the source codes for details.
    \item To reduce sequence leakage caused by evolutionary conservative sequences across multiple species, we use the ch-hit-est tool to calculate the sequence similarity between sequences from the OneKP database and downstream tasks. We adopt the similarity threshold of $80\%$ for ch-hit-est ~\citep{Li06cdhit} to eliminate sequences whose homogeneous sequences appeared in the OneKP database. Subsequently, we exploit the blastn~\citep{Altschul90Blast} tool to query potentially leaked sequences in downstream benchmark datasets and further alleviate the data leakage problem. The e-value has been set to $1$ for rigorous sequence filtering. 
\end{itemize}

\section{Detailed Benchmark Performance Report}
\label{app:detailed_performance_report}

\subsection{Evaluation Settings in Benchmarking}
In this experiment, we carefully selected a set of key hyperparameters to optimize model performance. Below are the main hyperparameter settings along with detailed explanations:

\begin{itemize}[leftmargin=*]
    \item \textbf{Dropout}: To prevent the model from overfitting during training, we set the Dropout value to 0.1, meaning that no random neuron dropout is applied during training. This choice was made based on our consideration of model stability and generalization ability.

    \item \textbf{Learning Rate}: We set the learning rate to 2e-5, which is a relatively small value to ensure stable convergence, especially in complex training tasks. A smaller learning rate helps to avoid drastic fluctuations during the training process, leading to more precise optimization.

    \item \textbf{Weight Decay}: We applied a weight decay of 0.01 to control model complexity and prevent overfitting. Weight decay is a regularization technique that effectively constrains the growth of model parameters, maintaining the model's generalization capability.

    \item \textbf{Adam Optimizer}: We used the Adam optimizer with its parameters set to $\beta_1=0.9$ and $\beta_2=0.999$. The Adam optimizer combines the benefits of momentum and adaptive learning rates, accelerating convergence and adapting to different gradient changes, thereby improving the efficiency and effectiveness of model training.

    \item \textbf{Learning Rate Scheduler}: We opted for a linear decay learning rate scheduler, allowing the learning rate to gradually decrease during training. This strategy helps the model make smaller adjustments as it approaches the optimal solution, ensuring a better convergence outcome.

    \item \textbf{Batch Size}: The batch size was set to 8. This relatively small batch size helps to efficiently train the model within limited memory resources, particularly when handling large-scale data, enabling a balance between model performance and computational resource usage. For long-sequence tasks ($\geq$ 4 kb) we reduce batch to 1-2 to fit 24 GB GPUs; epochs are capped at 10.

    \item \textbf{\# of Epochs}: We set the number of training epochs to 20. This setting ensures that the model can fully learn the features within the data while avoiding the negative effects of overtraining.

    \item \textbf{Early Stopping}: We implemented an early stopping mechanism, terminating the training early if the validation performance does not improve for 5 consecutive epochs. This mechanism effectively prevents model overfitting and saves training time.
\end{itemize}

It is important to note that for different tasks, some hyperparameter settings may be adjusted. To obtain accurate experimental results, please refer to the detailed parameter configurations in the compiled dataset specific to each task.

\subsection{Evaluation GFMs in Benchmarks}
\label{app:baselines}

To evaluate existing GFMs on the four integrated benchmark suites (RGB, PGB, GUE, GB), we adapted each public checkpoint to the \our{} interface and report their results in \pref{sec:experiments}. \pref{tab:fm_details} summarizes key statistics for all baselines.\footnote{%
`SNT' denotes \emph{single-nucleotide tokenisation}, one character per base.}%

\begin{table*}[t!]
  \centering
  \caption{Statistics of RNA and DNA foundation-model baselines.  Pre-training data sizes come verbatim from each paper and are therefore not directly comparable.}
  \resizebox{\linewidth}{!}{
    \begin{tabular}{lcccccc}
      \toprule
      \textbf{Model} & \textbf{Tokenisation} & \textbf{\#Params} & \textbf{Pre-train Size} & \textbf{Data Source} & \textbf{Species} & \textbf{Sequence Type}\\
      \midrule
      DNABERT-2               & BPE  & 117 M  & 32.5 B tokens  & 1000 Genomes       & Human+$135$ sp. & DNA \\
      NT-V2-100M             & k-mer & 96 M   & 300 B tokens  & Multi-source DNA   & $850$ sp.       & DNA \\
      HyenaDNA-Large         & SNT  & 47 M   & 3.2 B tokens  & Human GRCh38       & Human           & DNA \\
      Caduceus               & SNT  & 1.9 M  & 35 B tokens   & Human GRCh38       & Human           & DNA \\
      Agro-NT-1B             & k-mer & 985 M  & 472 B tokens  & Ensembl Plants     & 48 plants       & DNA \\
      SpliceBERT             & SNT  & 19 M   & 2 M seqs      & UCSC pre-mRNA      & Vertebrates     & pre-mRNA \\
      RNA-BERT               & SNT  & 0.5 M  & 4 069 families& Rfam               & Multi-sp.       & ncRNA \\
      RNA-MSM                & SNT  & 96 M   & 4 069 families& Rfam (MSA)         & Multi-sp.       & ncRNA \\
      RNA-FM                 & SNT  & 96 M   & 23 M seqs     & RNAcentral         & Multi-sp.       & ncRNA \\
      3UTRBERT               & k-mer& 86 M   & 20 k seqs     & GENCODE UTR        & Human           & mRNA 3'UTR \\
      \og{}    & SNT  & 186 M  & 54.2 B tokens & OneKP              & 1124 plants     & mRNA/CDS/UTR \\
      \bottomrule
    \end{tabular}}
  \label{tab:fm_details}
  \vspace{-6pt}
\end{table*}

ViennaRNA secondary-structure annotations were \emph{optionally} supplied for structure-aware variants of \our{}, but are \textbf{not} required by baseline models.

We exclude models without publicly-usable code or checkpoints (e.g.\ Uni-RNA, 5UTR-LM) and briefly \emph{summarize} each included FM below; see the original publications for full method details.

\begin{itemize}[leftmargin=*]
    \item ViennaRNA~\citep{Lorenz11}. ViennaRNA is a comprehensive genomic analysis tool that includes a diverse set of interfaces, such as RNAFold\footnote{\url{https://www.tbi.univie.ac.at/RNA/RNAfold.1.html}} and RNAInverse\footnote{\url{https://www.tbi.univie.ac.at/RNA/RNAinverse.1.html}} design. ViennaRNA serves as the baseline for RNA structure prediction and RNA design in our experiments.
    \item DNABERT2~\citep{Zhou23DNABERT2}. DNABERT2 is one of the latest DNA FMs which improves the performance of DNABERT. The main modification of DNABERT2 is the tokenization method, which was changed to BPE from k-mers.
    \item HyenaDNA~\citep{Nguyen23}. HyenaDNA is an autoregressive FM optimized for long-range genome data processing. HyenaDNA is based on the Hyena convolution architecture and capable of handling sequences up to $1$M bases in length.
    \item Caduceus~\citep{Schiff24caduceus}. Caduceus\footnote{\url{https://huggingface.co/kuleshov-group/caduceus-ps_seqlen-131k_d_model-256_n_layer-16}} is an advanced DNA language model built on the MambaDNA architecture, designed to address challenges in genomic sequence modeling, such as long-range token interactions and reverse complementarity (RC). 
    \item Nucleotide Transformer (NT) V2~\citep{Dalla23NT}. The NT FMs were trained on DNA data, including the human reference genome and multi-species DNA sequences. They aim to capture the complex patterns within nucleotide sequences for various genome modeling applications.
    \item Agricultural Nucleotide Transformer (Agro-NT)~\citep{Mendoza23}. Agro-NT is a large-scale DNA FM ($1$B parameters) akin to the Nucleotide Transformers but with a focus on plant DNA.
    \item SpliceBERT~\citep{ChenZD23}. It was trained on $2$M precursor messenger RNA (pre-mRNA) and specialised in RNA splicing of pre-mRNA sequences.
    \item 3UTRBERT~\citep{YangLP23}. This model was trained on $20$k 3'UTRs for 3'UTR-mediated gene regulation tasks. It uses k-mers tokenization instead of SNT.
    RNA-BERT~\citep{Akiyama22informative}. RNA-BERT is a BERT-style model pre-trained on a large corpus of non-coding RNA sequences. It uses masked language modeling (MLM) as its primary training objective. The model is designed to predict RNA structural alignments and can be fine-tuned for various RNA sequence classification and regression tasks
    \item RNA-MSM~\citep{ZhangLJGXLCSHXS24} RNA-MSM is an unsupervised RNA language model based on multiple sequence alignment (MSA). It is the first model of its kind to produce embeddings and attention maps that directly correlate with RNA secondary structure and solvent accessibility. RNA-MSM is particularly effective for tasks involving evolutionary relationships in RNA sequences.
    \item RNA-FM~\citep{Chen22} RNA-FM is a BERT-based RNA foundation model trained on a vast dataset of non-coding RNA sequences. The model excels in predicting RNA structure and function by leveraging masked language modeling (MLM) during pre-training. RNA-FM's training data is sourced from the RNAcentral database, providing it with extensive knowledge across diverse RNA species.
    \item \our. \our\ is the RNA genome FM that advocates the importance of sequence-structure alignment. Moreover, it is the first FM which addressed the \insilico{} RNA design task.
    \item \textbf{OmniGenome}: A FM dedicated to RNA genome modeling. This model leverages the computation-based structure to enhance the genome modeling ability and archives impressive performance on both RNA and DNA genomes.
\end{itemize}

\subsection{RNA Genomic Benchmark (RGB)} 
\label{sec:results-rgb}

The RGB comprises seven challenging single-nucleotide level RNA modeling tasks, designed to evaluate models' fine-grained capabilities in understanding RNA sequences, such as predicting RNA structures. These tasks include mRNA degradation rate prediction, single-nucleotide modification detection (SNMD), single-nucleotide modification regression (SNMR), and RNA secondary structure prediction tasks such as Archive2, Stralign, and bpRNA. Additionally, EternaV2 evaluates models on RNA design tasks.

\pref{tab:rgb-results} presents the performance of various GFMs on the RGB tasks. Overall, \og{} achieves the best performance across all tasks, highlighting its exceptional capability in RNA structure modeling. This superior performance can be attributed to \og's integration of structural information into its modeling process, which is particularly beneficial for tasks requiring secondary structure prediction.

\label{sec:rgb_exp}
\begin{table}[H]
	\centering
	\caption{The performance of \our\ and baseline models on the RGB, with results averaged based on five random seeds. ``N.A.'' means not available for predictive tasks.}
	\resizebox{.75\linewidth}{!}{
	\begin{tabular}{lccccccc}
		\toprule
		\multirow{2}[4]{*}{\textbf{Model}} & \textbf{mRNA}  &  \textbf{SNMD}  & \textbf{SNMR}  &  \textbf{Archive2} &  \textbf{Stralign} &  \textbf{bpRNA} &  \textbf{EternaV2} \\
		\cmidrule{2-8}          & RMSE  & AUC   & F1    & F1    & F1    & F1  & Accuracy \\
		\midrule
		ViennaRNA & N.A.    & N.A.    & N.A.    & $73.99$ & $74.09$ & $65.03$ & $33$\\
		MXFold2     & N.A.    & N.A.    & N.A.    & $90.09$ & $97.01$ & $64.99$ & N.A.\\
		Ufold           & N.A.    & N.A.    & N.A.    & $89.78$ & $95.76$ & $78.38$  & N.A.\\
		\midrule
		DNABERT2 & $0.8158$ & $49.94$ & $15.86$ & $55.73$ & $64.09$ & $33.77$ & $0$ \\
		HyenaDNA & $0.8056$ & $53.32$ & $39.80$  & $71.18$ & $91.24$ & $57.43$ & $0$\\
            Caduceus & $0.8026$ & $57.01$ & $39.59$ & $74.37$ & $92.28$ & $59.76$ & $0$\\
		NT-V2 & $0.7826$ & $50.49$ & $26.01$ & $68.36$ & $83.18$ & $56.95$ & $0$\\
		Agro-NT & $0.7830$ & $49.99$ & $26.38$ & $62.81$ & $72.54$ & $46.87$ & $0$\\
		SpliceBERT & $0.7340$ & $58.11$ & $46.44$ & $79.89$ & $93.81$ & $71.59$ & $3$\\
		3UTRBERT & $0.7772$ & $50.02$ & $24.01$ & $68.62$ & $88.55$ & $57.90$ & $0$\\
		RNABERT & $0.8087$ & $51.32$ & $29.14$ & $24.66$ & $83.68$ & $47.96$ & $0$\\
		RNA-MSM & $0.7321$ & $57.86$ & $45.22$ & $68.72$ & $91.15$ & $64.44$ & $2$\\
		RNA-FM & $0.7297$ & $59.02$ & $42.21$ & $82.55$ & $95.07$ & $78.16$ & $4$\\
		\og & $\mathbf{0.7121}$ & $\mathbf{64.13}$ & $\mathbf{52.44}$ & $\mathbf{91.89}$ & $\mathbf{98.21}$ & $\mathbf{83.18}$ & $\mathbf{84}$ \\
		\bottomrule
	\end{tabular}%
	}
	
	\label{tab:rgb-results}%
\end{table}%
In particular, \og{} significantly outperforms other models on the mRNA degradation rate prediction task, achieving an RMSE of $0.7121$, compared to the second-best RMSE of $0.7297$ by RNA-FM. Similarly, for the SNMD task, \og{} achieves an AUC of $64.13$, surpassing the second-best score of $59.02$ by RNA-FM. These results indicate that \og{} effectively captures single-nucleotide level variations, which are crucial in RNA function and regulation. Furthermore, in the secondary structure prediction tasks (Archive2, Stralign, bpRNA), \og{} demonstrates superior performance, highlighting its proficiency in modeling RNA secondary structures. This can be attributed to \og's incorporation of structural context during pretraining, which enhances its ability to understand and predict RNA folding patterns.
One limitation observed is that models not specifically designed for RNA tasks, such as DNABERT2 and HyenaDNA, perform poorly on RNA-specific tasks. This underscores the importance of tailoring GFMs to the specific characteristics of RNA sequences.

In summary, the RGB results highlight the critical role of structural modeling in RNA genomics and demonstrate the effectiveness of \og{} in capturing complex RNA features. Future GFMs may benefit from incorporating structural information to enhance performance on RNA-related tasks.

\subsection{Plant Genomic Benchmark (PGB)}
\label{sec:results-pgb}

The PGB comprises DNA-based tasks focused on plant biology. The sequences in PGB contain up to $6,000$ bases, presenting challenges for models in handling long genomic sequences.
\pref{tab:pgb-results} summarizes the performance of various GFMs on the PGB tasks. Resembling the results of RGB, \og{} achieves top-tier performance across most tasks, even though it was only trained on RNA. This suggests that \og{} generalizes well to DNA-based tasks, likely due to shared sequence motifs and structural similarities between RNA and DNA.

\begin{table*}[t!]
  \centering
  \caption{Performance of open-source GFMs on PGB, where the results are re-implemented based on our evaluation protocol. ``PolyA'' stands for Polyadenylation, ``Chrom Acc'' for Chromatin Accessibility, ``Prom Str'' for Promoter Strength, ``Term Str'' for Terminator Strength, ``Splice'' for Splice Site, ``Gene Exp'' for Gene Expression, and ``Enh Reg'' for Enhancer Region. }
  \setlength{\tabcolsep}{2pt} 
  \resizebox{.75\linewidth}{!}{
    \begin{tabular}{lcccccccc}
    \toprule
    \multirow{2}[1]{*}{\textbf{Model}} & \textbf{PolyA} & \textbf{LncRNA} & \textbf{Chrom Acc} & \textbf{Prom Str} & \textbf{Term Str} & \textbf{Splice} & \textbf{Gene Exp} & \textbf{Enhancer} \\
    \cmidrule{2-9}    & F1 & F1 & F1 & RMSE & RMSE & F1 & RMSE & F1 \\
    \midrule
    DNABERT2  & $41.35$ & $72.55$ & $61.49$ & $0.99$ & $0.24$ & $45.34$ & $14.78$ & $36.40$ \\
    HyenaDNA  & $83.11$ & $58.21$ & $52.20$ & $0.88$ & $0.26$ & $90.28$ & $14.79$ & $66.17$ \\
    Caduceus  & $70.89$ & $68.40$ & $64.53$ & $0.91$ & $0.26$ & $78.51$ & $14.72$ & $60.83$ \\	
    NT-V2     & $71.26$ & $73.08$ & $65.71$ & $0.81$ & $0.27$ & $95.05$ & $14.79$ & $73.89$ \\
    Agro-NT   & $78.89$ & $67.24$ & $63.27$ & $0.94$ & $0.78$ & $88.45$ & $15.56$ & $62.83$ \\
    SpliceBERT& $65.23$ & $71.88$ & $63.62$ & $0.75$ & $0.22$ & $96.45$ & $\mathbf{14.70}$ & $69.71$ \\
    3UTRBERT  & $76.48$ & $70.75$ & $63.71$ & $1.04$ & $0.36$ & $94.44$ & $14.87$ & $71.67$ \\
    RNA-BERT  & $78.54$ & $61.99$ & $48.94$ & $1.81$  & $0.38$  & $94.45$ & $14.89$ & $57.61$ \\
    RNA-MSM   & $84.25$ & $67.49$ & $53.52$ & $1.28$  & $0.28$  & $95.49$ & $14.87$ & $61.45$ \\
    RNA-FM    & $84.94$ & $68.75$ & $54.92$ & $0.95$  & $0.27$  & $95.95$ & $14.83$ & $57.14$ \\
    \og & $\mathbf{87.55}$ & $\mathbf{77.96}$ & $\mathbf{67.69}$ & $\mathbf{0.59}$ & $\mathbf{0.18}$ & $\mathbf{98.41}$ & $14.71$ & $\mathbf{79.77}$ \\
    \bottomrule
    \end{tabular}
  }
  \label{tab:pgb-results}
  
\end{table*}

In the PolyA task, \og{} achieves an F1 score of $87.55$, outperforming the second-best model, RNA-FM, which achieves $84.94$. Similarly, for the LncRNA task, \og{} attains an F1 score of $77.96$, significantly higher than the second-best score of $73.08$ by NT-V2.
\og{} excels in the Splice Site prediction task, achieving an F1 score of $98.41$, surpassing the second-best score of $96.45$ by SpliceBERT. This suggests that \og{} effectively captures sequence motifs important for splicing, which is crucial in gene expression regulation.
These results indicate that GFMs incorporating structural context, like \og{}, can generalize effectively across different genomic modalities (RNA and DNA) and species (plants). The strong performance of \og{} on DNA-based tasks suggests that structural modeling enhances the understanding of genomic sequences beyond the specific type of nucleic acid.
However, it's also observed that some models specifically designed for DNA tasks, such as NT-V2 and SpliceBERT, perform competitively on certain tasks. This underscores the importance of task-specific pretraining and the potential benefits of integrating both sequence and structural information in GFMs.

In summary, the PGB results highlight the potential for cross-modal generalization in GFMs and the value of incorporating structural context to enhance performance on diverse genomic tasks.

\subsection{Genomic Understanding Evaluation (GUE)} 
\label{sec:results-gue}

The GUE is a multi-species benchmark like RGB and PGB, but focuses on the non-plant genomes. The sequences in GUE range in length and complexity, providing a robust assessment of GFMs' abilities to generalize across species and genomic tasks.
\pref{tab:gue-results} presents the performance of various GFMs on the GUE tasks. While \og{} does not achieve the highest performance on all tasks, it consistently delivers competitive results, demonstrating strong cross-species generalization despite being primarily trained on RNA data.

\begin{table*}[htbp]
  \centering
  \caption{Performance of open-source GFMs on GUE, where the results are re-implemented based on our evaluation protocol. The performance for each task is the average macro F1 score in all sub-datasets.}
  \resizebox{\linewidth}{!}{
    \begin{tabular}{lccccccc}
    \toprule
    \multirow{2}[1]{*}{\textbf{Model}} & \textbf{Yeast EMP} & \textbf{Mouse TF-M} & \textbf{Virus CVC} & \textbf{Human TF-H} & \textbf{Human PD} & \textbf{Human CPD} & \textbf{Human SSP} \\
    \cmidrule{2-8}
    & F1 & F1 & F1 & F1 & F1 & F1 & F1 \\
    \midrule
    DNABERT-2 & $75.85$ & $\mathbf{86.23}$ & $58.23$ & $81.80$ & $90.17$ & $82.57$ & $85.21$ \\
    HyenaDNA  & $73.08$ & $73.44$ & $27.59$ & $77.62$ & $91.19$ & $84.31$ & $83.34$ \\
    Caduceus  & $73.49$ & $78.18$ & $27.49$ & $79.56$ & $89.13$ & $85.09$ & $81.82$ \\ 
    NT-V2     & $74.93$ & $78.10$ & $32.71$ & $79.12$ & $90.87$ & $84.70$ & $84.13$ \\
    SpliceBERT& $77.66$ & $84.97$ & $47.17$ & $\mathbf{82.77}$ & $\mathbf{92.24}$ & $83.96$ & $\mathbf{93.81}$ \\
    3UTRBERT  & $71.89$ & $71.46$ & $34.84$ & $74.85$ & $82.37$ & $\mathbf{90.51}$ & $81.95$ \\
    RNA-BERT  & $60.14$ & $59.83$ & $21.08$ & $67.48$ & $79.87$ & $76.25$ & $44.75$ \\
    RNA-MSM    & $64.99$ & $79.15$ & $51.81$ & $78.72$ & $91.28$ & $85.42$ & $84.24$ \\
    RNA-FM     & $74.41$ & $78.24$ & $52.22$ & $79.27$ & $92.18$ & $86.05$ & $84.76$  \\
    \og       & $\mathbf{78.51}$ & $84.72$ & $\mathbf{64.41}$ & $81.73$ & $90.04$ & $85.22$ & $90.39$ \\
    \bottomrule
    \end{tabular}%
    }
    
  \label{tab:gue-results}%
\end{table*}%

In the Yeast EMP task, \og{} achieves the highest F1 score of $78.51$, slightly outperforming SpliceBERT of $77.66$. For the Virus CVC task, \og{} also achieves the best performance with an F1 score of $74.72$, indicating its strong ability to model viral genomic sequences.
However, for tasks like Human TF-H and Human SSP, models like SpliceBERT and DNABERT2 achieve higher scores. This suggests that these models may be better optimized for human genomic sequences or specific tasks like splice site prediction.
The results on GUE highlight the challenges in developing GFMs that generalize across different species and genomic tasks. While \og{} demonstrates strong cross-species performance, there is variability depending on the specific task and species.
These findings suggest that combining the strengths of different GFMs or developing ensemble methods could be a fruitful direction for future research. Additionally, incorporating more diverse training data and task-specific fine-tuning may enhance the performance of GFMs across a broader range of tasks.

\subsection{Genomic Benchmarks (GB)} 
\label{sec:results-gb}

GB is a collection of DNA genome datasets aimed at evaluating the performance of models on sequence classification tasks involving regulatory elements such as promoters, enhancers, and open chromatin regions across different species including humans, mice, and roundworms.
\pref{tab:gb-results} shows the performance of various GFMs. The tasks are denoted by their species and regulatory elements, and the acronyms are explained in \pref{app:gb}.

\begin{table*}[htbp]
  \centering
  \caption{Performance of open-source GFMs on GB, where the results are re-implemented based on our evaluation protocol.. The performance (macro F1) for each task is the average macro F1 score across all sub-datasets.}
  \resizebox{.72\linewidth}{!}{
    \begin{tabular}{lccccccccc}
    \toprule
    \multirow{2}[1]{*}{\textbf{Model}} & \textbf{DEM} & \textbf{DOW} & \textbf{DRE} & \textbf{DME} & \textbf{HCE} & \textbf{HEE} & \textbf{HRE} & \textbf{HNP} & \textbf{HOR} \\
    \cmidrule{2-10}      
    & \textbf{F1} & \textbf{F1} & \textbf{F1} & \textbf{F1} & \textbf{F1} & \textbf{F1} & \textbf{F1} & \textbf{F1} & \textbf{F1} \\
    \midrule
    DNABERT-2  & $92.67$ & $95.17$ & $43.77$ & $77.21$ & $\mathbf{75.58}$ & $80.66$ & $78.14$ & $85.80$  & $68.03$ \\
    HyenaDNA   & $88.21$ & $94.13$ & $70.11$ & $76.44$ & $70.38$ & $79.58$ & $96.33$ & $85.99$ & $67.03$ \\
    Caduceus   & $92.13$ & $94.74$ & $72.03$ & $75.61$ & $70.20$ & $76.47$ & $79.16$ & $84.36$ & $63.17$ \\
    NT-V2      & $91.66$ & $94.32$ & $\mathbf{78.20}$  & $\mathbf{81.72}$ & $71.98$ & $79.85$ & $93.30$  & $85.30$  & $68.53$ \\
    SpliceBERT & $\mathbf{94.72}$ & $\mathbf{96.42}$ & $72.29$ & $74.70$  & $73.50$  & $79.60$  & $95.23$ & $\mathbf{89.57}$ & $68.89$ \\
    3UTRBERT   & $89.50$  & $90.22$ & $74.35$ & $80.14$ & $70.23$ & $76.33$ & $\mathbf{98.47}$ & $82.49$ & $66.78$ \\
    RNA-BERT   & $76.56$  & $62.17$ & $50.11$ & $60.79$ & $66.69$ & $63.29$ & $46.57$ & $73.80$ & $56.59$ \\
    RNA-MSM    & $79.38$  & $93.71$ & $54.13$ & $75.90$ & $69.79$ & $78.07$ & $94.87$ & $84.28$ & $63.93$  \\
    RNA-FM     & $91.53$  & $95.49$ & $74.77$ & $79.74$ & $71.62$ & $80.03$ & $95.72$ & $87.14$ & $\mathbf{69.38}$  \\
    \og       & $94.16$ & $93.49$ & $77.17$ & $80.34$ & $73.51$ & $\mathbf{82.23}$ & $95.66$ & $87.87$ & $68.97$ \\
    \bottomrule
    \end{tabular}%
  }

  \label{tab:gb-results}%
\end{table*}

In the DEM and DOW tasks, SpliceBERT achieves the highest F1 scores, with \og{} closely following in DEM and RNA-FM in DOW. For the DRE task, NT-V2 achieves the best performance with an F1 score of $78.20$, with \og{} performing closely.
In the HEE task, \og{} attains the highest F1 score of $82.23$, surpassing the second-best score of $80.66$ by DNABERT-2. This indicates \og's effectiveness in modeling human enhancer regions.
From a global perspective, these results demonstrate that while different GFMs excel in specific tasks, \og{} consistently performs well across various genomic benchmarks, highlighting its versatility. The performance variations across models suggest that task-specific features and training data significantly impact model efficacy.
A limitation observed is that GFMs primarily trained on RNA data, like RNA-BERT and RNA-MSM, lost on DNA-based tasks. This underscores the importance of training data relevance and the potential need for multimodal pretraining strategies.

In conclusion, the GB results emphasize the need for GFMs that can generalize across different genomic tasks and species. Integrating structural information, as done in \og{}, appears to enhance model performance on complex genomic tasks.

\subsection{BEACON Results}

We have completed data organization and the compilation of benchmark tasks. However, we are currently unable to provide results because the experimental outcomes cannot be reproduced. Our next steps involve verifying the custom implementation of the evaluation metrics and checking the integrity of the dataset. Additionally, there are still missing dataset in the BEACON benchmark. We have submitted an issue to the authors and are awaiting a response.

\subsection{Overall Discussion}

Our comprehensive evaluation across four genomic benchmarks reveals that \og{} consistently achieves top-tier performance, particularly excelling in tasks that involve structural modeling of RNA sequences. The integration of structural information in \og{} enhances its ability to capture complex sequence features, which is advantageous across diverse genomic tasks.
While \og{} demonstrates strong performance even on DNA-based tasks, models specifically tailored to certain tasks or species, such as SpliceBERT and DNABERT2, sometimes outperform \og{} in those specific contexts. This suggests that task-specific or species-specific pretraining can provide benefits, and there is potential for combining the strengths of different models.

The absence of results for certain models on some benchmarks (e.g., RNA-BERT, RNA-MSM, and RNA-FM on GUE) highlights the challenges in benchmarking GFMs across diverse datasets. Differences in model architectures, pretraining data, and tokenization strategies can impact a model's applicability to specific tasks. Future work should focus on developing unified evaluation protocols and improving the interoperability of GFMs. An important consideration is the need for detailed descriptions of the models evaluated, including their architectures, pretraining data, and key features. This information is crucial for understanding the factors contributing to their performance and for reproducing results.

Overall, our comprehensive benchmarking highlights the importance of integrating structural information into GFMs and suggests that models capable of capturing both sequence and structural features offer improved performance across a range of genomic tasks. This work provides valuable insights for the development of next-generation GFMs and underscores the need for continued efforts in benchmarking to drive advancements in genomic modeling.

\section{Integrated Genomic Foundation Models in \our{}}
\label{app:gfm_details}

\begin{figure}[h!]
\centering
\includegraphics[width=.7\linewidth]{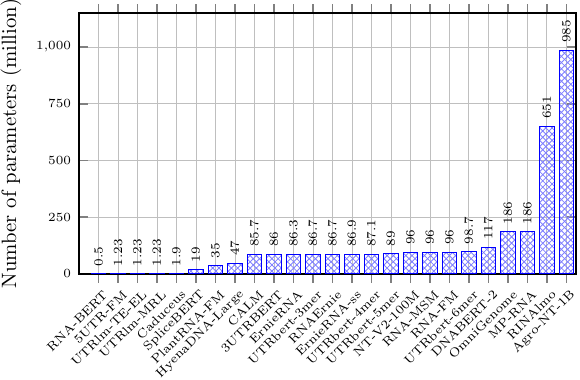}
\caption{Overview of GFM parameter scales adapted in \our{}. The models span from approximately 0.5 million to nearly 1 billion parameters, showcasing the scalability of the benchmarking framework. Our own model, OmniGenome, is included with 186 million parameters.}
\label{fig:model_scalability}
\end{figure}

As illustrated in \pref{fig:model_scalability}, \our{} has been utilized to benchmark a diverse array of GFMs (26 types with 31 models in total) that vary significantly in scale. The tested models range from compact architectures, such as RNA-BERT (0.5 million parameters) and Caduceus (1.9 million parameters), to substantially larger models like DNABERT-2 (117 million), our own OmniGenome (186 million), and Agro-NT-1B, which approaches 1 billion (985 million) parameters. This breadth demonstrates the capability of \our{} to handle and rigorously evaluate GFMs across a wide spectrum of complexities and sizes, which is crucial for comprehensive and fair benchmarking.

The \our{} platform is designed for continuous growth, with ongoing efforts to incorporate new and emerging GFMs. The following list details the models currently integrated, categorized broadly by scale and primary application domain. Each entry provides key characteristics and relevant citations.

\begin{itemize}[leftmargin=*,nosep]

\item RNA-BERT. {\small(0.5 M; MLM on 4,069 ncRNA families) } A pioneering lightweight BERT encoder, RNA-BERT introduced nucleotide-level masking for effective ncRNA representation~\citep{Akiyama22informative}.

\item 5UTR-FM. {\small(1.23 M; 18M UTRs, auxiliary structure labels) } This foundation model targets the regulatory grammar of \textbf{5'-UTRs}. It excels at predicting translation efficiency by leveraging extensive UTR pre-training and structural information~\citep{Chu24}.

\item UTR-LM (TE–EL).{\small(1.23 M) } A variant of the UTR-LM series, specifically focused on co-predicting \textbf{T}ranslation \textbf{E}fficiency and mRNA \textbf{E}xpression \textbf{L}evel from 5'-UTR sequences~\citep{Chu24}.

\item UTR-LM (MRL).{\small(1.23 M) } Another UTR-LM variant, this model is trained to predict \textbf{M}ean \textbf{R}ibosome \textbf{L}oad, serving as a proxy for global protein output based on 5'-UTR features~\citep{Chu24}.

\item Caduceus. {\small(1.9 M; MambaDNA for human chromosomes) } The first \textbf{MambaDNA} state-space model, Caduceus introduces reverse-complement (RC) equivariant gating, enabling efficient processing of long human chromosomal sequences~\citep{Schiff24caduceus}.

\item SpliceBERT. {\small(19 M; 2M vertebrate pre-mRNA) } Pre-trained on a vast dataset of vertebrate pre-mRNA, SpliceBERT achieves state-of-the-art splice-site detection using single-nucleotide tokenization~\citep{ChenZD23}.

\item PlantRNA-FM. {\small(35 M; \textit{Arabidopsis} \& 20+ crops, sequence \& structure) } An interpretable encoder designed for plant systems, PlantRNA-FM jointly leverages sequence and structural information to predict translation control mechanisms in \textit{Arabidopsis} and over 20 other crop species~\citep{Yu2024interpretable}.

\item HyenaDNA-L. {\small(47 M; Autoregressive, Hyena filters) } This autoregressive DNA language model employs Hyena operators, allowing it to handle sequences up to 1 Mb with remarkable memory efficiency~\citep{Nguyen23}.

\item CALM. {\small(85.7 M; Codon-aware, ESM-based) } The Codon-aware Language Model (\textsc{CaLM}) adapts ESM layers to operate on nucleotide triplets (codons), thereby improving the analysis of protein-coding regions~\citep{Outeiral24CaLM}.

\item 3UTRBERT. {\small(86 M; BERT for 3'-UTRs) } An encoder focusing on 3'-UTRs, 3UTRBERT provides family-aware embeddings crucial for understanding post-transcriptional regulation tasks~\citep{YangLP23}.

\item UTRbert-{3,4,5,6}mer. {\small(86–87 M) } This suite comprises four 3UTRBERT checkpoints, each utilizing a different k-mer tokenization strategy (3-mer to 6-mer). They are valuable for investigating the impact of tokenization granularity on UTR analysis~\citep{YangLP23}. 

\item ErnieRNA. {\small(86.3 M; Structure-enhanced, base-pair priors) } An RNA language model enhanced with structural information, ErnieRNA injects base-pair priors during masked language modeling (MLM) pre-training~\citep{Yin24}.

\item RNAErnie. {\small(86.7 M; Motif-aware, motif-level masking) } This model employs motif-aware pre-training, performing random masking at the \textit{motif} level to better capture cis-regulatory elements within RNA sequences~\citep{WangBLLMKX24}.

\item ErnieRNA-ss. {\small(86.9 M) } A fine-tuned variant of ErnieRNA, specializing in RNA secondary-structure (\textbf{ss}) annotation tasks.

\item NT-V2. {\small(96 M; 300B DNA tokens, 850 species, hybrid k-mer) } The second-generation \textbf{Nucleotide Transformer}, NT-V2 was trained on an extensive dataset of 300 billion DNA tokens from 850 species, utilizing a hybrid k-mer tokenization approach~\citep{Dalla23NT}.

\item RNA-MSM. {\small(96 M; MSA-based evolutionary context) } The first RNA language model to leverage multiple-sequence alignments (MSAs), RNA-MSM couples evolutionary context with token predictions for enhanced understanding~\citep{ZhangLJGXLCSHXS24}.

\item RNA-FM. {\small(96 M; 23M ncRNA sequences) } A large-scale encoder pre-trained on 23 million non-coding RNA sequences, RNA-FM has demonstrated superior performance over traditional tools like RNAfold for structure prediction tasks~\citep{Chen22}.

\item UTRbert-6mer. {\small(98.7 M) } The checkpoint within the UTRbert family employing the most granular 6-mer tokenization, particularly effective for discovering long motifs in UTRs.

\item DNABERT-2. {\small(117 M; BPE, 32B tokens, 136 species) } An upgrade to the classic DNABERT, DNABERT-2 uses byte-pair encoding (BPE) and was pre-trained on 32 billion tokens from 136 species~\citep{Zhou23DNABERT2}.

\item OmniGenome. {\small(186 M; Multi-modal plant RNA/DNA, 54B tokens) } Our flagship multi-modal foundation model, OmniGenome is pre-trained on 54 billion plant RNA and DNA tokens. It demonstrates state-of-the-art (SoTA) coverage across multiple benchmark suites (see Sec.~\ref{sec:experiments}).

\item MP-RNA. {\small(186 M; Multi-Phyla, explicit structure pre-training) } The \underline{M}ulti-\underline{P}hyla RNA foundation model (MP-RNA) incorporates explicit structure information during pre-training, achieving over 40\% performance gains on RGB tasks~\citep{MPRNA24}.

\item RiNALMo. {\small(651 M; 36M ncRNA sequences, Flash-Attention-2) } Currently the \emph{largest} publicly available RNA language model, RiNALMo (650 M parameters) was trained on 36 million ncRNA sequences and utilizes Flash-Attention-2 for efficiency~\citep{RiNALMo24}.

\item Agro-NT-1B. {\small(985 M; Billion-scale DNA LM for 48 plant genomes) } A billion-parameter scale DNA language model, Agro-NT-1B is specialized for 48 edible plant genomes and inherits the k-mer vocabulary from NT-V2~\citep{Mendoza23}.

\end{itemize}

\vspace{0.5em}
\noindent\textbf{Ongoing Integration and Adaptability.}
The integration of new GFMs into \our{} is a streamlined process, facilitated by its flexible adapter interfaces. For \emph{most} models, adaptation primarily involves configuring two main components: the tokenizer and the model's forward prediction function. This typically requires only minor code modifications (often $\approx$100 lines), enabling a rapid turnaround time of \textbf{2–3 days} for onboarding new GFMs. Work is continuously in progress to incorporate the latest advancements in the field; preliminary wrappers for several new models have already passed our unit-test suite and are awaiting large-scale benchmark evaluations. This ensures \our{} remains a current and comprehensive platform for GFM assessment.

\section{Public Leaderboard}
\label{app:leaderboard}

\begin{figure}[t!]
    \centering
    \includegraphics[width=\linewidth]{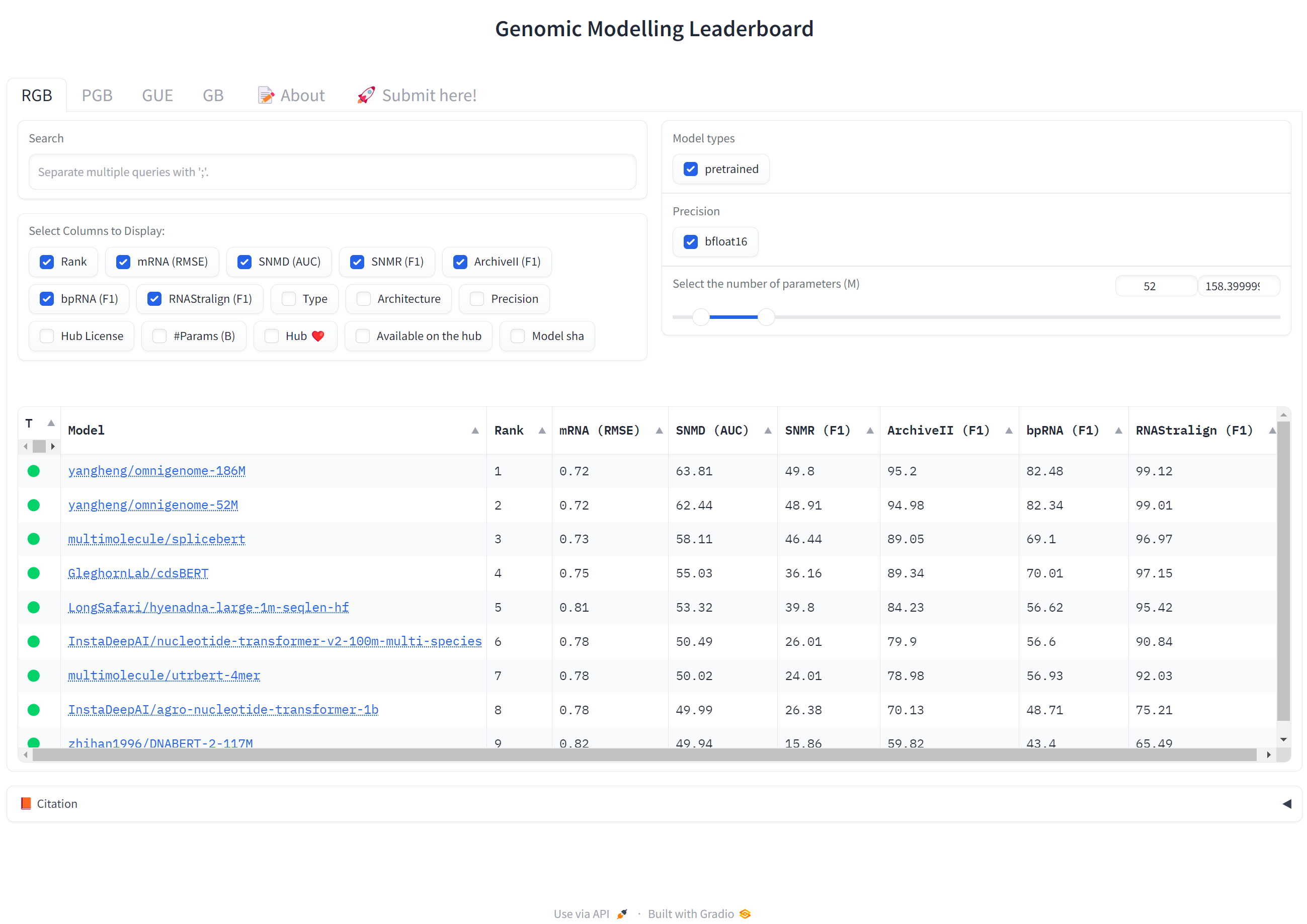}
    \caption{Screenshot of the interactive web interface for the public leaderboard, illustrating suite selection, filtering controls, and customizable metric display.}
    \label{fig:leaderboard}
\end{figure}

To promote transparency and reproducibility in genomic foundation modeling, we have released a dynamic, publicly accessible leaderboard alongside this manuscript. The current interface, shown in \pref{fig:leaderboard}, is built on our OmnigenBench platform and provides the following key features:

\begin{itemize}[leftmargin=*,nosep]
  \item \textbf{Multi‐Suite Coverage:} Users can switch among the four major benchmark suites (RGB, PGB, GUE, GB) via the top‐level tabs, allowing direct comparison of model performance on distinct task collections.
  \item \textbf{Interactive Filtering:} A flexible search bar accepts model names or keywords, while sidebar controls enable filtering by model type (pretrained vs.\ fine-tuned), numeric precision (e.g., bfloat16), and model scale (parameter count slider).  
  \item \textbf{Customizable Metrics Display:} Checkboxes let users choose which columns to display—rank, task-specific metrics (e.g., mRNA RMSE, SNMD AUC, SNMR F1, ArchiveII F1, bpRNA F1, RNAStralign F1), as well as model metadata such as architecture, hub availability, and license.  
  \item \textbf{Real-Time Ranking Updates:} Underlying OmnigenBench pipelines automatically re‐compute ranks whenever new submissions are processed, ensuring that the leaderboard reflects the latest community contributions.  
  \item \textbf{Submission Portal:} A “Submit here!” button links to a standardized result‐upload API, where contributors package their evaluation outputs along with a Docker image or YAML environment spec. Automated sanity checks validate input format and metric integrity before the new entry is incorporated.  
\end{itemize}

\noindent
We are actively extending the leaderboard to integrate the latest GFMs—particularly those published after our initial release—and to refine the user experience. Planned enhancements include:

\begin{itemize}[leftmargin=*,nosep]
  \item \textbf{Expanded Task Coverage:} Addition of emerging genomic benchmarks (e.g., Plant Genomic Benchmark, epigenetic marking tasks) and support for multi‐omics datasets.  
  \item \textbf{Drill-Down Analytics:} Clickable cells will reveal per‐task performance distributions, confidence intervals, and training details (e.g., data splits, seed settings).  
  \item \textbf{Versioning and Provenance:} Each model entry will record a Git SHA and timestamp, enabling exact replication of results and rollback to prior evaluations.  
  \item \textbf{Programmatic Access:} A RESTful API will allow users to query leaderboard data for downstream analysis, visualization, or integration into continuous‐integration workflows.  
\end{itemize}

\noindent
By providing this living, community‐driven resource, we aim to accelerate progress in genomic foundation models and foster an open benchmarking culture that underscores fairness, rigor, and collaborative innovation.

\section{Tutorials}
\label{app:tutorials}

This appendix provides practical examples demonstrating the usage of the \our{} framework for various genomic tasks. These tutorials illustrate core functionalities and highlight the framework's utility in addressing key challenges in computational genomics, moving beyond basic model application to enabling robust research workflows. The examples correspond to the Jupyter notebooks available in the project's repository\footnote{\url{https://github.com/yangheng95/OmniGenBench/tree/master/examples}}.

\subsection{Automated Benchmarking with AutoBench}
\label{app:tutorial_autobench}

Evaluating and comparing the performance of the rapidly growing number of Genomic Foundation Models (GFMs) is crucial for advancing the field, yet it presents significant challenges due to variations in datasets, tasks, and evaluation protocols. The \texttt{AutoBench} module within \our{} addresses this by providing a standardized and automated pipeline for rigorous GFM assessment. This tutorial demonstrates its configuration and execution.

First, initialize the \texttt{AutoBench} class, specifying the root directory of the desired benchmark suite (e.g., \texttt{RGB} for RNA Genome Benchmark, \texttt{GUE} for general DNA tasks), the GFM identifier (local path or Hugging Face name), and the target compute device.

\begin{tcolorbox}[colback=gray!5!white, colframe=gray!75!black, title=Benchmark Preparation, fonttitle=\bfseries]
\begin{verbatim}
from omnigenome import AutoBench
import autocuda

# Define benchmark root and model path
root = 'RGB' # Example: RNA Genome Benchmark suite
model_name_or_path = 'anonymous8/OmniGenome-52M'
device = autocuda.auto_cuda() # Automatically select available GPU

# Initialize AutoBench
auto_bench = AutoBench(
    bench_root=root,
    model_name_or_path=model_name_or_path,
    device=device,
    overwrite=True,
)
\end{verbatim}
\label{code:autobench}
\end{tcolorbox}

Next, execute the benchmark run using the \texttt{run} method. While \texttt{AutoBench} uses predefined, community-accepted configurations for each task to ensure fair comparison, users can override specific hyperparameters (e.g., \texttt{epochs}, \texttt{batch\_size}, \texttt{seeds}) for experimental purposes or resource adaptation.

\begin{tcolorbox}[colback=gray!5!white, colframe=gray!75!black, title=Automated Benchmarking, fonttitle=\bfseries]
\begin{verbatim}
# Define run-specific parameters (optional override)
batch_size = 8
epochs = 10
seeds = [42, 43, 44] # Multiple seeds for robustness assessment

# Run the benchmark across all tasks in the specified suite
auto_bench.run(epochs=epochs, batch_size=batch_size, seeds=seeds)
\end{verbatim}
\end{tcolorbox}

The results are systematically logged, providing detailed metrics for each task across different random seeds. This automated approach significantly reduces manual effort, enhances reproducibility, and facilitates transparent comparison of GFM capabilities, which is essential for identifying state-of-the-art models and guiding future development.

\subsection{Fine-tuning for RNA Secondary Structure Prediction}
\label{app:tutorial_ssp_finetune}

RNA secondary structure (SSP), the pattern of base pairings within an RNA molecule, is fundamental to its function, stability, and interactions with other molecules. Predicting SSP accurately from the primary sequence remains a key challenge in computational biology. This tutorial demonstrates how to leverage the power of pre-trained GFMs by fine-tuning them for this specific token-level prediction task using \our{}.

The process begins by importing necessary framework components and initializing a task-specific model (\texttt{OmniGenomeModelForTokenClassification}) built upon a chosen GFM backbone. A suitable tokenizer, potentially specialized for nucleotides (\texttt{OmniSingleNucleotideTokenizer}), is also required.

\begin{tcolorbox}[colback=gray!5!white, colframe=gray!75!black, title=SSP Model Intialization, fonttitle=\bfseries]
\begin{verbatim}
from omnigenome import (
    OmniGenomeDatasetForTokenClassification, ClassificationMetric,
    OmniTokenizer, OmniGenomeModelForTokenClassification,
    Trainer, ModelHub
)
import torch
import autocuda

# Define model path and task-specific label mapping (dot-bracket)
model_name_or_path = "anonymous8/OmniGenome-186M"
label2id = {"(": 0, ")": 1, ".": 2}

# Initialize tokenizer and model with classification head
tokenizer = OmniTokenizer.from_pretrained(model_name_or_path)
ssp_model = OmniGenomeModelForTokenClassification(
    model_name_or_path, tokenizer=tokenizer, label2id=label2id
)
device = autocuda.auto_cuda()
ssp_model.to(device)
\end{verbatim}
\end{tcolorbox}

Datasets containing sequences and their corresponding known structures (e.g., from bpRNA) are loaded using \texttt{OmniGenomeDatasetForTokenClassification}. This class handles tokenization and alignment of labels to tokens, preparing the data for training.

\begin{tcolorbox}[colback=gray!5!white, colframe=gray!75!black, title=Training Preparation, fonttitle=\bfseries]
\begin{verbatim}
# Define dataset paths and parameters
train_file = "Archive2/train.json"
# ... define test_file, valid_file ...
max_length = 512
batch_size = 8

# Load datasets (assuming JSON format: {"seq": "...", "label": "..."})
train_set = OmniGenomeDatasetForTokenClassification(
    data_source=train_file, tokenizer=tokenizer,
    label2id=label2id, max_length=max_length
)
# ... load valid_set, test_set ...
# Create DataLoaders
train_loader = torch.utils.data.DataLoader(
    train_set, batch_size=batch_size, shuffle=True
)
# ... create valid_loader, test_loader ...
\end{verbatim}
\end{tcolorbox}

Appropriate evaluation metrics (e.g., F1-score, Matthews correlation coefficient for structured prediction) are selected using \texttt{ClassificationMetric}. The \texttt{Trainer} class orchestrates the fine-tuning process, managing the training loop, optimization, evaluation, and checkpointing.

\begin{tcolorbox}[colback=gray!5!white, colframe=gray!75!black, title=RNA Structure Prediction Fine-tuning, fonttitle=\bfseries]
\begin{verbatim}
# Define metrics suitable for structure prediction
compute_metrics = [
    # Base accuracy
    ClassificationMetric().accuracy_score, 
    # F1 is often key for SSP
    ClassificationMetric(average="macro").f1_score, 
    # MCC useful for imbalanced classes
    ClassificationMetric().matthews_corrcoef, 
]

# Setup optimizer and Trainer
learning_rate = 2e-5; weight_decay = 1e-5
optimizer = torch.optim.AdamW(
    ssp_model.parameters(),
    lr=learning_rate,
    weight_decay=weight_decay
    )
epochs = 10; seed = 42
trainer = Trainer(
    model=ssp_model, train_loader=train_loader,
    eval_loader=valid_loader, test_loader=test_loader, 
    batch_size=batch_size, epochs=epochs,
    optimizer=optimizer, compute_metrics=compute_metrics, 
    seed=seed, device=device
)

# Run training and evaluation
metrics = trainer.train()
\end{verbatim}
\end{tcolorbox}

This fine-tuning approach allows the GFM to specialize its learned representations for the nuances of RNA folding, often achieving higher accuracy than traditional physics-based or purely statistical methods, especially for complex or novel structures. The trained model can be saved, shared via the \texttt{ModelHub}, and used for inference on new RNA sequences.

\subsection{Zero-Shot RNA Secondary Structure Prediction}
\label{app:tutorial_ssp_zeroshot}

While fine-tuning yields high accuracy, GFMs pre-trained on vast sequence datasets (including structural information implicitly or explicitly) may possess inherent capabilities for tasks like SSP even without specific fine-tuning. This "zero-shot" capability is valuable for rapid analysis or when task-specific labeled data is limited. This tutorial demonstrates zero-shot SSP prediction using \our{}.

Load a pre-trained model potentially suitable for this task (e.g., one trained with objectives related to structure or using relevant datasets). Crucially, the model class should provide a high-level inference method, such as \texttt{fold}, for direct structure prediction.

\begin{tcolorbox}[colback=gray!5!white, colframe=gray!75!black, title=SSP Model Initialization, fonttitle=\bfseries]
\begin{verbatim}
import torch
import autocuda
from transformers import AutoTokenizer
from omnigenome import OmniGenomeForTokenClassification

# Load a potentially zero-shot capable pre-trained model
model_path = "anonymous8/OmniGenome-186M" # Example model path
ssp_model = OmniGenomeForTokenClassification.from_pretrained(model_path)
ssp_model.to(autocuda.auto_cuda())
ssp_model.eval()
tokenizer = AutoTokenizer.from_pretrained(model_path)
\end{verbatim}
\end{tcolorbox}

Utilize the model's specialized inference method (\texttt{fold}) to directly predict the secondary structure from a sequence string. This abstracts away the tokenization, prediction, and decoding steps involved in typical token classification inference.

\begin{tcolorbox}[colback=gray!5!white, colframe=gray!75!black, title=Zero-Shot RNA Secondary Structure Prediction, fonttitle=\bfseries]
\begin{verbatim}
# Example RNA sequence
sequence = "GAAAAAAAAGGGGAGAAAUCCCGCCCGAAAGGGCGCCCAAAGGGC"

# Predict structure directly using the model's high-level API
# This relies on the model class 
# having implemented this specific functionality
predicted_structure_list = ssp_model.fold(sequence)
if predicted_structure_list:
    print("Zero-shot predicted structure:", predicted_structure_list[0])
\end{verbatim}
\end{tcolorbox}

Zero-shot prediction offers significant advantages in speed and data requirements compared to fine-tuning. While its accuracy might be lower than a fine-tuned model, it provides a powerful baseline and is useful for large-scale exploratory analysis. Comparing zero-shot results with established algorithms like \texttt{ViennaRNA} (as shown in the notebook) helps gauge the GFM's intrinsic structural understanding gained during pre-training.

\subsection{Generating RNA Embeddings}
\label{app:tutorial_rna_embedding}

GFMs learn rich, context-dependent representations of sequences. Extracting these representations as fixed-size vectors (embeddings) provides a powerful way to encode biological sequences for various downstream machine learning tasks, often capturing more nuanced information than traditional sequence similarity metrics. This tutorial demonstrates embedding generation using \texttt{OmniGenomeModelForEmbedding}.

Initialize the embedding model using a pre-trained GFM backbone. The choice of backbone influences the nature of the learned representations.

\begin{tcolorbox}[colback=gray!5!white, colframe=gray!75!black, title=Embedding Model Initialization, fonttitle=\bfseries]
\begin{verbatim}
from omnigenome import OmniGenomeModelForEmbedding

# Initialize using a pre-trained model. The model's architecture
# dictates the embedding properties.
model_name = "anonymous8/OmniGenome-186M" # Example GFM
embedding_model = OmniGenomeModelForEmbedding(model_name)
\end{verbatim}
\end{tcolorbox}

The model can encode single sequences or batches, producing vector representations (typically derived from the hidden states of the model, e.g., the CLS token or mean-pooled token states).

\begin{tcolorbox}[colback=gray!5!white, colframe=gray!75!black, title=RNA Embedding, fonttitle=\bfseries]
\begin{verbatim}
# Example RNA sequences
rna_sequences = ["AUGGCUACG", "CGGAUACGGC", "UGGCCAAGUC"]

# Encode a batch of sequences
# Output shape depends on pooling strategy 
# (e.g., [batch_size, hidden_dim])
embeddings = embedding_model.batch_encode(rna_sequences)

# Encode a single sequence
embedding = embedding_model.encode_single_sequence("AUGGCUACG")
\end{verbatim}
\end{tcolorbox}

These embeddings encapsulate learned features and can be used directly for tasks like similarity search (calculating cosine similarity between embedding vectors often provides a measure of functional or structural similarity), sequence clustering, or as input features for simpler downstream classifiers or regressors. The ability to represent complex biological sequences as dense vectors is a cornerstone application of foundation models in genomics.

\begin{tcolorbox}[colback=gray!5!white, colframe=gray!75!black, title=Embedding Similarity Calculation, fonttitle=\bfseries]
\begin{verbatim}
# Save/load embeddings for later use
embedding_model.save_embeddings(rna_embeddings, "rna_embeddings.pt")
loaded_embeddings = embedding_model.load_embeddings("rna_embeddings.pt")

# Compute similarity based on embeddings
similarity = embedding_model.compute_similarity(
    loaded_embeddings[0], 
    loaded_embeddings[1]
)
print(f"Embedding-based similarity: {similarity:.4f}")
\end{verbatim}
\end{tcolorbox}

\subsection{Computational RNA Sequence Design}
\label{app:tutorial_rna_design}

Designing RNA sequences (a.k.a., Inverse Design) that adopt a specific target secondary structure is a important challenge in synthetic biology and RNA therapeutics (e.g., mRNA vaccine optimization). This inverse folding problem involves searching the vast sequence space for candidates meeting structural constraints. This tutorial demonstrates how \our{} can facilitate this using GFMs~\cite{YangL24omnigenome}, potentially combined with optimization algorithms.

Initialize a specialized model class, \texttt{OmniGenomeModelForRNADesign}, which might incorporate structure prediction capabilities or fitness functions suitable for design.

\begin{tcolorbox}[colback=gray!5!white, colframe=gray!75!black, title=RNA Inverse Design, fonttitle=\bfseries]
\begin{verbatim}
from omnigenome import OmniGenomeModelForRNADesign

# Initialize the RNA design model
# This might wrap a GFM with structure prediction capabilities
# or a design-specific objective
model = OmniGenomeModelForRNADesign("anonymous8/OmniGenome-186M")
\end{verbatim}
\end{tcolorbox}

Define the target structure and execute the design process using \texttt{run\_rna\_design}. This method likely employs an optimization strategy (like the genetic algorithm mentioned in the notebook) guided by a fitness function that assesses how well a candidate sequence folds into the target structure (perhaps evaluated internally using the GFM or an external tool like \texttt{ViennaRNA}). Parameters control the search process.

\begin{verbatim}
# Define target structure (dot-bracket notation)
target_structure = "(((....)))"

# Run RNA design using an optimization algorithm (e.g., Genetic Algorithm)
# Parameters like mutation_ratio, num_population, 
# num_generation tune the search
best_sequences = model.run_rna_design(
    structure=target_structure,
    mutation_ratio=0.5,    # Rate of mutation in GA
    num_population=100,  # Size of the sequence population per generation
    num_generation=100   # Number of optimization iterations
)

# Output the best candidate sequences found
print("Designed RNA sequences:", best_sequences)
\end{verbatim}
This computational approach allows exploring sequence possibilities guided by learned models of sequence-structure relationships, potentially identifying novel solutions for synthetic biology applications faster than experimental screening or simpler computational methods. Verifying the designs by folding the output sequences with standard tools is a common validation step.

\subsection{Sequence Augmentation via Masked Language Modeling}
\label{app:tutorial_rna_augmentation}

Training robust deep learning models often requires large and diverse datasets. When real data is limited, augmentation techniques can generate synthetic variations to improve model generalization. This tutorial demonstrates sequence augmentation using a Masked Language Model (MLM) approach within \our{}, leveraging a GFM's understanding of sequence patterns to create plausible variants.

Initialize \texttt{OmniGenomeModelForAugmentation} with a pre-trained MLM (like a BERT-style GFM). Key parameters are the `noise\_ratio` (fraction of sequence tokens to mask) and `instance\_num` (how many augmented versions to generate per input).

\begin{tcolorbox}[colback=gray!5!white, colframe=gray!75!black, title=Augmentation Model Initialization, fonttitle=\bfseries]
\begin{verbatim}
from omnigenome import OmniGenomeModelForAugmentation

# Initialize the augmentation model using a pre-trained MLM
model = OmniGenomeModelForAugmentation(
    model_name_or_path="anonymous8/OmniGenome-186M",
    noise_ratio=0.2,  # Mask 20%
    max_length=1026,  # Max sequence length
    instance_num=3    # Generate 3 variants per original sequence
)
\end{verbatim}
\end{tcolorbox}

Apply augmentation to single sequences or process entire datasets from files. The model masks tokens randomly based on the `noise\_ratio` and uses the MLM to predict replacements, generating new sequence instances.

\begin{tcolorbox}[colback=gray!5!white, colframe=gray!75!black, title=RNA Sequence Augmentation based on MLM, fonttitle=\bfseries]
\begin{verbatim}
\begin{verbatim}
# Augment a single DNA/RNA sequence
sequence = "ATCTTGCATTGAAG" # Example sequence
augmented_sequences = model.augment_sequence(sequence)
# Output will be a list if instance_num > 1
print(f"Augmented variants: {augmented_sequences}")

# Augment all sequences in a JSON file and save to a new file
input_file = "test.json" # Assumes {"seq": "..."} per line
output_file = "augmented_sequences.json"
model.augment_from_file(input_file, output_file)
print(f"Augmented dataset saved to {output_file}")
\end{verbatim}
\end{tcolorbox}

This MLM-based augmentation is considered more sophisticated than simple random mutations, as the GFM fills masked positions based on learned sequence context, potentially generating more biologically plausible or challenging variations for model training, thus enhancing robustness against sequence variability or noise.

\section{Interpretability Cases for Genomic Foundation Models}
\label{app:interpretability}

To assess whether GFMs learn biologically relevant representations and internal mechanisms, we conduct three interpretability case studies: the sequence motif preservation analysis in~\pref{sec:seq_motif_preservation}, feature-embedding analysis for antimicrobial resistance prediction~\pref{app:feature_embedding_analysis}, and attention inspection for RNA secondary structure~\pref{app:attention_inspection}.

\subsection{Feature-Embedding Analysis}
\label{app:feature_embedding_analysis}

\paragraph{Motivation.}
For tasks such as antimicrobial resistance (AMR) prediction, a GFM should ideally map resistant (ARG) and non-resistant (non-ARG) sequences to distinct regions in its representation space. Visualizing this latent geometry allows us to gauge whether pre-training already encodes functional cues related to AMR and how much additional separation task-specific fine-tuning provides.

\paragraph{Experimental Design.}
We collated DNA sequence data from key ARG databases (CARD, MEGARes, ResFinder), ensuring sequences were labeled as ARG or Non-ARG. For each of the three RNA language models, \og{}, RNA-MSM, and RNA-FM, we extracted sequence representations under three distinct model states: random initialization, after pre-training, and after task-specific fine-tuning on AMR prediction. We use RNA models for this analysis because we aim to study the ability of GFMs to learn cross-modal genomic representations (i.e., utilizing RNA GFM to decipher DNA modelling). Specifically, the last hidden state representation for each sequence was obtained and mean-pooled across the sequence dimension to generate a fixed-length high-dimensional embedding vector.

To evaluate the models' ability to discriminate between ARG and Non-ARG classes, these high-dimensional embeddings were subjected to density-based clustering using DBSCAN. The quality of the resulting clusters was quantitatively assessed using the Silhouette Coefficient (SC, higher is better, indicating better separation and cohesion) and the Davies-Bouldin Index (DB, lower is better, indicating more distinct and compact clusters). For visualization of the high-dimensional embedding space, we first applied Principal Component Analysis (PCA) for initial dimensionality reduction, followed by t-SNE (perplexity = 30, 1,000 iterations) to project the data into a 2-D space, allowing for intuitive comparison of the representational capabilities of the different models and training stages. It is important to note that the SC and DB scores are calculated on the original high-dimensional embeddings before t-SNE projection, as t-SNE can distort global structure.

\begin{figure}[t!]
  \centering
  \begin{minipage}{0.32\textwidth}\centering
    \includegraphics[width=\linewidth]{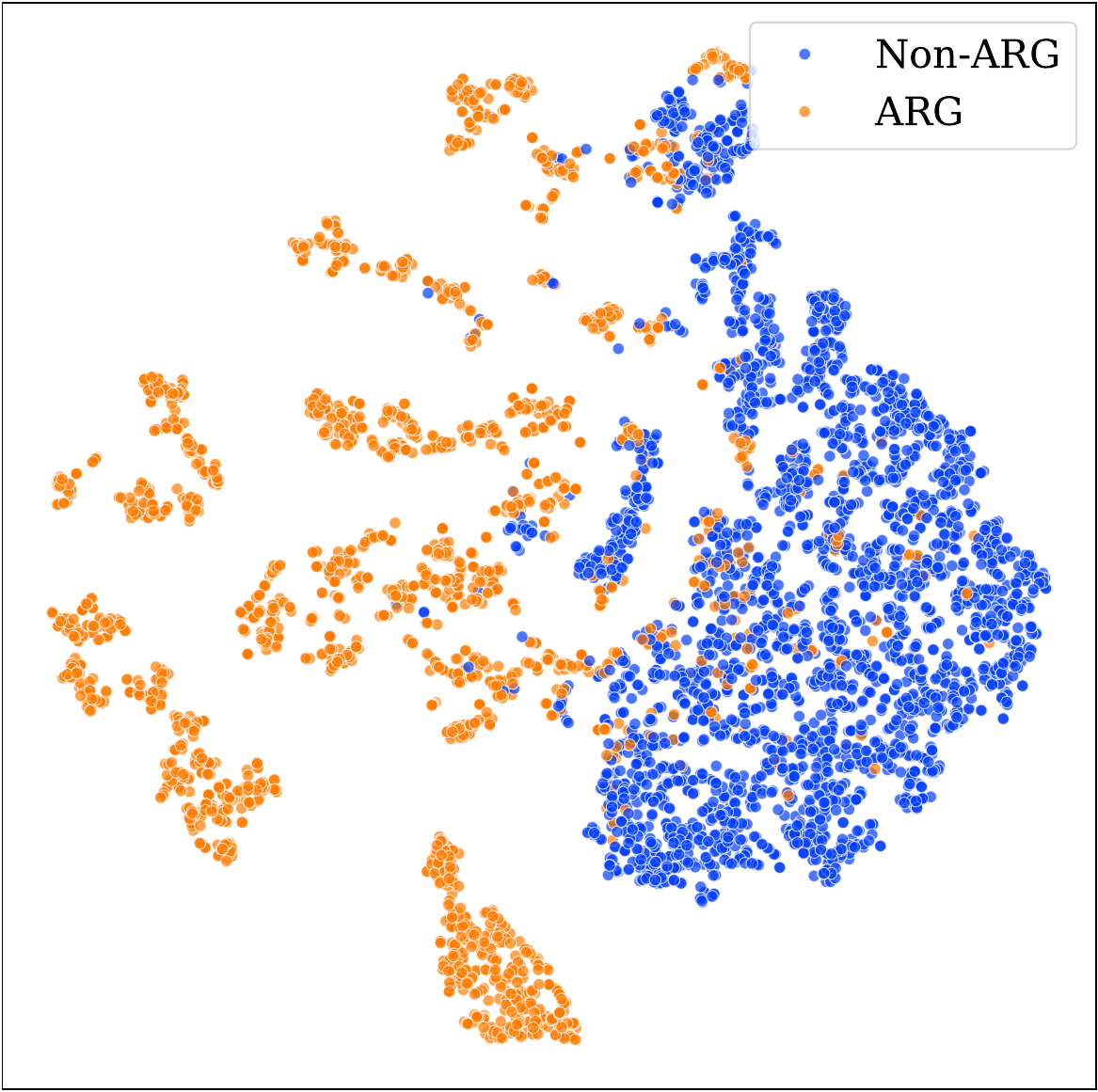}\par
    \caption*{(a) \og{} Finetuning \\ (SC, DB) = (0.74, 0.31)} %
  \end{minipage}\hfill
  \begin{minipage}{0.32\textwidth}\centering
    \includegraphics[width=\linewidth]{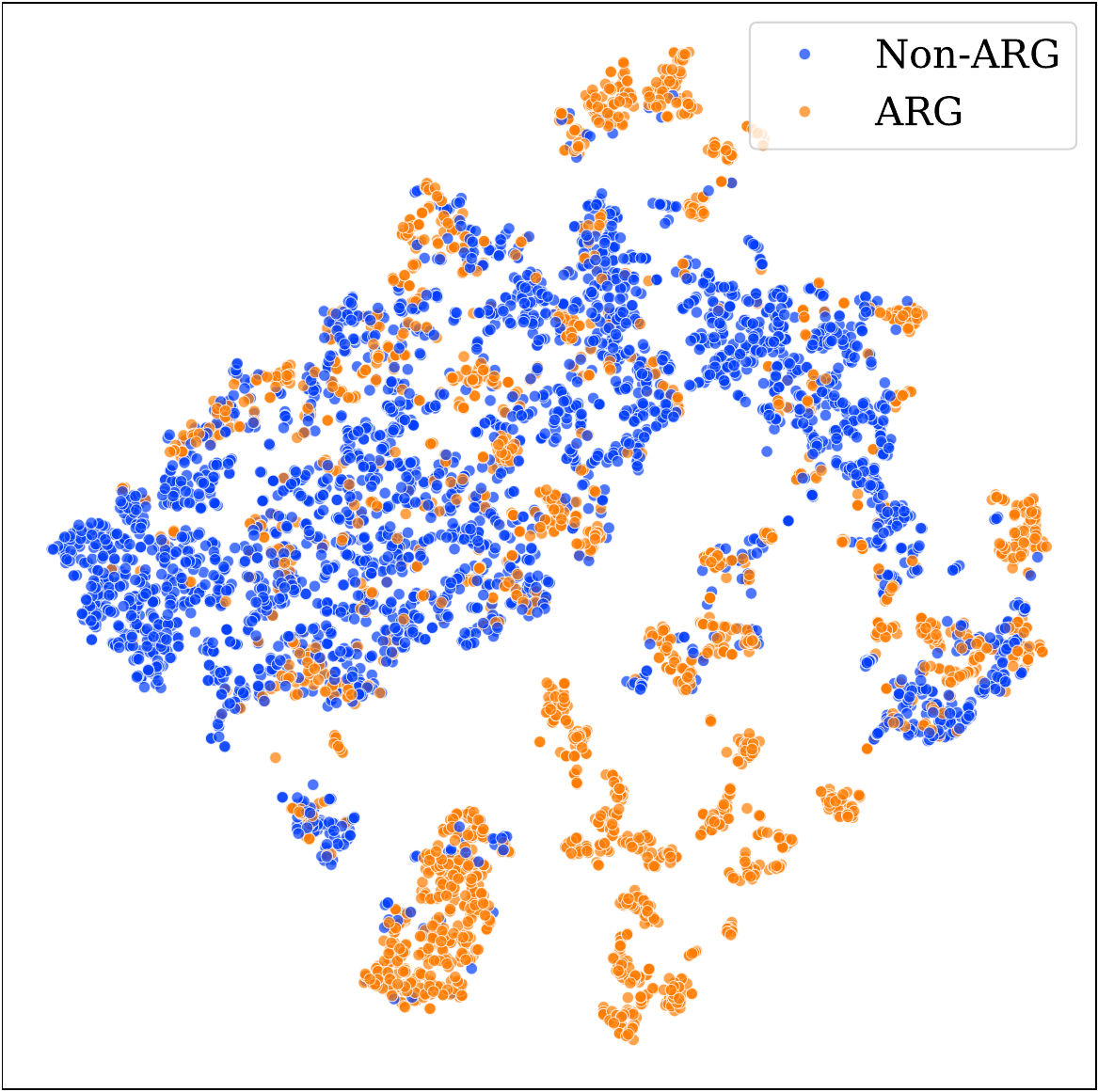}\par
    \caption*{(b) \og{} Pretraining \\ (SC, DB) = (0.64, 0.43)}
  \end{minipage}\hfill
  \begin{minipage}{0.32\textwidth}\centering
    \includegraphics[width=\linewidth]{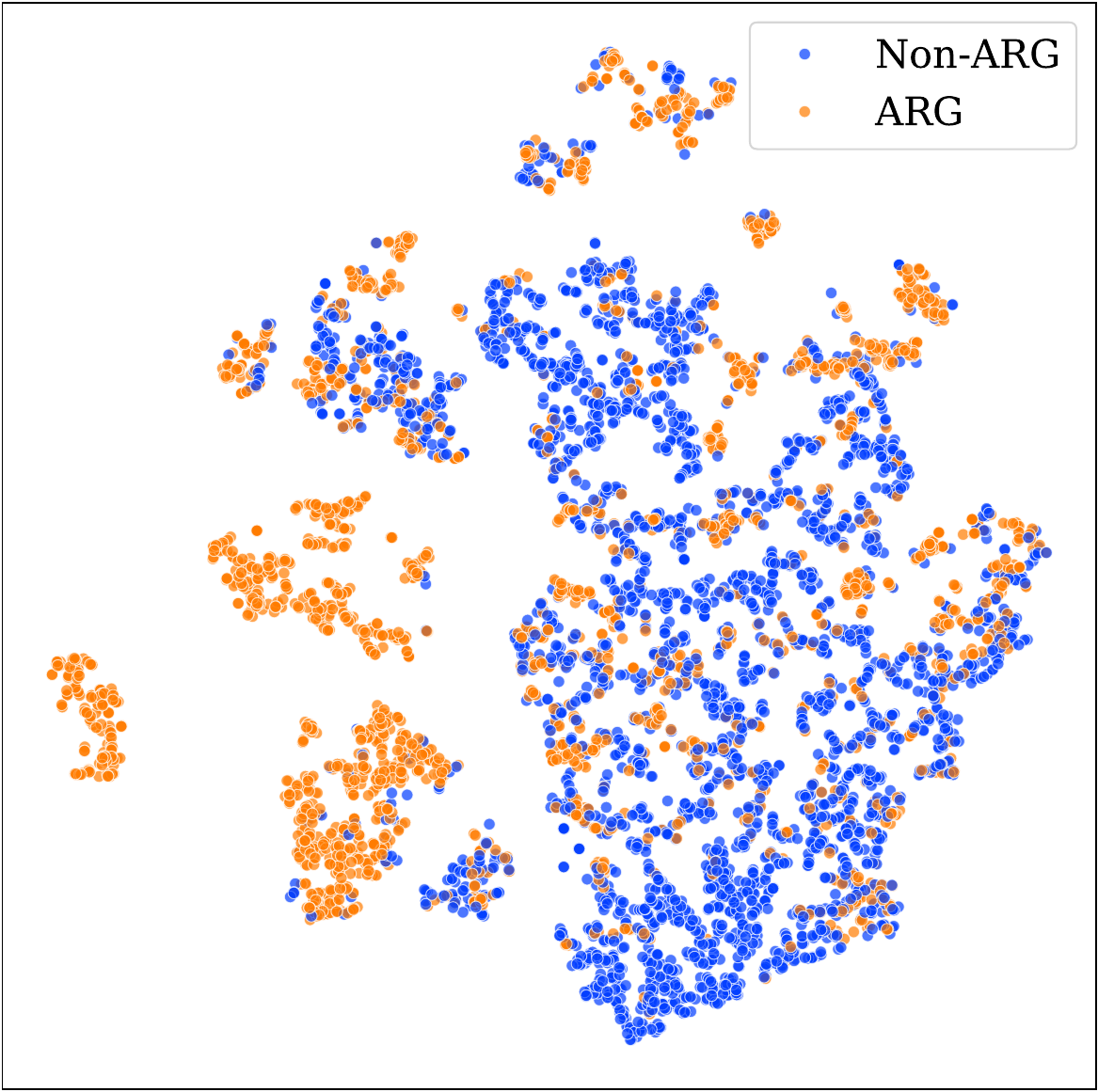}\par
    \caption*{(c) \og{} No Training \\ (SC, DB) = (-0.10, 0.60)}
  \end{minipage}\vspace{0.5em}

  \begin{minipage}{0.32\textwidth}\centering
    \includegraphics[width=\linewidth]{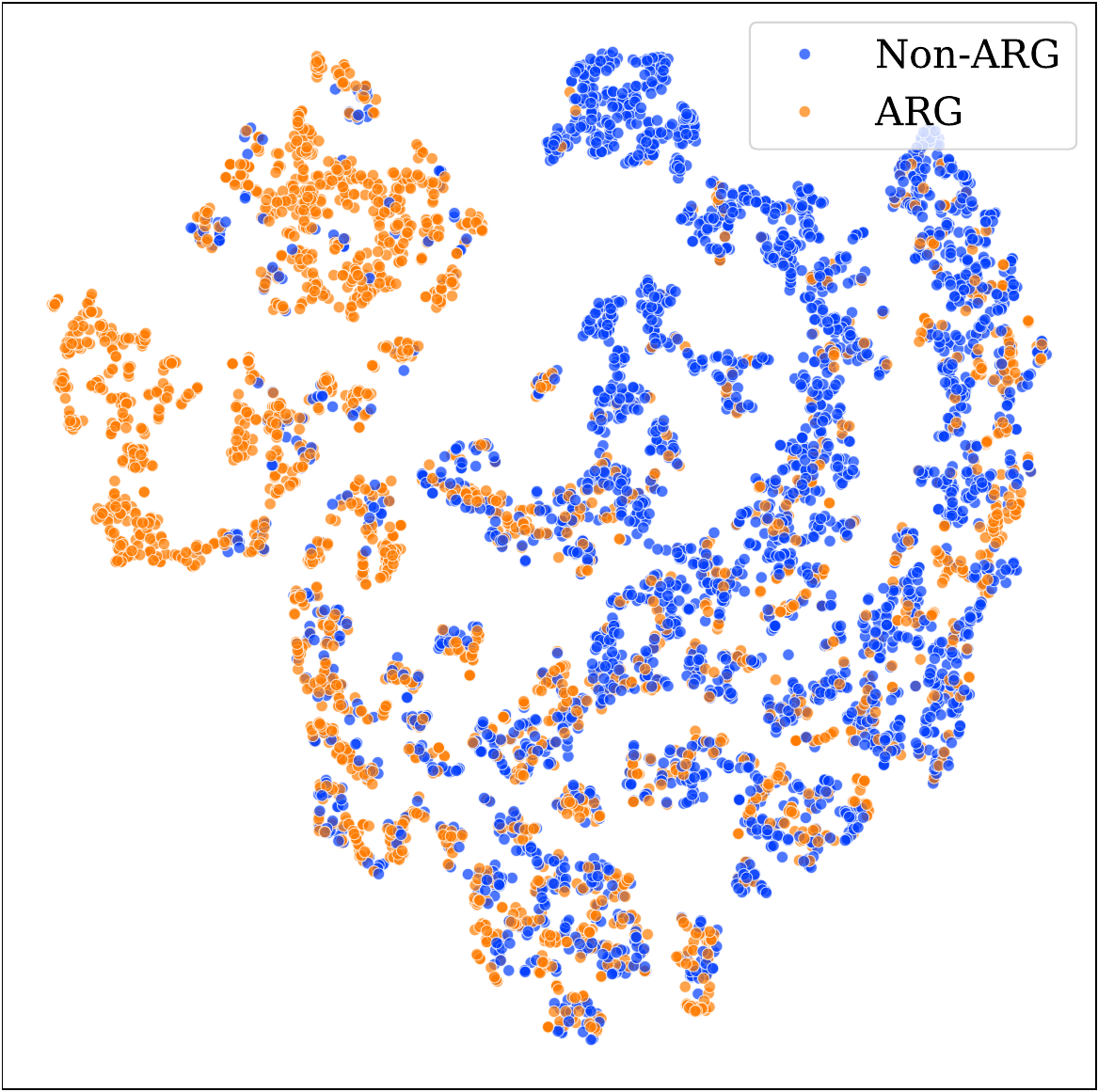}\par
    \caption*{(d) RNA-MSM Finetuning \\ (SC, DB) = (0.81, 0.24)}
  \end{minipage}\hfill
  \begin{minipage}{0.32\textwidth}\centering
    \includegraphics[width=\linewidth]{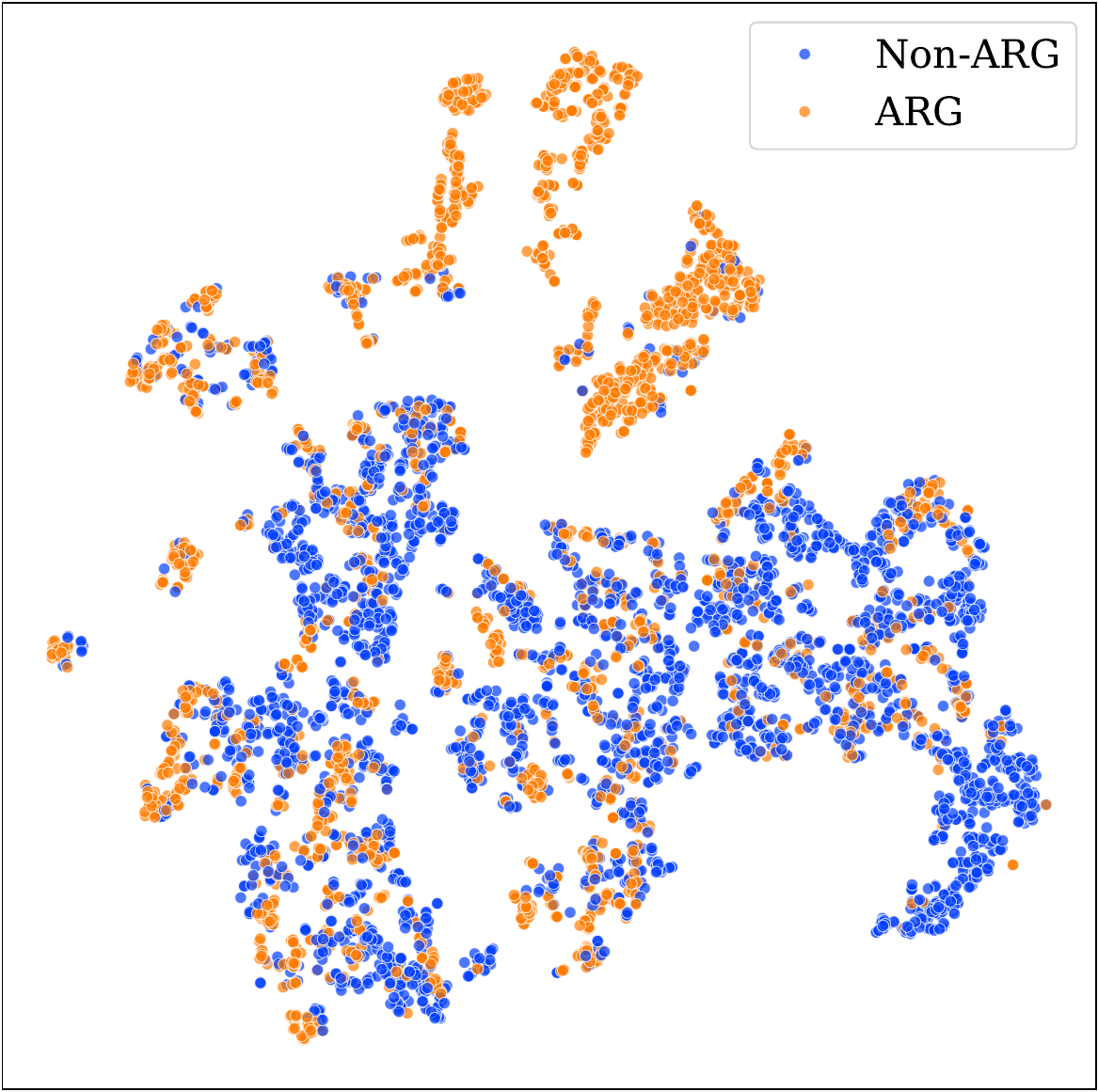}\par
    \caption*{(e) RNA-MSM Pretraining \\ (SC, DB) = (-0.02, 0.61)}
  \end{minipage}\hfill
  \begin{minipage}{0.32\textwidth}\centering
    \includegraphics[width=\linewidth]{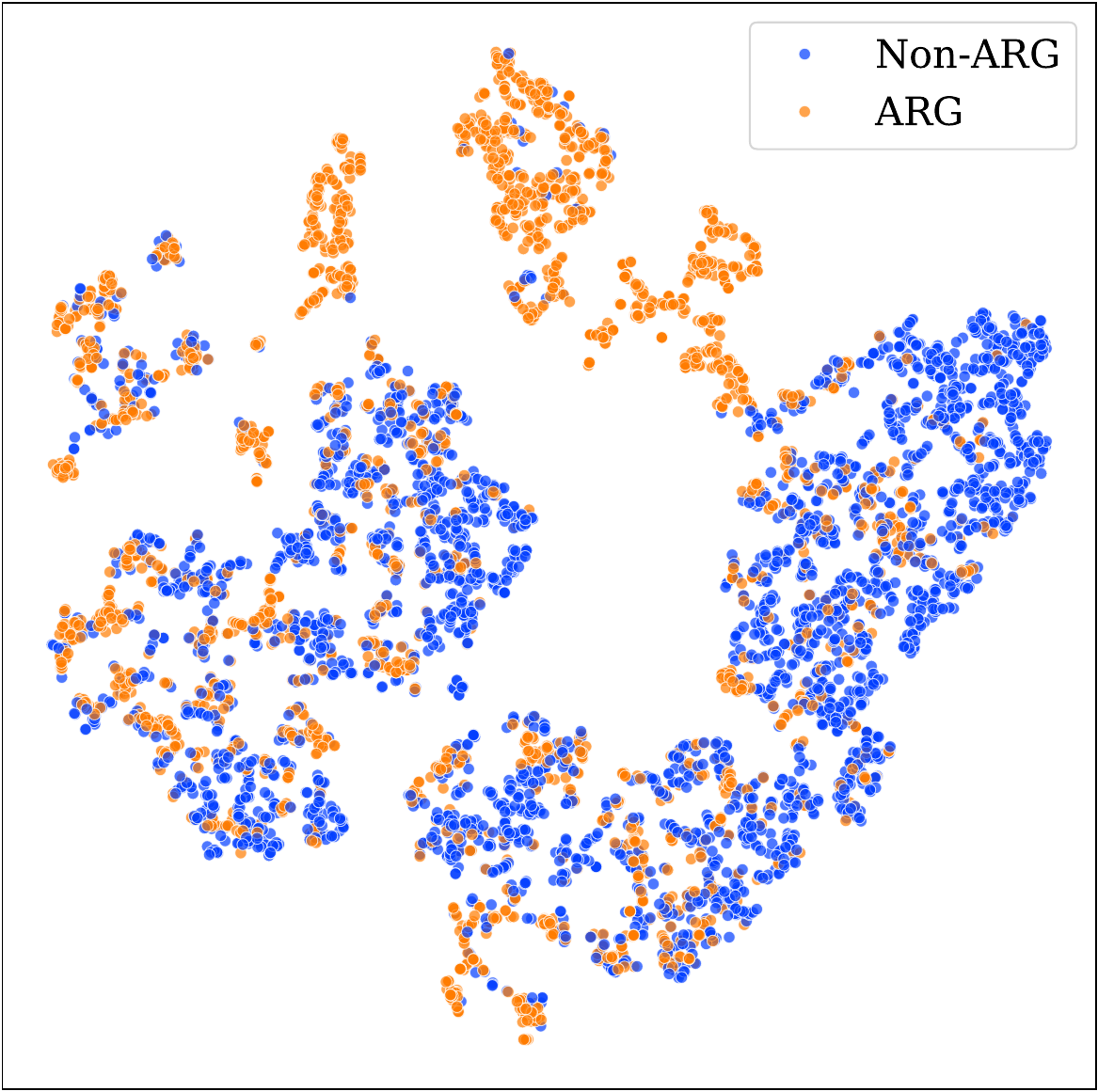}\par
    \caption*{(f) RNA-MSM No Training \\ (SC, DB) = (0.10, 0.52)} %
  \end{minipage}\vspace{0.5em}

  \begin{minipage}{0.32\textwidth}\centering
    \includegraphics[width=\linewidth]{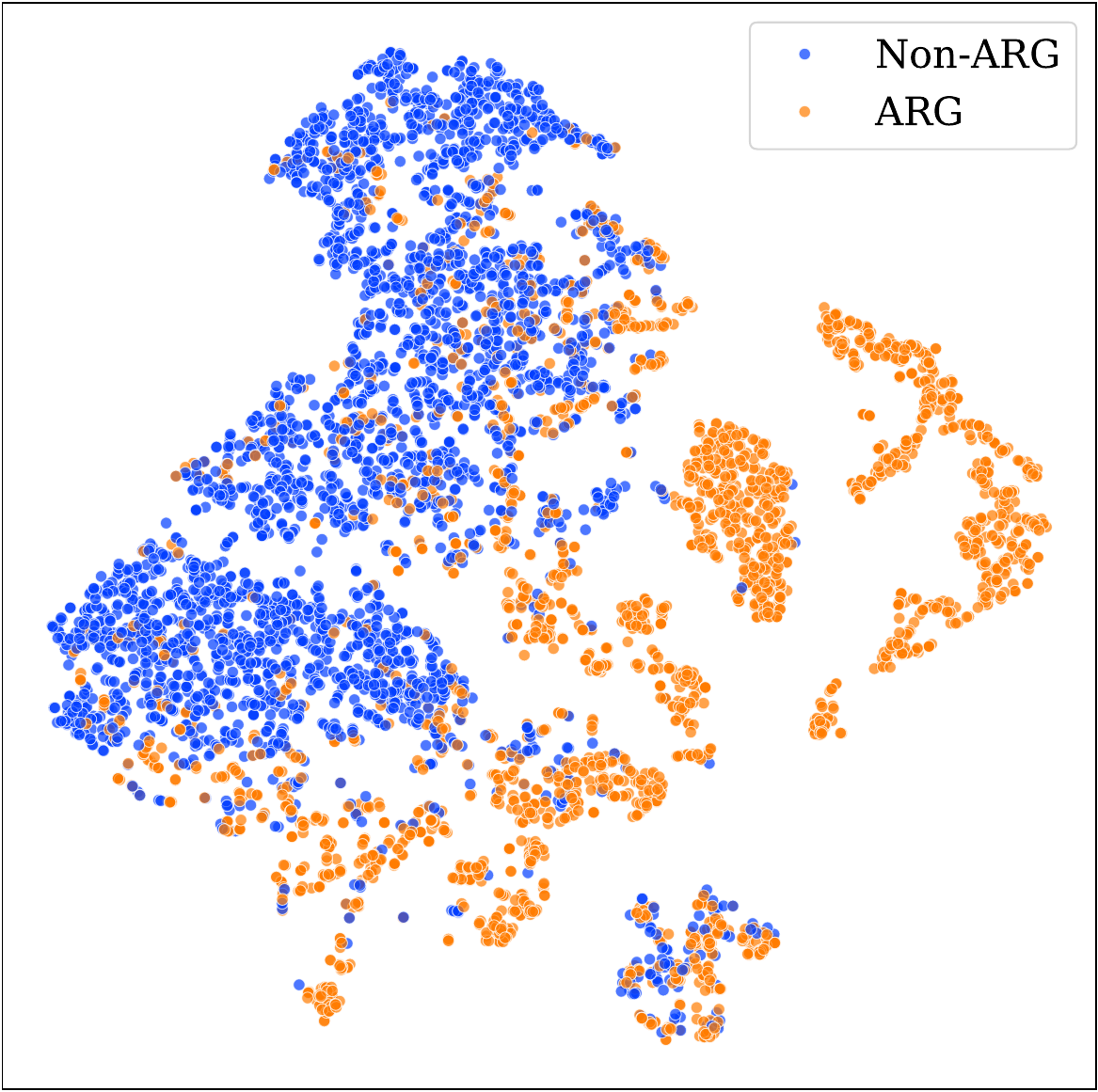}\par %
    \caption*{(g) RNA-FM Finetuning \\ (SC, DB) = (0.84, 0.19)}
  \end{minipage}\hfill
  \begin{minipage}{0.32\textwidth}\centering
    \includegraphics[width=\linewidth]{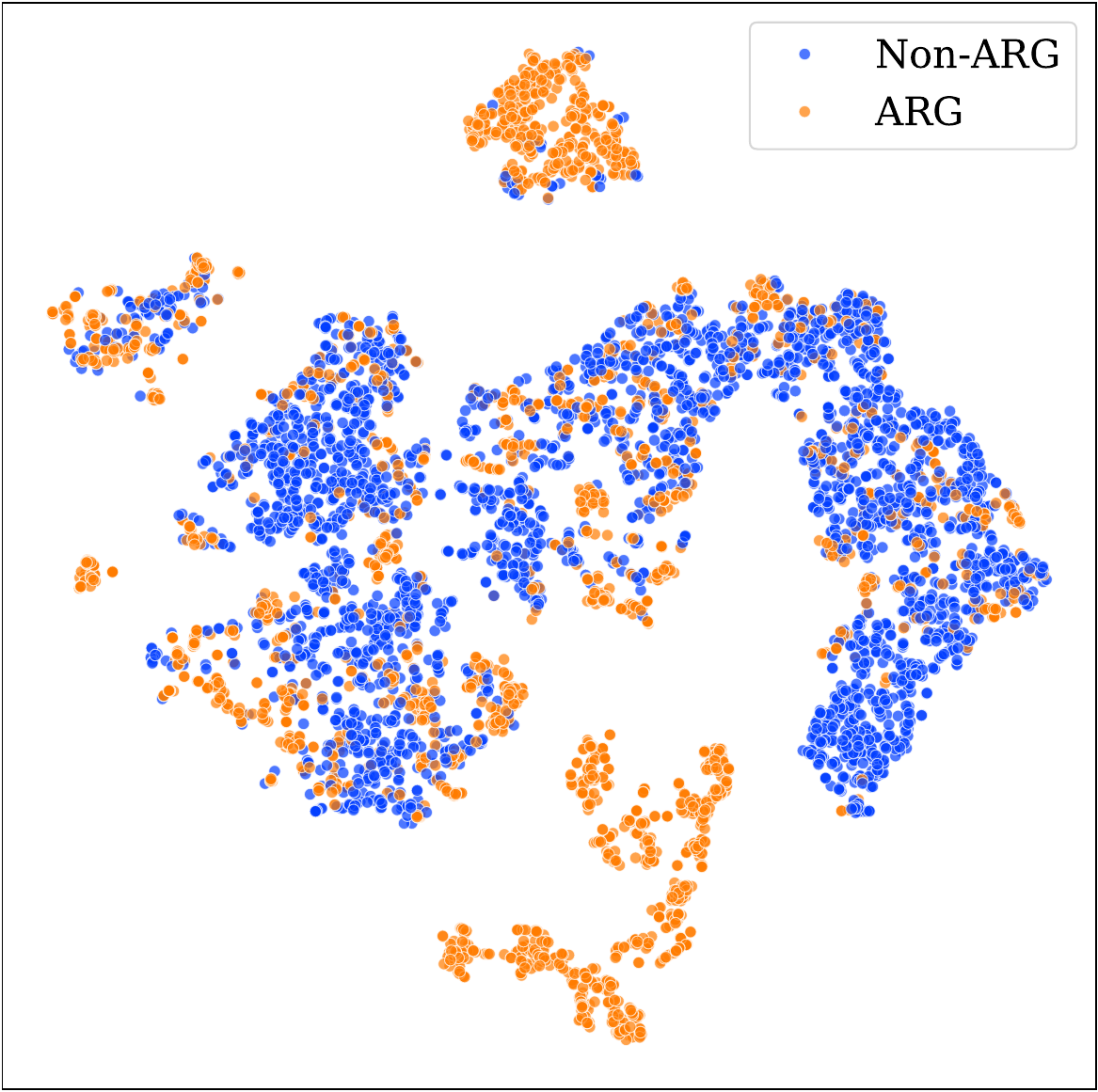}\par %
    \caption*{(h) RNA-FM Pretraining \\ (SC, DB) = (0.04, 0.64)}
  \end{minipage}\hfill
  \begin{minipage}{0.32\textwidth}\centering
    \includegraphics[width=\linewidth]{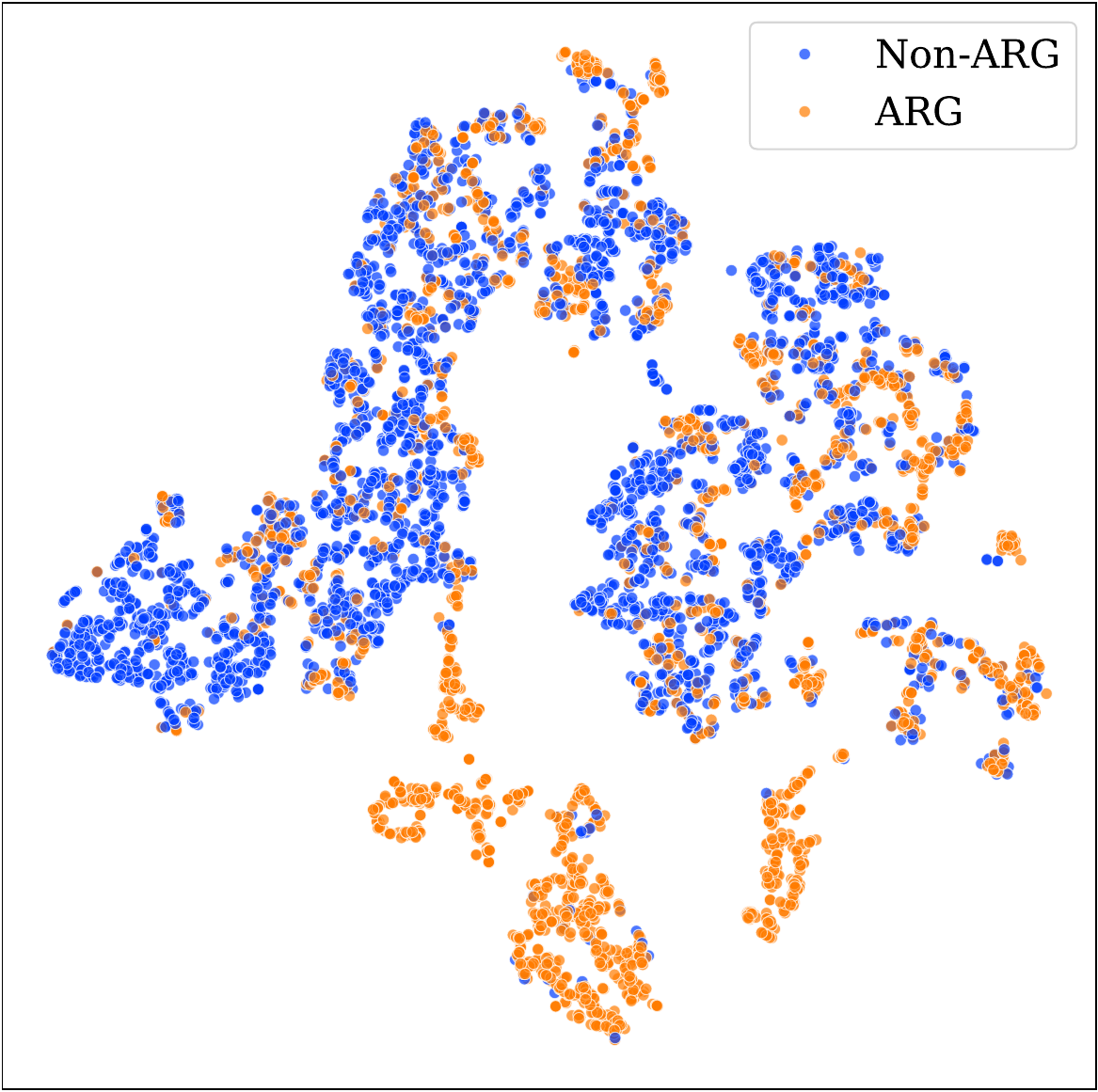}\par %
    \caption*{(i) RNA-FM No Training \\ (SC, DB) = (-0.12, 0.58)}
  \end{minipage}\vspace{0.5em}

  \caption{t-SNE visualization of GFM-derived sequence embeddings for antimicrobial resistance gene (ARG) classification. Each point represents a DNA fragment embedded by the corresponding model and training stage, colored by its class: orange for ARG and blue for Non-ARG. Models shown are \og{} (top row, a-c), RNA-MSM (middle row, d-f), and RNA-FM (bottom row, g-i), each under three training regimes: task fine-tuning, pre-training only, and random initialization (no training). Quantitative clustering scores (Silhouette coefficient, Davies-Bouldin index) are displayed below each sub-plot. Clearer clustering and separation are observed in finetuned and pretrained models, with \og{} exhibiting the most discriminative embeddings.}
\label{fig:tsne_combined1} %
\end{figure}

\paragraph{Results.}
\pref{fig:tsne_combined1} displays the t-SNE visualizations of sequence embeddings for the three models across the three training stages. The Silhouette Coefficient (SC) and Davies-Bouldin Index (DB) for each condition are reported directly on the subplots.

\begin{itemize}[nosep,leftmargin=*]
\item \textbf{\og{} (Finetuned)} (\pref{fig:tsne_combined1}a) demonstrates excellent class separation, forming distinct and compact clusters for ARG (orange) and Non-ARG (blue) sequences. This is quantitatively supported by a high SC of 0.74 and a low DB index of 0.31, indicating that fine-tuning has enabled the model to effectively learn AMR-specific discriminative features.
\item \textbf{\og{} (Pretrained)} (\pref{fig:tsne_combined1}b) already exhibits substantial separation between the ARG and Non-ARG classes (SC = 0.64, DB = 0.43). This strong performance post-pre-training suggests that \og{}'s general pre-training regimen on diverse RNA/DNA sequences successfully captures latent signals relevant to antimicrobial resistance, providing a significant head-start for downstream tasks.
\item \textbf{\og{} (Random)} (\pref{fig:tsne_combined1}c), as anticipated, shows no meaningful separation, with intermingled ARG and Non-ARG sequences and poor clustering metrics (SC = -0.10, DB = 0.60).
\item \textbf{RNA-MSM (Finetuned)} (\pref{fig:tsne_combined1}d) yields well-separated clusters after fine-tuning, with strong quantitative scores  (SC = 0.81, DB = 0.24), comparable to RNA-FM and outperforming \og{} on these metrics.
\item \textbf{RNA-MSM (Pretrained)} (\pref{fig:tsne_combined1}e) 
demonstrates minimal class separation in its pretrained state (SC = -0.02, DB = 0.61), suggesting its pre-training does not inherently capture AMR-relevant features as effectively as \og{}.
\item \textbf{RNA-MSM (Random)} (\pref{fig:tsne_combined1}f) shows some slight tendency for ARG sequences to cluster but overall poor metrics (SC = 0.10, DB = 0.52).
\item \textbf{RNA-FM (Finetuned)} (\pref{fig:tsne_combined1}g) 
surprisingly achieves the best quantitative clustering among all finetuned models (SC = 0.84, DB = 0.19), forming very distinct clusters. This suggests that while its pre-training might lack strong AMR-related signals, its architecture is highly amenable to learning discriminative features when provided with task-specific supervision.
\item \textbf{RNA-FM (Pretrained)} (\pref{fig:tsne_combined1}h) 
shows very little inherent structure related to AMR classification (SC = 0.04, DB = 0.64), indicating its pre-training does not effectively separate these classes.
\item \textbf{RNA-FM (Random)} (\pref{fig:tsne_combined1}i) 
also shows poor separation (SC = -0.12, DB = 0.58).
\end{itemize}

\paragraph{Conclusion.}
The analysis of feature embeddings reveals distinct learning dynamics across the GFMs. Task-specific fine-tuning significantly improves the ability of all models to separate ARG and Non-ARG sequences, with RNA-MSM and RNA-FM achieving particularly high Silhouette Coefficients and low Davies-Bouldin Index scores post-finetuning, indicating very well-defined clusters. Notably, \og{}'s pre-training phase already instills a strong capability to distinguish between these classes, as evidenced by its superior SC and DB scores in the pretrained state compared to RNA-MSM and RNA-FM. This suggests that \og{}'s pre-training strategy is more effective at capturing inherent, functionally relevant genomic signals for AMR even before task-specific adaptation. While RNA-MSM and RNA-FM show excellent clustering after fine-tuning, their pretrained embeddings do not exhibit the same level of innate class separation as \og{}. These findings underscore the importance of both robust pre-training objectives for capturing generalizable biological signals and the capacity of models to adapt effectively during fine-tuning. The geometry of these learned representations serves as an insightful diagnostic for GFM capabilities.

\subsection{Attention Representation Inspection}
\label{app:attention_inspection}

\paragraph{Motivation.}
RNA function is governed by its secondary structure, i.e.\ the set of canonical
base pairs that form stems, loops and, occasionally, long-range
(pseudoknot-like) contacts.
If a GFM truly internalises these constraints,
high self-attention weights should coincide with the known base pairs.
Visualising attention maps therefore offers an intuitive lens on the
\emph{reasoning path} learned by a model.

\paragraph{Experimental design.}
We analyse two RNA sequences from the bpRNA-1m benchmark:

\begin{minipage}[t]{0.48\linewidth}
\footnotesize
\textbf{Example 1}\\[-0.4em]
\begin{verbatim}
  Sequence:  GCGCCCAAGGGUGCACGCCGGUCAGCGAGGUUUCGCCCACCGGCGUCUUUGUC
  Structure: ...............(((((((..((((....))))..)))))))........
\end{verbatim}
\end{minipage}

\begin{minipage}[t]{0.48\linewidth}
\footnotesize
\textbf{Example 2}\\[-0.4em]
\begin{verbatim}
  Sequence:  UGAAAGACACGGGUAGUGAGAUAUGGAUUUUUCAUCUCUAUAUUCGUGUCUUUC
  Structure: .((((((((.((((.((((..(...........)..)).)).))))))))))))
\end{verbatim}
\end{minipage}

\begin{figure}[t!]
  \centering
  \begin{minipage}{0.32\textwidth}\centering
    \includegraphics[width=\linewidth]{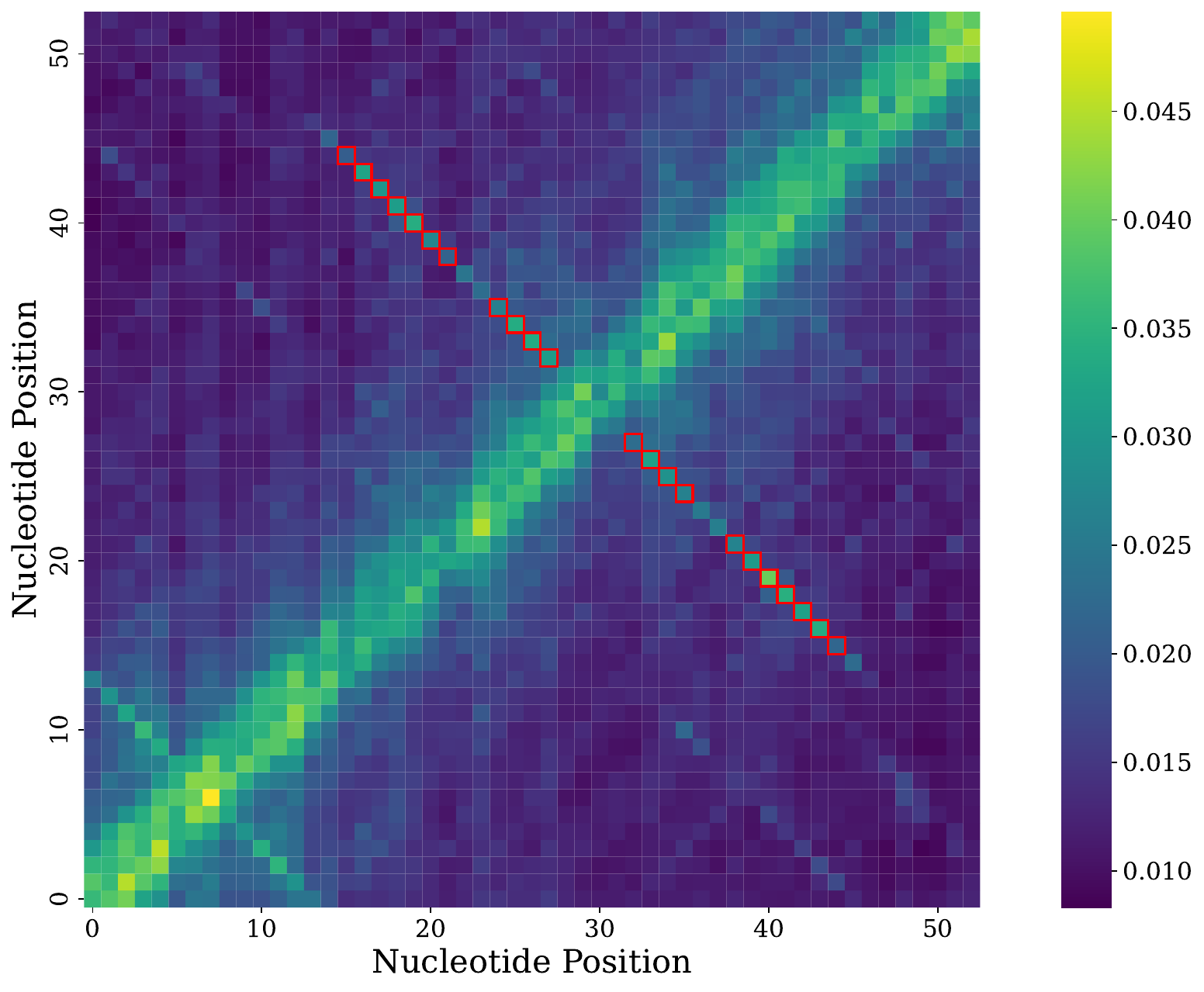}\par
    \caption*{(a)\ \og{} Finetuning}
  \end{minipage}\hfill
  \begin{minipage}{0.32\textwidth}\centering
    \includegraphics[width=\linewidth]{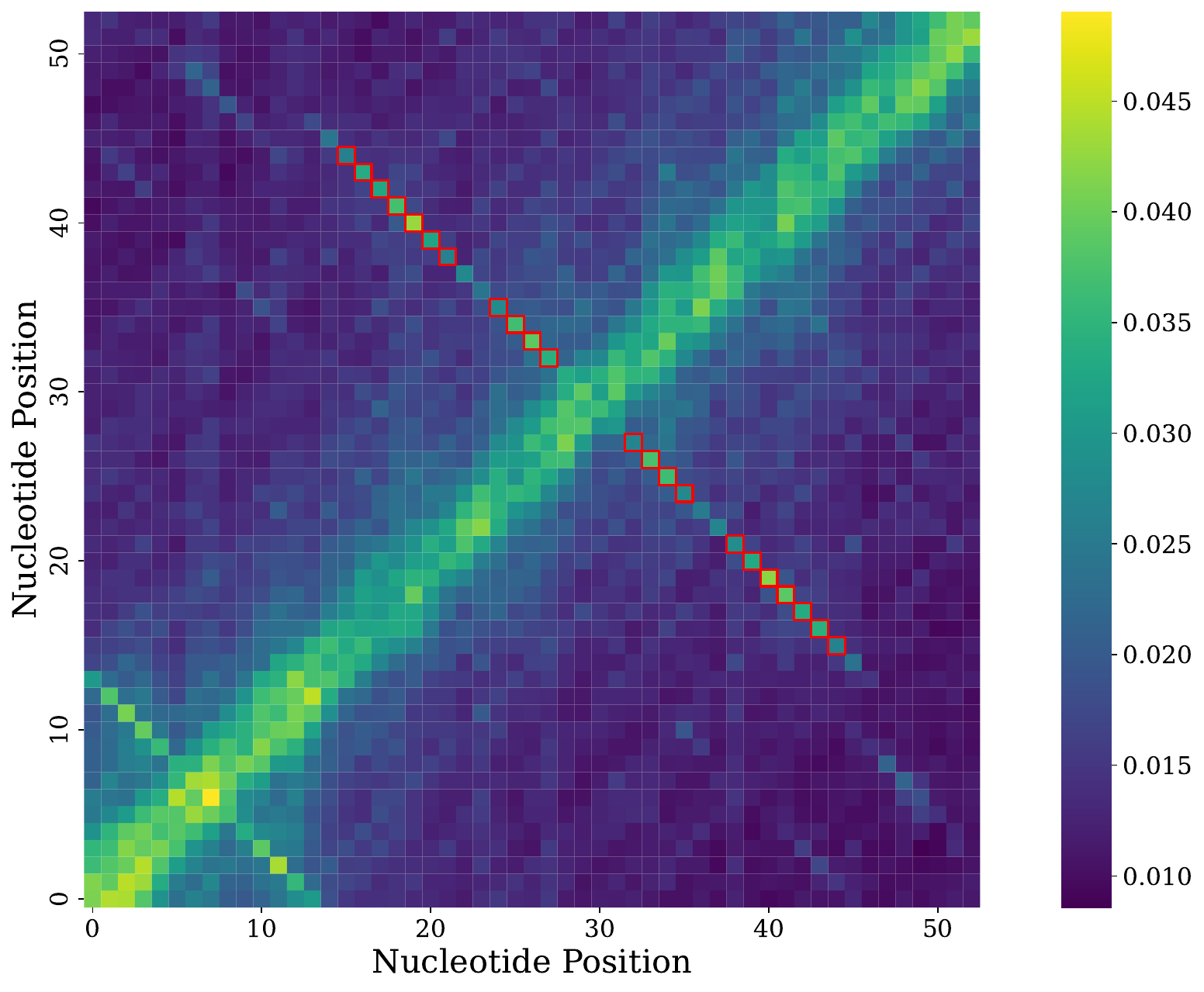}\par
    \caption*{(b)\ \og{} Pretraining}
  \end{minipage}\hfill
  \begin{minipage}{0.32\textwidth}\centering
    \includegraphics[width=\linewidth]{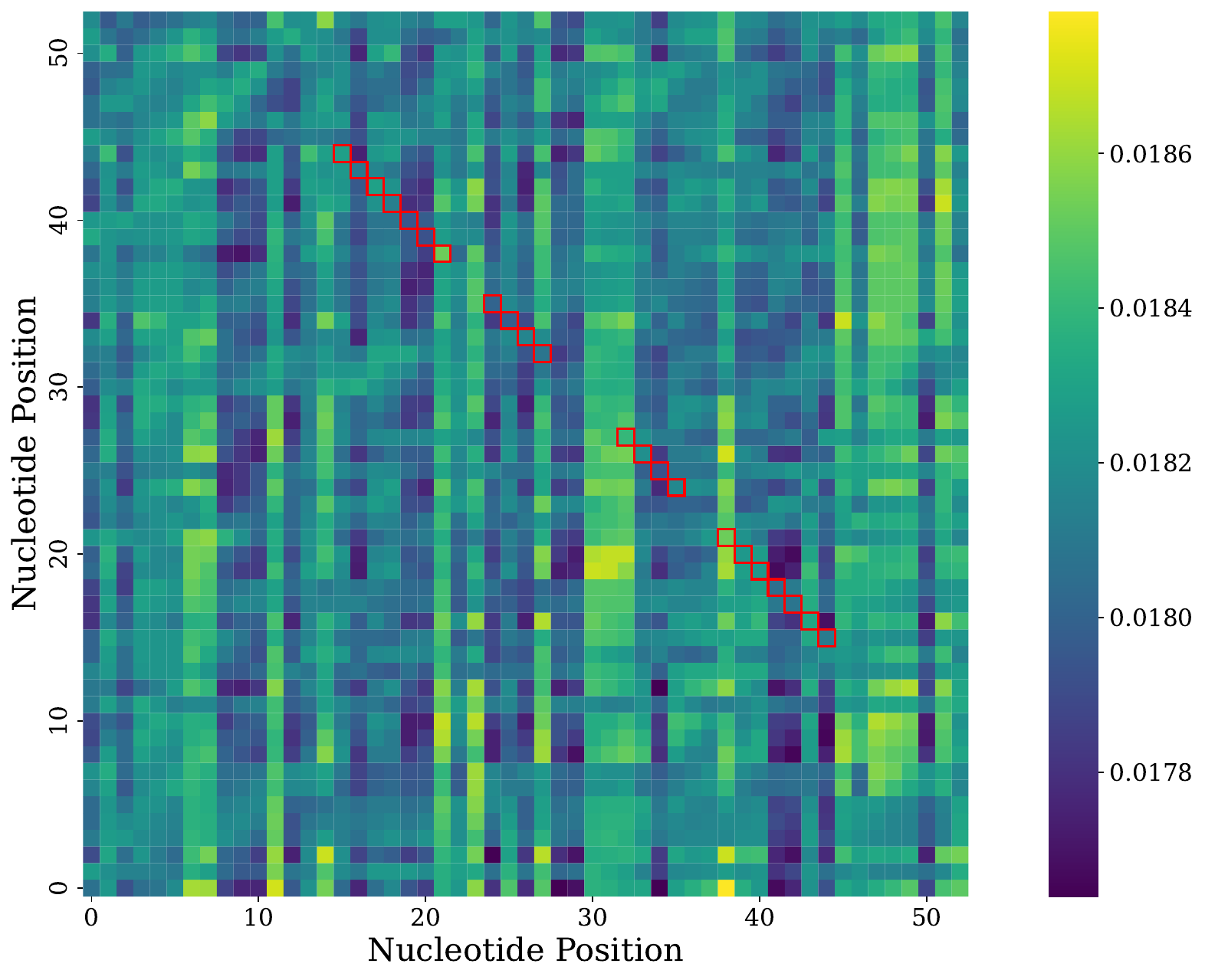}\par
    \caption*{(c)\ \og{} No Training}
  \end{minipage}\vspace{0.5em}

  \begin{minipage}{0.32\textwidth}\centering
    \includegraphics[width=\linewidth]{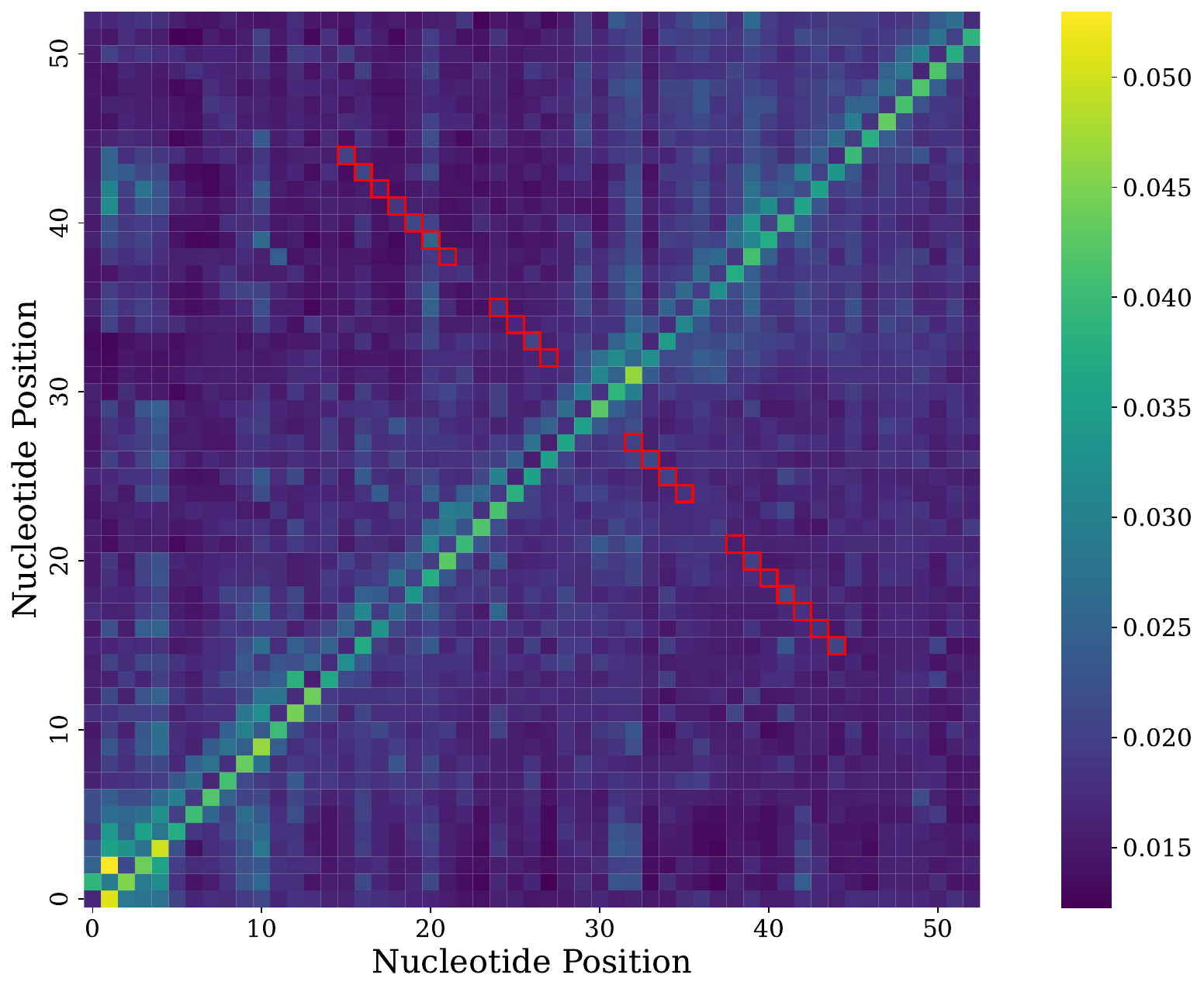}\par
    \caption*{(d)\ RNA-MSM Finetuning}
  \end{minipage}\hfill
  \begin{minipage}{0.32\textwidth}\centering
    \includegraphics[width=\linewidth]{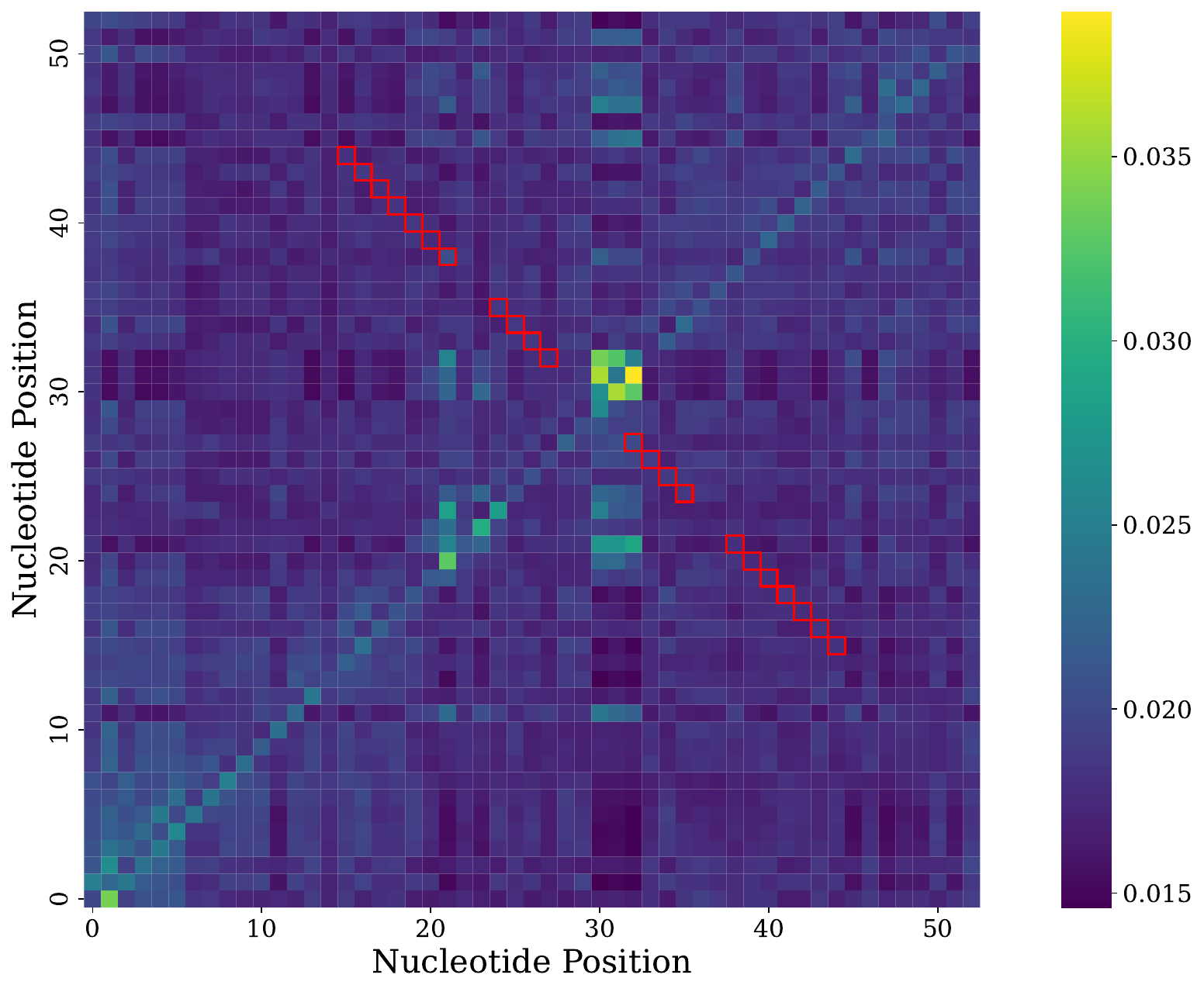}\par
    \caption*{(e)\ RNA-MSM Pretraining}
  \end{minipage}\hfill
  \begin{minipage}{0.32\textwidth}\centering
    \includegraphics[width=\linewidth]{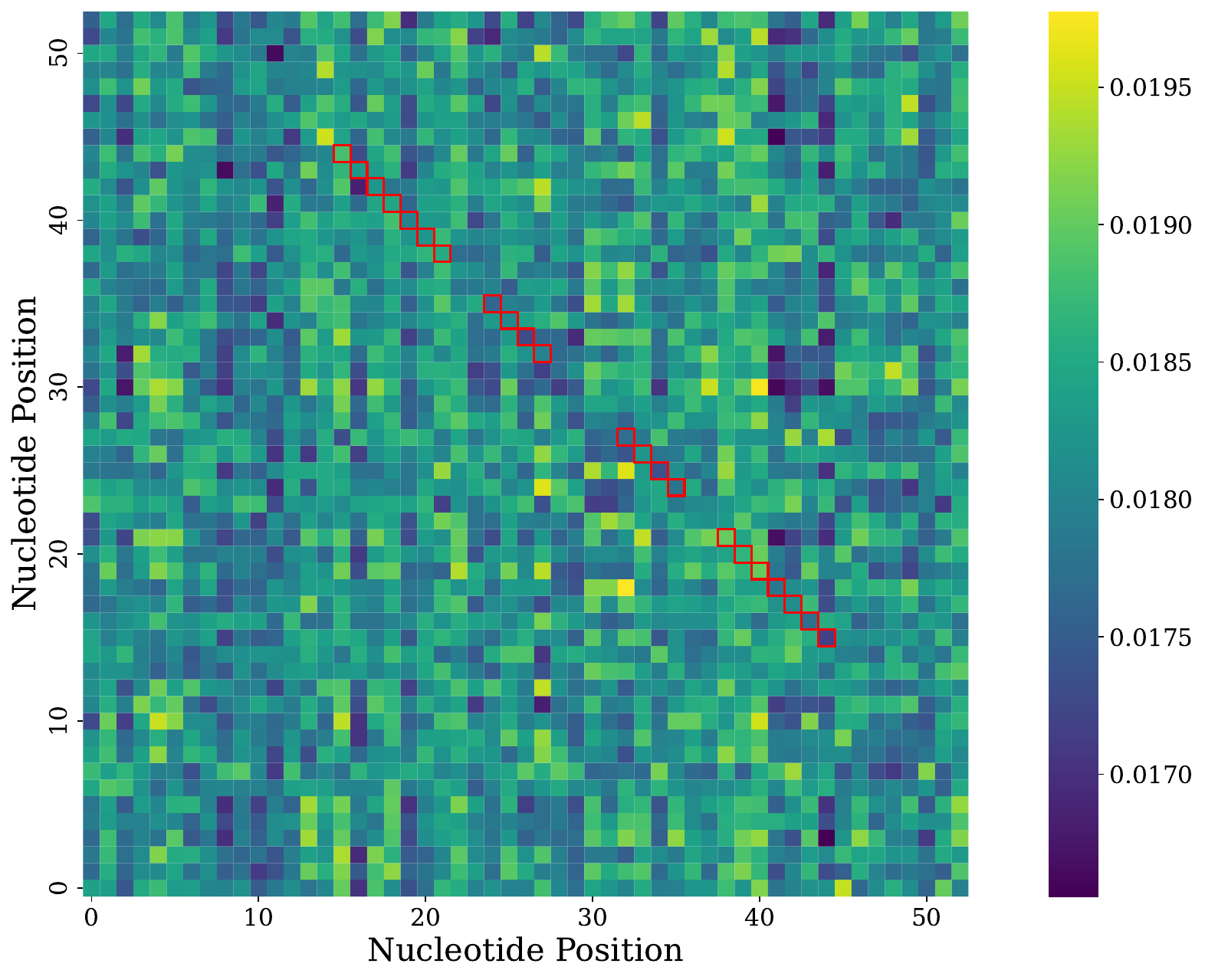}\par
    \caption*{(f)\ RNA-MSM No Training}
  \end{minipage}\vspace{0.5em}

    \begin{minipage}{0.32\textwidth}\centering
    \includegraphics[width=\linewidth]{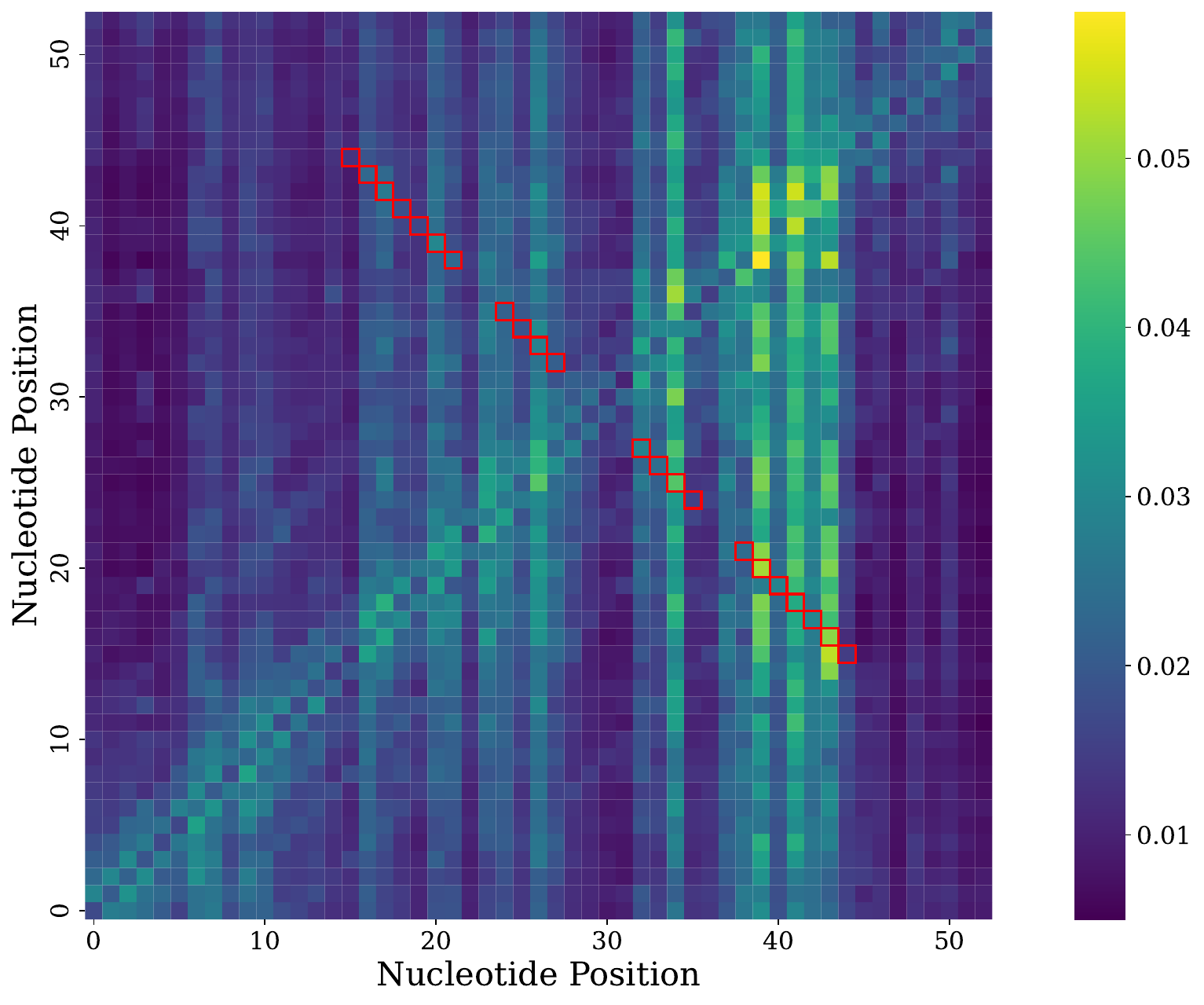}\par
    \caption*{(g)\ RNA-FM Finetuning}
  \end{minipage}\hfill
  \begin{minipage}{0.32\textwidth}\centering
    \includegraphics[width=\linewidth]{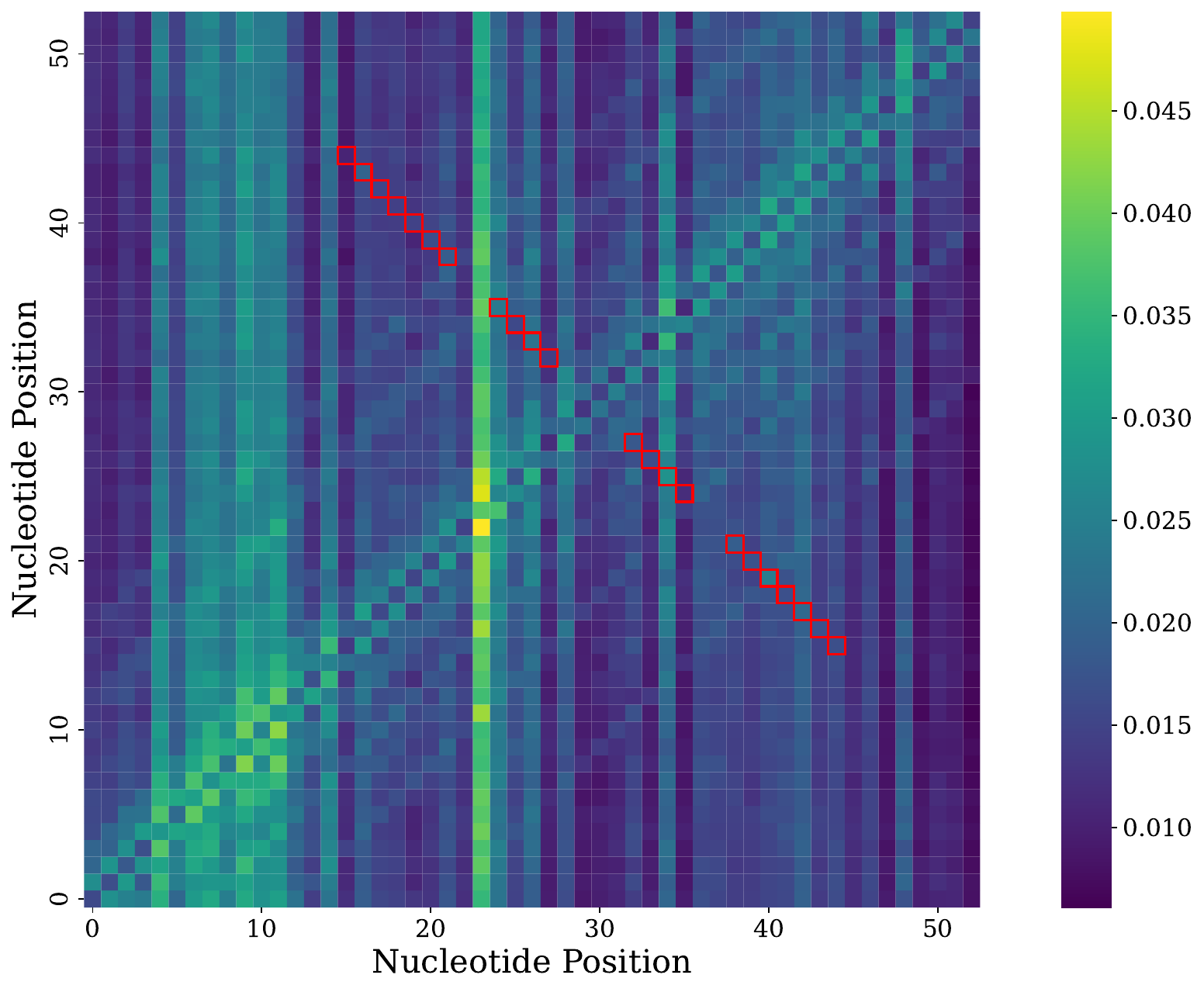}\par
    \caption*{(h)\ RNA-FM Pretraining}
  \end{minipage}\hfill
  \begin{minipage}{0.32\textwidth}\centering
    \includegraphics[width=\linewidth]{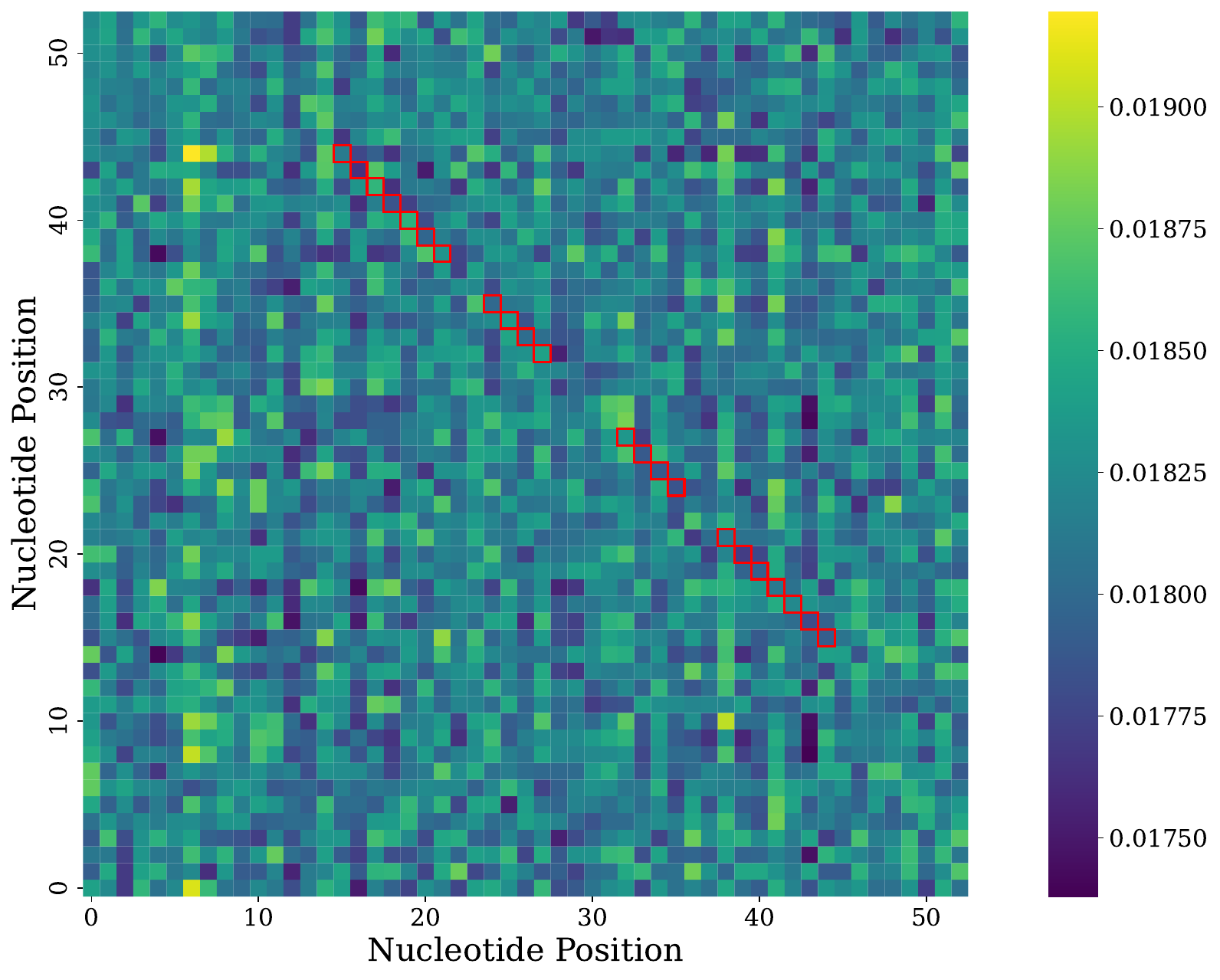}\par
    \caption*{(i)\ RNA-FM No Training}
  \end{minipage}\vspace{0.5em}
    \caption{Attention maps (Example 1) for RNA sequences from the bpRNA-1m benchmark. Each heatmap shows the average self-attention weights in the last encoder layer.
    Red squares denote ground-truth base pairs from the reference RNA secondary structure.
    \og{} (top row) demonstrates strong alignment with biologically meaningful contacts after pretraining and finetuning, whereas RNA-MSM (middle row) shows weaker or scattered patterns.
    Randomly initialized models (rightmost column) exhibit noisy or structure-less attention.
    }
  \label{fig:attn_combined1}
\end{figure}

\begin{figure}[t!]
  \centering
  \begin{minipage}{0.32\textwidth}\centering
    \includegraphics[width=\linewidth]{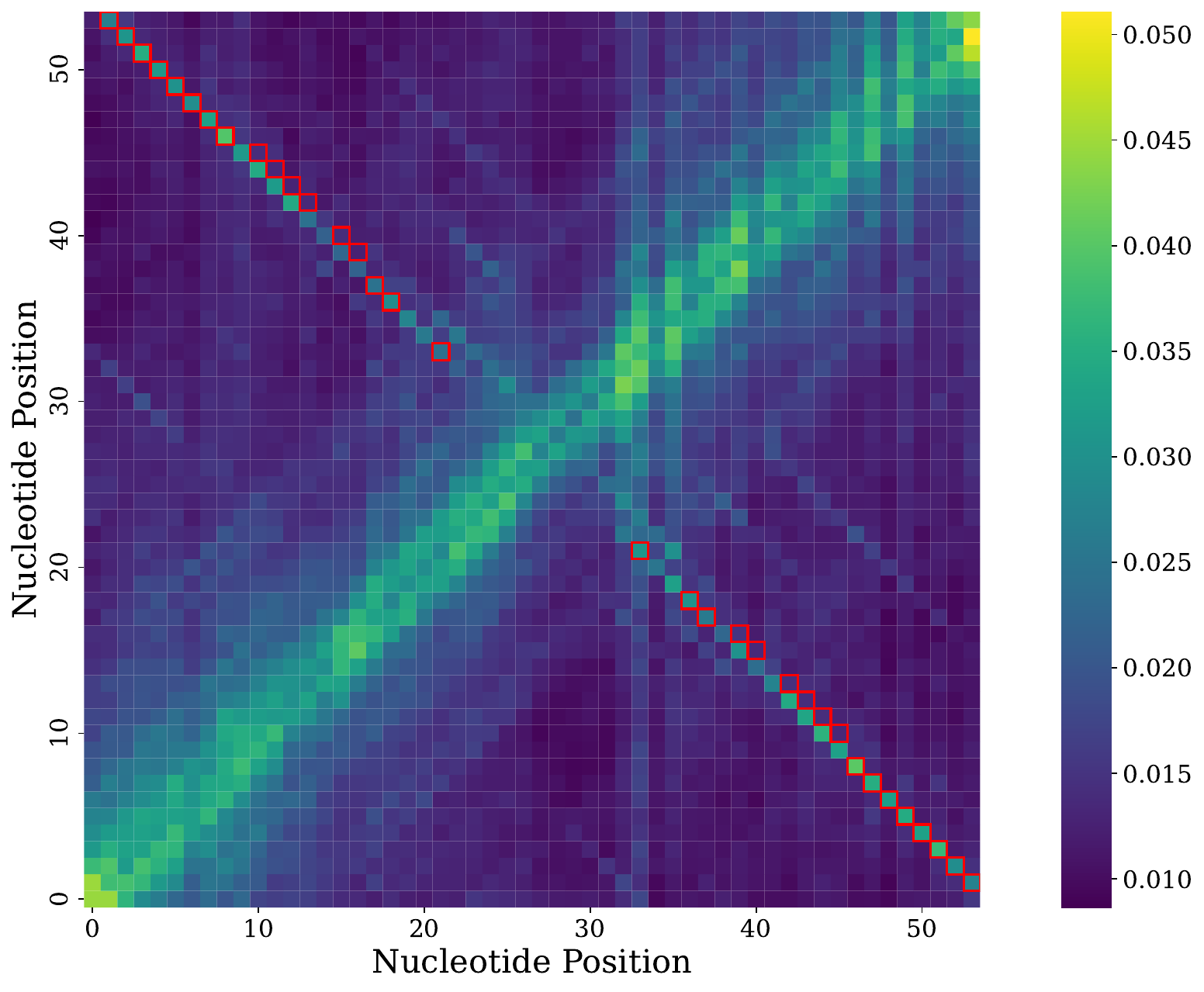}\par
    \caption*{(a)\ \og{} Finetuning}
  \end{minipage}\hfill
  \begin{minipage}{0.32\textwidth}\centering
    \includegraphics[width=\linewidth]{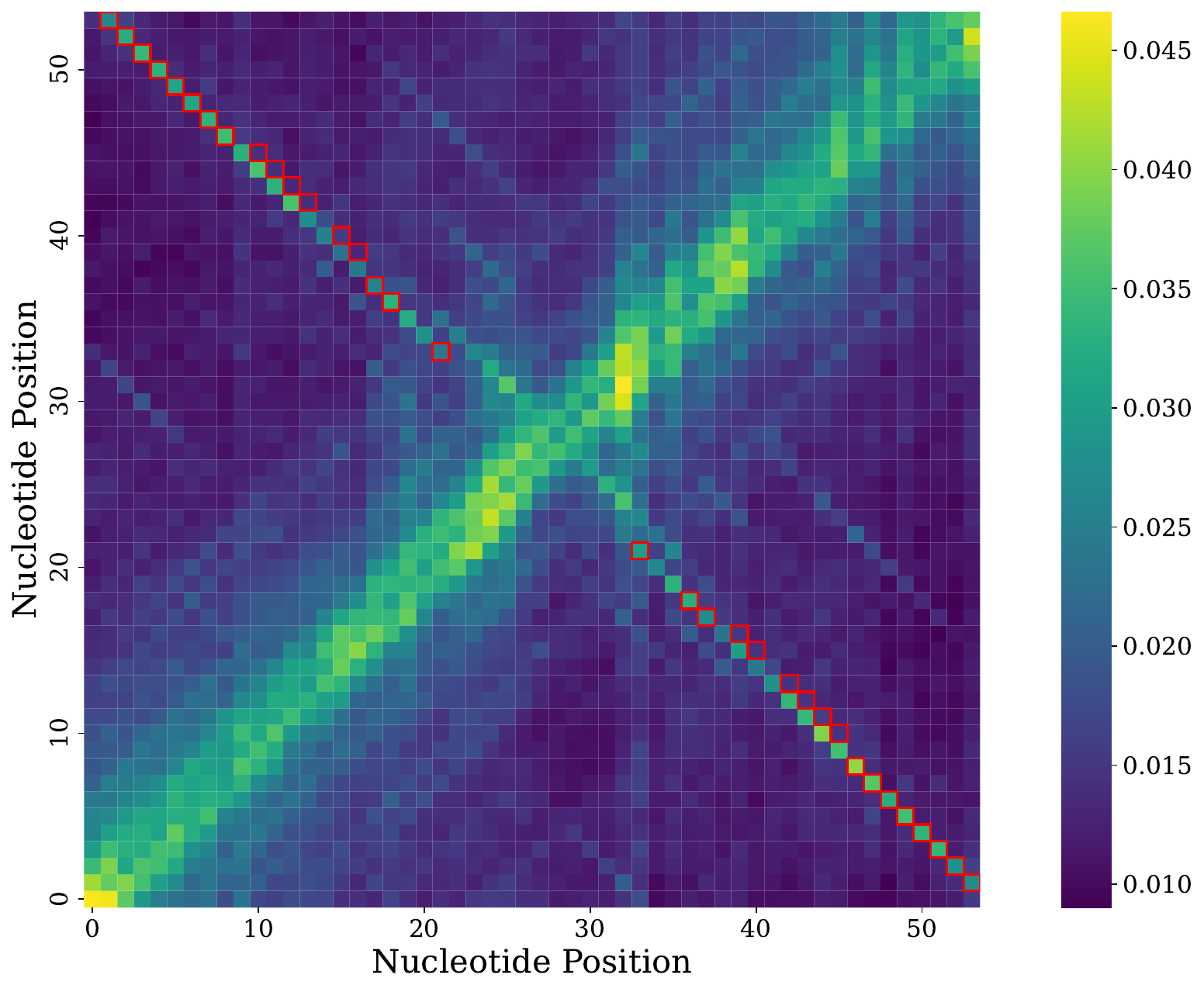}\par
    \caption*{(b)\ \og{} Pretraining}
  \end{minipage}\hfill
  \begin{minipage}{0.32\textwidth}\centering
    \includegraphics[width=\linewidth]{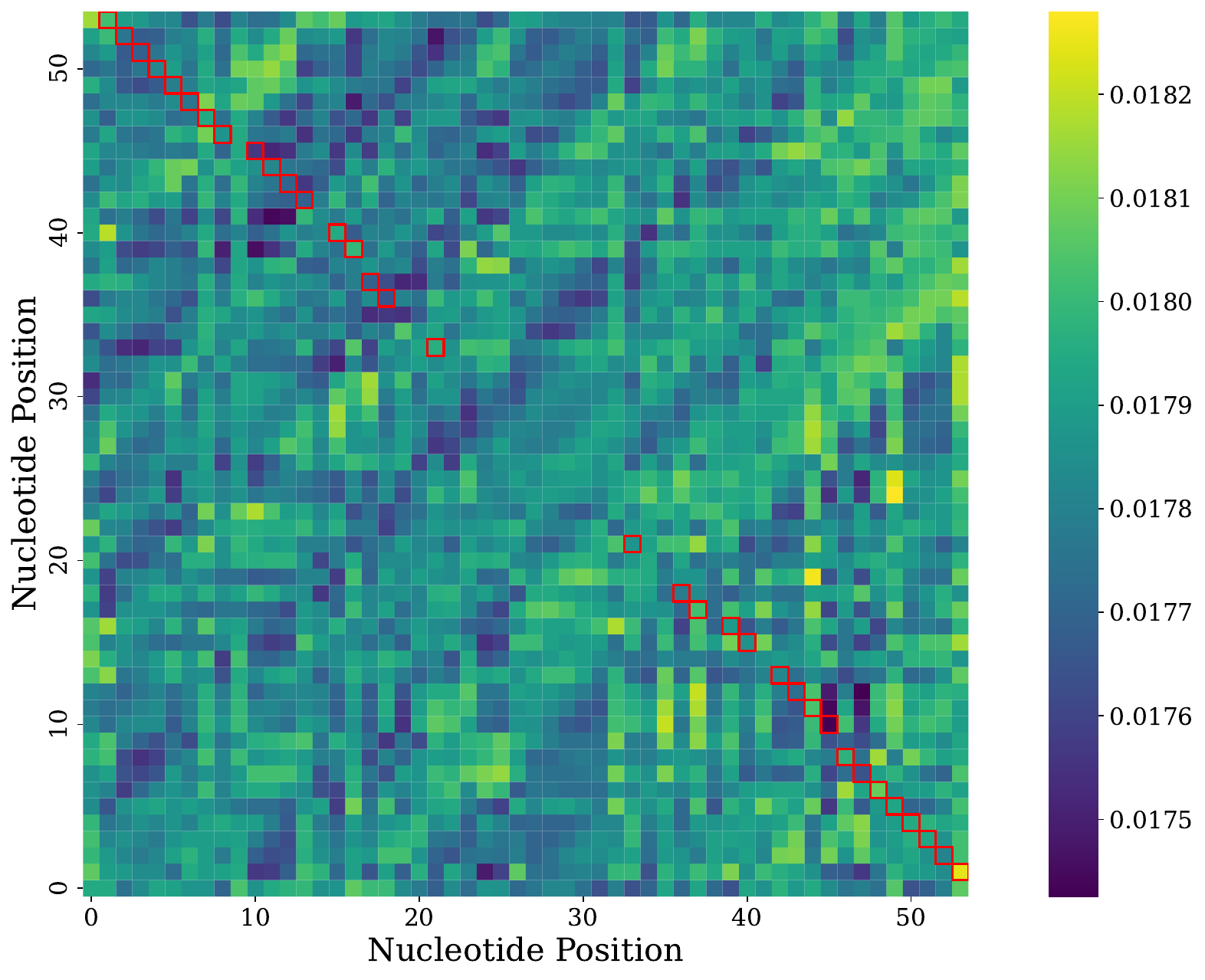}\par
    \caption*{(c)\ \og{} No Training}
  \end{minipage}\vspace{0.5em}

  \begin{minipage}{0.32\textwidth}\centering
    \includegraphics[width=\linewidth]{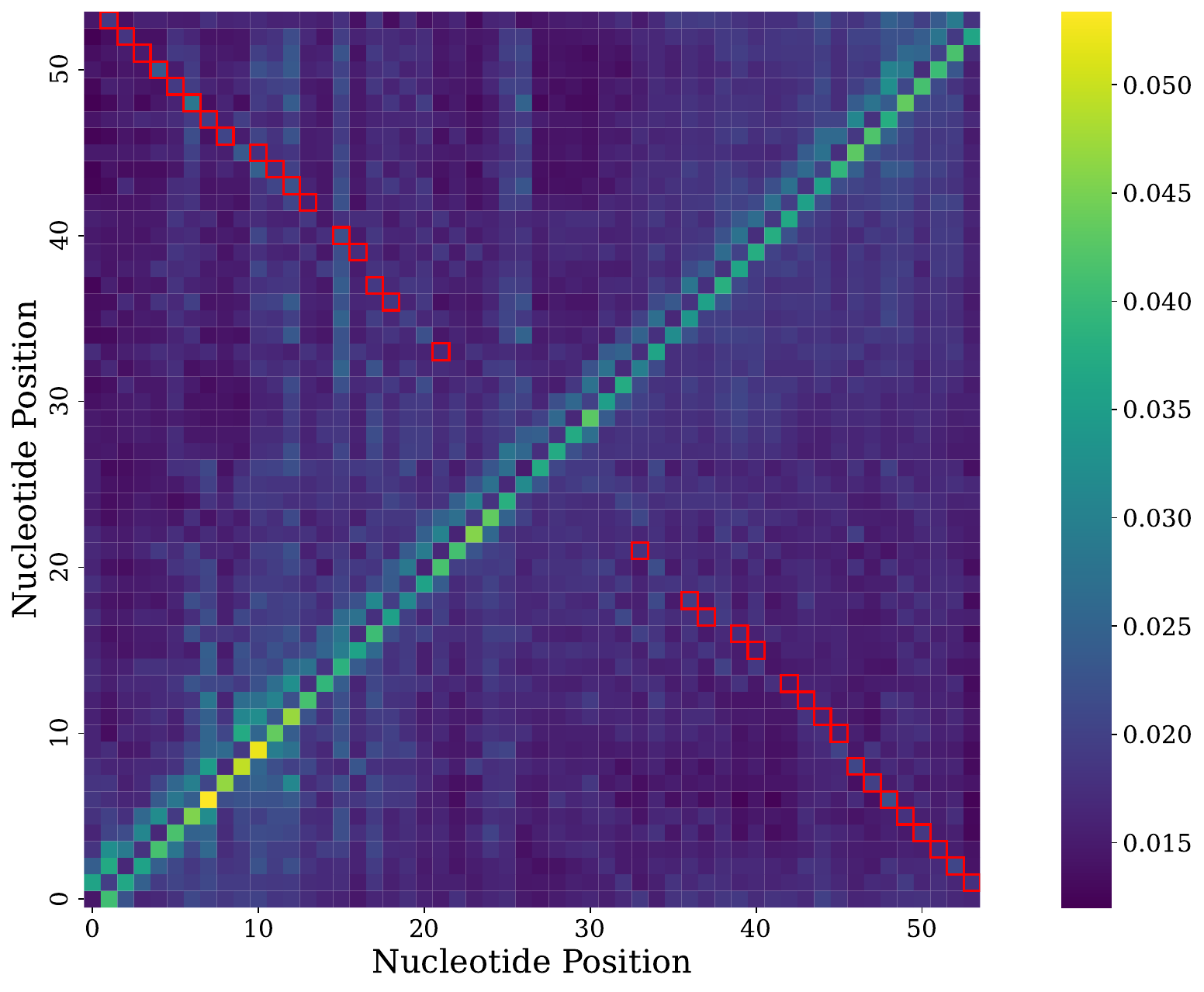}\par
    \caption*{(d)\ RNA-MSM Finetuning}
  \end{minipage}\hfill
  \begin{minipage}{0.32\textwidth}\centering
    \includegraphics[width=\linewidth]{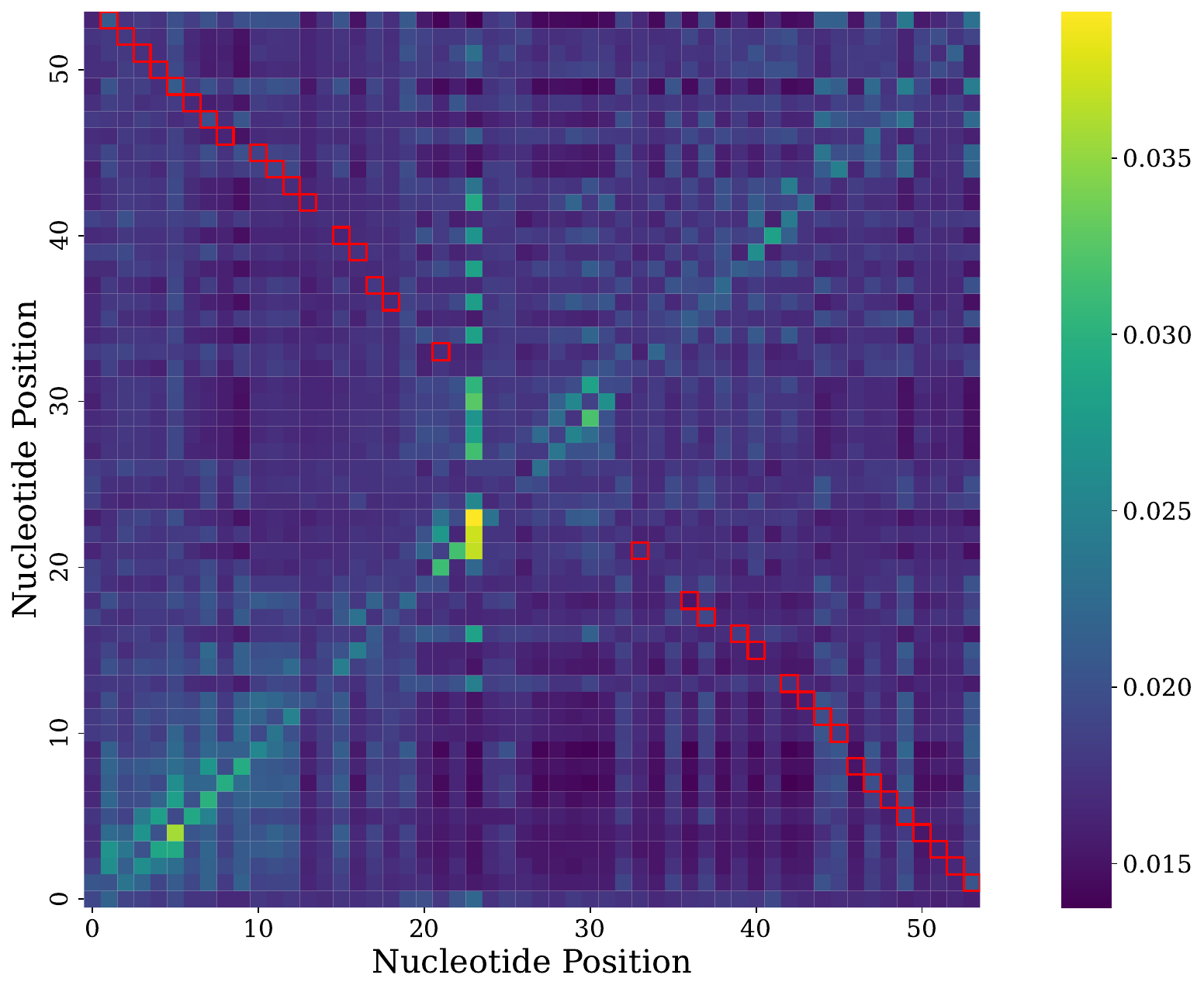}\par
    \caption*{(e)\ RNA-MSM Pretraining}
  \end{minipage}\hfill
  \begin{minipage}{0.32\textwidth}\centering
    \includegraphics[width=\linewidth]{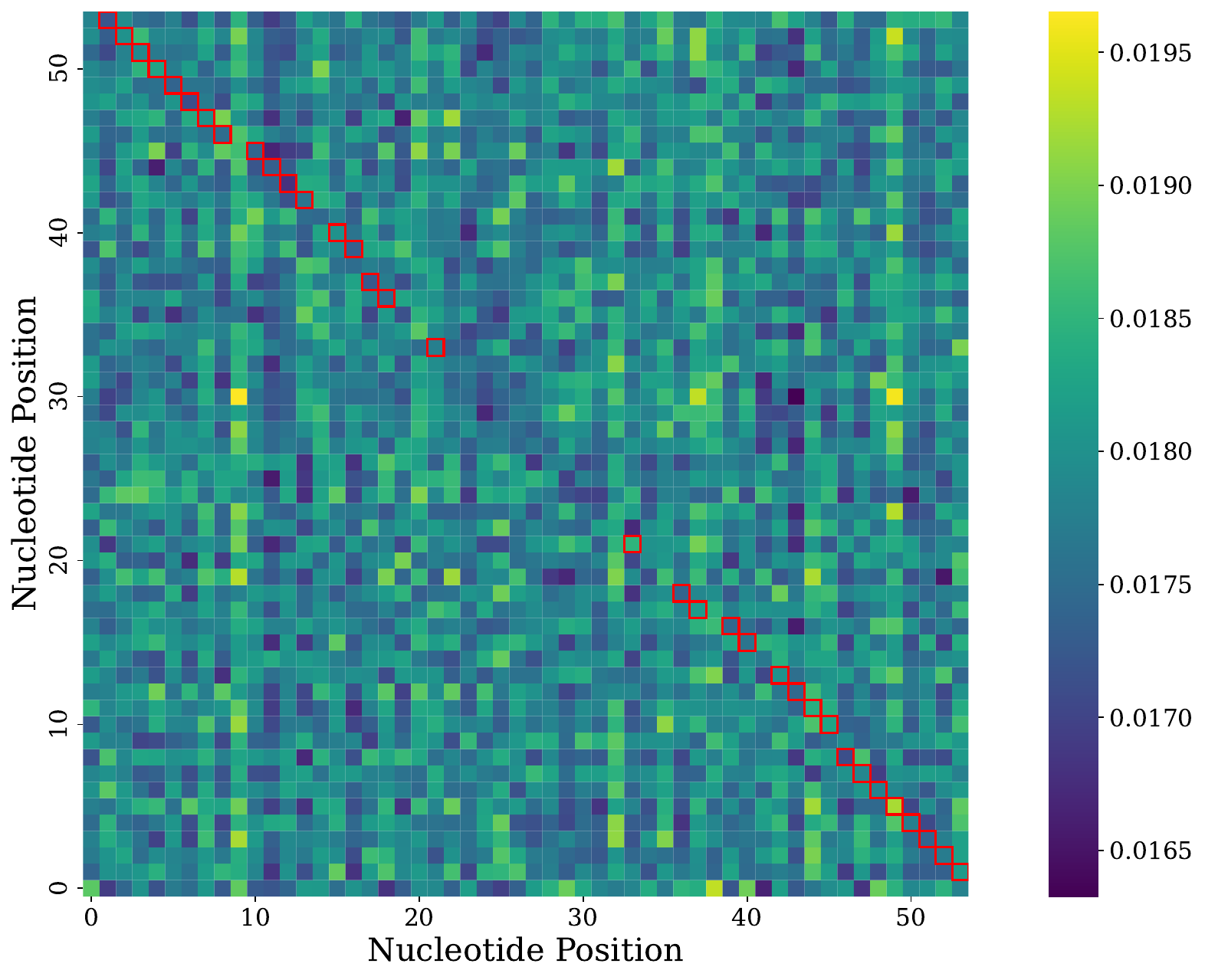}\par
    \caption*{(f)\ RNA-MSM No Training}

  \end{minipage}\vspace{0.5em}
    \begin{minipage}{0.32\textwidth}\centering
    \includegraphics[width=\linewidth]{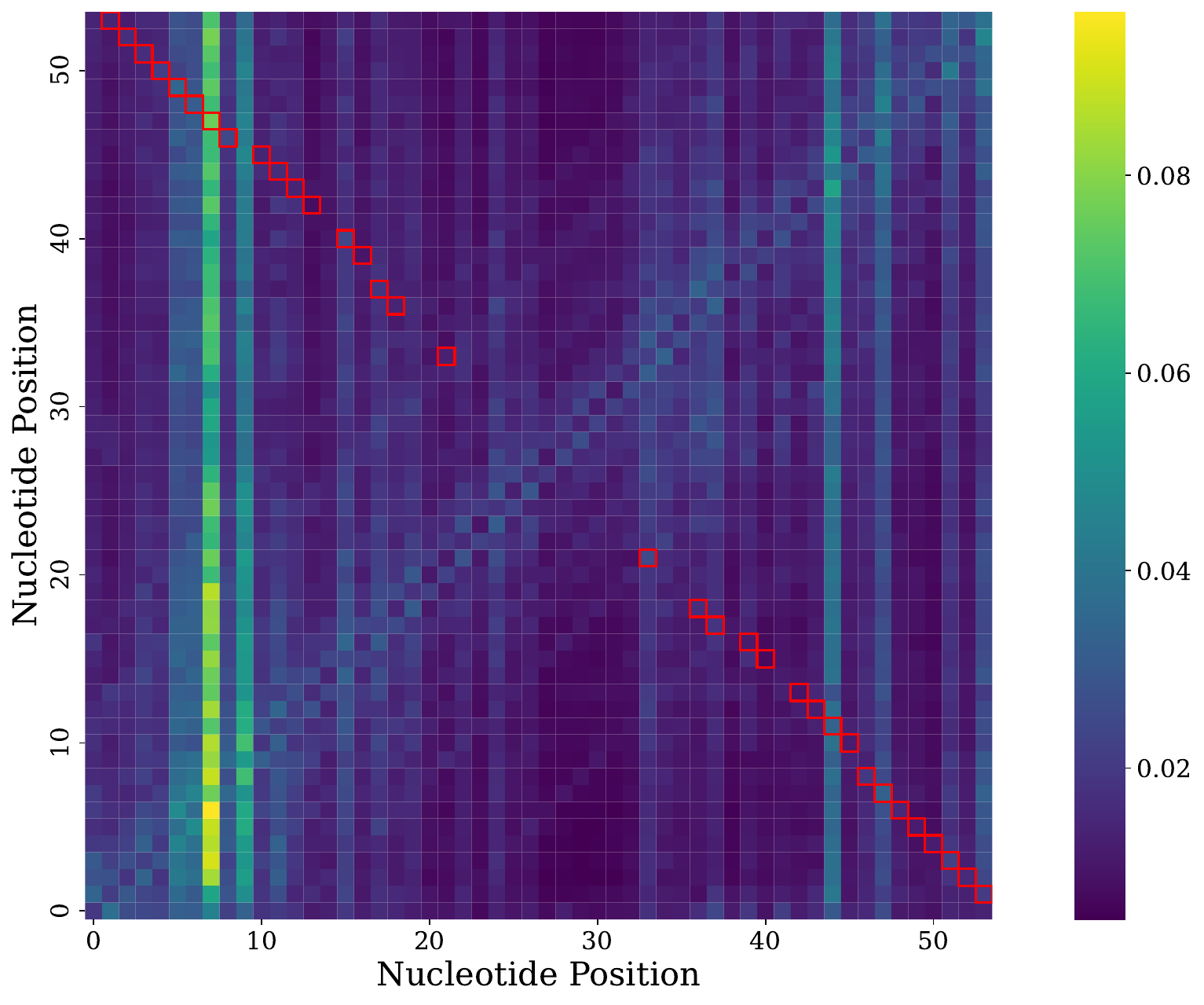}\par
    \caption*{(g)\ RNA-FM Finetuning}
  \end{minipage}\hfill
  \begin{minipage}{0.32\textwidth}\centering
    \includegraphics[width=\linewidth]{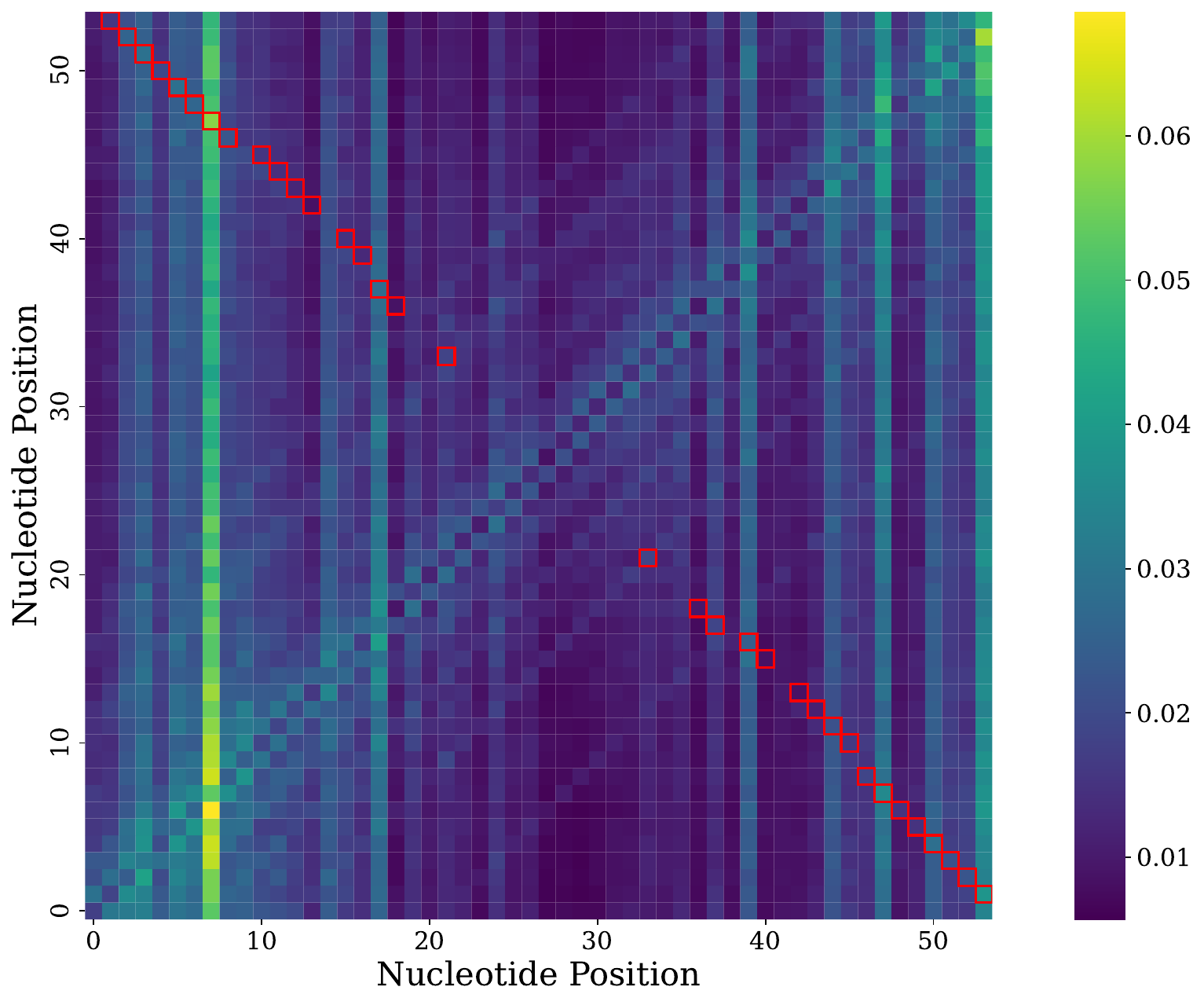}\par
    \caption*{(h)\ RNA-FM Pretraining}
  \end{minipage}\hfill
  \begin{minipage}{0.32\textwidth}\centering
    \includegraphics[width=\linewidth]{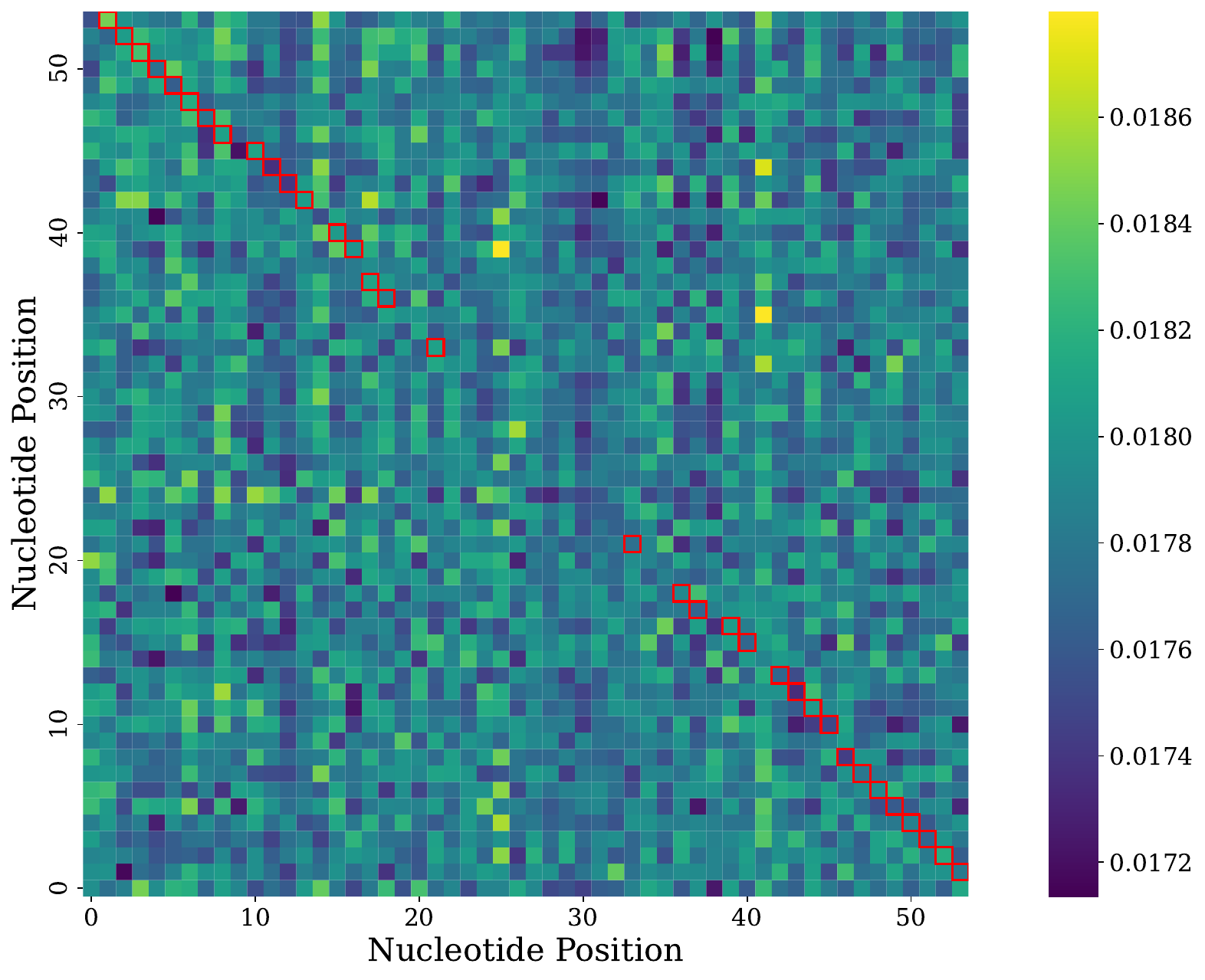}\par
    \caption*{(i)\ RNA-FM No Training}
  \end{minipage}\vspace{0.5em}
  
  \caption{Attention maps (Example 2) for another RNA sequence from the bpRNA-1m dataset. As in~\pref{fig:attn_combined1}, each attention map is overlaid with ground-truth contacts (red squares). \og{} (top row) again attends to meaningful structural regions, even in pretraining alone, while RNA-MSM (middle row) requires fine-tuning to weakly capture stems. Both models with random initialization fail to highlight any coherent secondary structure.}
  \label{fig:attn_combined2}
\end{figure}

For three GFMs, \textbf{\og{}}, \textbf{RNA-MSM}, and
\textbf{RNA-FM}, we plot the average final-layer attention under
three checkpoints: \emph{random}, \emph{pre-trained} (no task labels), and
\emph{fine-tuned} on an RNA task.
Red squares mark ground-truth base pairs (contact distance
$\le8$ Å).  The resulting heat-maps are shown in
\pref{fig:attn_combined1} and \pref{fig:attn_combined2}.

\paragraph{Results.}

\begin{itemize}[leftmargin=*]
\item \textbf{\og{}.}
      Even \emph{before} task supervision (panels\,b) the model attends
      strongly to both local stems and the long off-diagonal contacts in
      Example 2, indicating that its structure-aware pre-training
      already encodes canonical pairing rules.  Fine-tuning (panels\,a)
      further sharpens these tracks, producing attention patterns that
      visually trace the reference contact map.

\item \textbf{RNA-MSM.}
      The random checkpoint is essentially uniform noise (panels\,f).
      Pre-training (panels\,e) introduces sparse, patchy focus—some stem
      residues are highlighted, but many true pairs receive little
      weight.  Fine-tuning (panels\,d) strengthens the main diagonal
      stems of Example 1, yet long-range contacts remain weak or
      absent, suggesting limited inductive bias for global structure.

\item \textbf{RNA-FM.}
      Pre-training (panels\,h) yields clearer vertical/horizontal stripes
      than RNA-MSM, hinting at moderate structural awareness.  After
      fine-tuning (panels\,g) the model emphasises most stem pairs and
      begins to highlight a subset of distal contacts, though the signal
      is less continuous than \og{}'s and occasionally spreads to
      unpaired regions.

\item \textbf{Random baselines.}
      All three architectures with random parameters (panels\,c, f, i)
      display near-uniform, uninformative attention, confirming that the
      structure-aligned patterns above are learned rather than
      architectural artefacts.
\end{itemize}

\paragraph{Conclusion.}
Attention inspection corroborates our other interpretability findings.
\og{}'s explicit, structure-aware pre-training guides the model to
focus on native base pairs even without downstream labels, and
fine-tuning refines this latent alignment into relatively precise contact
patterns.  RNA-FM benefits from pre-training but retains some
off-target noise; RNA-MSM relies more heavily on labelled data and still
struggles with long-range interactions.  In practice, GFMs whose
pre-training embeds secondary-structure priors—such as \og{}, and
to a lesser extent RNA-FM—offer more biologically faithful internal
reasoning and are preferable for structure-sensitive RNA applications.

\subsection{Development Environment}
\label{app:environment}

The benchmark experiments based on \our{} were conducted on a dedicated Linux computation node, equipped with $2$ \textsc{Nvidia} RTX $4090$ GPUs. For distributed model training, we employed version $4.44.0$ of the Transformers library alongside version $0.28.3$ of the Accelerate library. Our implementation framework of choice for \our\ was PyTorch, specifically version $2.1.0$. The ViennaRNA version is $2.6.4$ in our experiments. While some existing code was adapted for the modules within \our, the majority of the codebase, such as genomic sequences preprocessing, model pre-training, objective functions, and experiments, was meticulously crafted from scratch.

\section{Ethical Considerations}

The advancement GFMs, including powerful systems like AlphaFold and other large-scale models (e.g., Evo), necessitates a careful examination of their ethical implications. While GFMs offer unprecedented capabilities for generating, predicting, and interpreting genomic sequences, thereby accelerating breakthroughs in genetic engineering, synthetic biology, and therapeutic development, they also introduce significant ethical challenges that require proactive governance.

We realize some key ethical dimensions in the work, including:

\begin{itemize}[leftmargin=*,nosep]
    \item \textbf{Biosecurity and Dual-Use Risks:} The capacity of GFMs to design and manipulate genetic material at scale, while beneficial for research, also presents dual-use risks. There is a potential for misuse by malicious actors to create synthetic pathogens or other biological threats, thereby impacting global bio-safety. Addressing this requires robust safety protocols, stringent access controls for particularly powerful model capabilities, and mechanisms for auditing model usage and research outputs \cite{carnegie2024,iscience2023}.

    \item \textbf{Equity in Access and Benefit Sharing:} The development and deployment of GFMs could exacerbate existing inequalities. While open-sourcing models like \og{} promote transparency and broader access, there is a risk that well-resourced entities (e.g., large pharmaceutical corporations) will disproportionately benefit, potentially monopolizing discoveries or creating treatments with prohibitive costs. This could widen global health disparities and limit access for researchers and patients in lower-resource settings. Intellectual property frameworks also need careful consideration to ensure fair benefit-sharing \cite{ashg2021,frontiers2023}.

    \item \textbf{Environmental Impact and Ecological Stability:} The enhanced ability to engineer organisms could lead to unintended ecological consequences, such as the disruption of natural ecosystems, loss of biodiversity, or the emergence of invasive or harmful synthetic species. Furthermore, the significant computational resources required for training and deploying large-scale GFMs contribute to an increased carbon footprint, an environmental cost that must be carefully weighed against the scientific and societal benefits \citep{iucn2019,sciencedirect2023}.
\end{itemize}

\section{Societal Impact}
 The societal impact of GFMs is substantial, with applications ranging from personalized medicine to environmental management.

\subsection*{Positive Societal Contributions}

\begin{itemize}[leftmargin=*,nosep]
    \item \textbf{Advancements in Healthcare and Personalized Medicine:} GFMs can significantly accelerate biomedical research by enabling more accurate prediction of disease risk, identification of novel biomarkers, and the design of targeted therapies. This can lead to breakthroughs in precision medicine, tailoring treatments to individual genetic profiles and improving outcomes for complex diseases, including cancer and rare genetic disorders \cite{pmc2022,mdpi2022}.

    \item \textbf{Innovations in Agriculture and Food Security:} In agriculture, GFMs can contribute to the development of crops with enhanced yields, improved nutritional content, and greater resilience to pests, diseases, and climate change. This has the potential to bolster global food security and promote sustainable agricultural practices \cite{wired2024,liebert2023}.

    \item \textbf{Environmental Science and Conservation:} GFMs can aid in understanding ecosystem dynamics, tracking biodiversity, and developing strategies for conservation and bioremediation by analyzing environmental DNA and RNA \cite{iucn2019}.
\end{itemize}

\subsection*{Challenges and Mitigation Strategies}

\begin{itemize}[leftmargin=*,nosep]
    \item \textbf{Risk of Exacerbating Disparities:} A primary concern is that unequal access to GFM technologies and the insights they generate could widen socio-economic and global health divides. Entities with substantial computational resources and technical expertise may gain a significant advantage, potentially leading to an inequitable distribution of benefits \cite{ashg2021,frontiers2023}.

    \item \textbf{Ecological Considerations:} While offering benefits for agriculture and environmental science, the application of GFM-driven genetic modifications requires cautious oversight to prevent unintended negative impacts on ecological balance and biodiversity \cite{iucn2019,sciencedirect2023}.
\end{itemize}

\section{Limitations}
The GFM benchmarking may not reflect the accurate performance in biology reality, we attribute the limitations of benchmarking to two major aspects:
\begin{itemize}[leftmargin=*]

\item \textbf{Lack of \invivo{} Data}: One of the critical limitations of GFMs lies in the absence of \invivo{} verified genome data. While GFMs perform well in \insilico{} environments, where computational models and simulations are used to predict biological processes, these models are rarely validated against \invivo{} data, which refers to experimental data obtained from living organisms. This presents a significant challenge for accurately translating model predictions to real-world biological applications. To be more specific, the complexity of biological systems, including interactions within cells, tissues, and organisms, often introduces variables that are not fully captured in computational simulations. For example, gene regulation, environmental factors, and cellular responses to genetic modifications may behave differently in living organisms than predicted by models trained on \insilico{} data. As a result, GFMs might not fully capture the biological complexity, leading to discrepancies between predicted and actual outcomes.

\item \textbf{Model Scale Constraints}: The second major limitation is the model scales in benchmarking. As GFMs become larger and more sophisticated, their performance improves, but this scaling comes at a significant cost. Training as well as benchmarking large-scale GFMs requires immense computational resources, including high-performance GPUs or TPUs, massive memory allocation, and extensive storage for datasets. The cost of acquiring and maintaining this infrastructure can be prohibitive for many research institutions or companies, limiting access to cutting-edge GFMs.

\end{itemize}

\newpage
\section*{NeurIPS Paper Checklist}

\begin{enumerate}

\item {\bf Claims}
    \item[] Question: Do the main claims made in the abstract and introduction accurately reflect the paper's contributions and scope?
    \item[] Answer: \answerYes{} %
    \item[] Justification: 

\item {\bf Limitations}
    \item[] Question: Does the paper discuss the limitations of the work performed by the authors?
    \item[] Answer: \answerYes{} %
    \item[] Justification: 

\item {\bf Theory assumptions and proofs}
    \item[] Question: For each theoretical result, does the paper provide the full set of assumptions and a complete (and correct) proof?
    \item[] Answer: \answerNA{} %
    \item[] Justification: 

    \item {\bf Experimental result reproducibility}
    \item[] Question: Does the paper fully disclose all the information needed to reproduce the main experimental results of the paper to the extent that it affects the main claims and/or conclusions of the paper (regardless of whether the code and data are provided or not)?
    \item[] Answer: \answerYes{} %
    \item[] Justification: 

\item {\bf Open access to data and code}
    \item[] Question: Does the paper provide open access to the data and code, with sufficient instructions to faithfully reproduce the main experimental results, as described in supplemental material?
    \item[] Answer: \answerYes{} %
    \item[] Justification: 

\item {\bf Experimental setting/details}
    \item[] Question: Does the paper specify all the training and test details (e.g., data splits, hyperparameters, how they were chosen, type of optimizer, etc.) necessary to understand the results?
    \item[] Answer: \answerYes{} %
    \item[] Justification: 

\item {\bf Experiment statistical significance}
    \item[] Question: Does the paper report error bars suitably and correctly defined or other appropriate information about the statistical significance of the experiments?
    \item[] Answer: \answerNo{} %
    \item[] Justification: The evaluation scale is tremendous for repreating in GFMs

\item {\bf Experiments compute resources}
    \item[] Question: For each experiment, does the paper provide sufficient information on the computer resources (type of compute workers, memory, time of execution) needed to reproduce the experiments?
    \item[] Answer: \answerYes{} %
    \item[] Justification: 
    
\item {\bf Code of ethics}
    \item[] Question: Does the research conducted in the paper conform, in every respect, with the NeurIPS Code of Ethics \url{https://neurips.cc/public/EthicsGuidelines}?
    \item[] Answer: \answerYes{} %
    \item[] Justification: 

\item {\bf Broader impacts}
    \item[] Question: Does the paper discuss both potential positive societal impacts and negative societal impacts of the work performed?
    \item[] Answer: \answerYes{} %
    \item[] Justification: 
    
\item {\bf Safeguards}
    \item[] Question: Does the paper describe safeguards that have been put in place for responsible release of data or models that have a high risk for misuse (e.g., pretrained language models, image generators, or scraped datasets)?
    \item[] Answer: \answerNA{} %
    \item[] Justification: 

\item {\bf Licenses for existing assets}
    \item[] Question: Are the creators or original owners of assets (e.g., code, data, models), used in the paper, properly credited and are the license and terms of use explicitly mentioned and properly respected?
    \item[] Answer: \answerYes{} %
    \item[] Justification: 

\item {\bf New assets}
    \item[] Question: Are new assets introduced in the paper well documented and is the documentation provided alongside the assets?
    \item[] Answer: \answerYes{} %
    \item[] Justification: 

\item {\bf Crowdsourcing and research with human subjects}
    \item[] Question: For crowdsourcing experiments and research with human subjects, does the paper include the full text of instructions given to participants and screenshots, if applicable, as well as details about compensation (if any)? 
    \item[] Answer: \answerNA{} %
    \item[] Justification: 

\item {\bf Institutional review board (IRB) approvals or equivalent for research with human subjects}
    \item[] Question: Does the paper describe potential risks incurred by study participants, whether such risks were disclosed to the subjects, and whether Institutional Review Board (IRB) approvals (or an equivalent approval/review based on the requirements of your country or institution) were obtained?
    \item[] Answer: \answerNA{} %
    \item[] Justification: 

\item {\bf Declaration of LLM usage}
    \item[] Question: Does the paper describe the usage of LLMs if it is an important, original, or non-standard component of the core methods in this research? Note that if the LLM is used only for writing, editing, or formatting purposes and does not impact the core methodology, scientific rigorousness, or originality of the research, declaration is not required.
    \item[] Answer: \answerNA{} %
    \item[] Justification: 

\end{enumerate}

\end{document}